\documentclass[aps, rmp, a4paper, twocolumn]{revtex4-1} %showpacs
\usepackage{bbm, amsmath, amssymb, amsthm, bm, textcomp,graphicx,pbox}
\usepackage[main=english]{babel}
\usepackage{hyperref}
\usepackage[utf8]{inputenc}
\usepackage[T1]{fontenc}
\usepackage[normalem]{ulem}

\usepackage{rotating,multirow,colortbl}
\usepackage[table]{xcolor}

\definecolor{Gray}{gray}{0.85}
\definecolor{LightCyan}{rgb}{0.88,1,1}
\newcolumntype{a}{>{\columncolor{Gray}}c}
\newcolumntype{b}{>{\columncolor{white}}c}

\graphicspath{ {./Figs/} }

\def\bra#1{\mathinner{\langle{#1}|}}
\def\ket#1{\mathinner{|{#1}\rangle}}

\def\prjct#1{\mathinner{|{#1}\rangle}\!\!\mathinner{\langle{#1}|}}
\newcommand{\coh}[2]{\mathinner{|{#1}\rangle}\!\!\mathinner{\langle{#2}|}}
\newcommand{\braket}[2]{\langle #1  |#2\rangle}
\def\text#1{\textrm{#1}}
\def\t#1{\textrm{#1}}

\def\cH{\mathcal{H}}
\newcommand\ii{\mathrm{i}}
\newcommand{\be}{\begin{equation}}
\newcommand{\ee}{\end{equation}}
\theoremstyle{definition}
\newtheorem{Exp}{Example}
\newcommand{\alive}{$| \mathcal{A} \rangle $ }
\newcommand{\dead}{$| \mathcal{D} \rangle $ }
\newcommand{\da}{$| \mathcal{A} \rangle + | \mathcal{D} \rangle $ }

\begin{document}
\title{Macroscopic quantum states: measures, fragility and implementations}

\author{Florian Fr\"owis$^1$}
\thanks{These authors contributed equally.}
\author{Pavel Sekatski$^2$}
\thanks{These authors contributed equally.}
\author{Wolfgang D\"ur$^2$}
\email{Electronic address: macro.quant.rmp@gmail.com}
\author{Nicolas Gisin$^1$}
\author{Nicolas Sangouard$^3$}
\affiliation{$^1$Group of Applied Physics - Université de Genève - CH-1211 Geneva - Switzerland}
\affiliation{$^2$Institut für Theoretische Physik - Universit\"at Innsbruck - Technikerstra{\ss}e 21a - A-6020 Innsbruck - Austria}
\affiliation{$^3$Quantum Optics Theory Group - Department of Physics - University of Basel - CH-4056 Basel - Switzerland}

\begin{abstract}Large-scale quantum effects have always played an important role in the foundations of quantum theory. With recent experimental progress and the aspiration for quantum enhanced applications, the interest in macroscopic quantum effects has been reinforced. In this review, we critically analyze and discuss measures aiming to quantify various aspects of macroscopic quantumness. We survey recent results on the difficulties and prospects to create, maintain and detect macroscopic quantum states. The role of macroscopic quantum states in foundational questions as well as practical applications is outlined. Finally, we present past and on-going experimental advances aiming to generate and observe macroscopic quantum states.
\end{abstract}
\date{\today}

\maketitle
\tableofcontents
%%%%%%%%%%%%%%%

\section{Introduction}
\label{sec:introduction}

With recent progresses in experimental physics, it is nowadays possible to investigate quantum effects such as interference and entanglement in larger and larger systems. Experimentalists  bring mechanical oscillators to the quantum regime, set interferometers with single giant molecules, produce superposition states with many atoms, photons and high superconducting currents, or reveal entanglement in many-body systems.\footnote{See Sec.~\ref{sec:implementations} for references and further details.}
Starting with \citet{Leggett_Macroscopic_1980}, many physicists came up with measures to compare these experiments, that is, to quantify how macroscopic and quantum a state is.
 Such measures allow one to characterize sets of states and to study systematically the requirements to observe the quantum features of macroscopic states. From a fundamental point of view, this helps to gain insight into the quantum-to-classical transition and to investigate the limits of quantum theory. From a more applied perspective, this is useful to reveal general mechanisms for quantum enhancement in applications such as quantum computing and metrology. \\

 It should be emphasized, though, that identifying key features of macroscopic quantumness is highly controversial. Intuitively, any approach should distinguish a genuine macroscopic quantum effect from accumulated microscopic effects. However, already the precise meaning of these and similar words is unclear and disputed, also because they are heavily loaded with emotions and prejudice.
 The example from \citet{Schrodinger_Present_1935} of a cat in superposition of being dead and alive is a paradigmatic starting point for many considerations and experiments with different physical systems. But we face many open questions. The first issue involves the role of the physical system (atoms, electrons, photons, etc.) and the different degrees of freedom. In all cases, a reduction of complexity, often to a single degree of freedom, is considered in order to keep essential properties while making it theoretically tractable and bringing it closer to experimental reality. While on an abstract level states might be isomorphic, there is no consensus whether a superposition state with different spin values or different positions of the wave packet can equally well be called a macroscopic superposition.
 The latter for instance is affected by gravitational collapse and allows to test proposed modifications of quantum mechanics, while the former is not. Another issue is how particle number, distance and mass enter in the assessment. How can we compare the spatial superposition of a single atom being 1 m apart to a Bose-Einstein condensate with one million atoms where the center-of-mass is separated by 1 $\mu$m? To address these issues, one needs to formalize the observation that ``dead and alive'' are more than two orthogonal states in a Hilbert space but are somehow ``macroscopically distinct''. Furthermore, it is unclear how to take into account loss of coherence (i.e., purity). Is there a way to deal with reduced visibility when scaling-up the system size? Finally, the quest does not end with macroscopic superpositions of two states. Generalizations to arbitrary quantum states including mixed states are important to further abstract the problem and to apply the theoretical concepts directly to experiments.

 In this review we summarize and discuss proposals to measure quantum states (or entire experiments) concerning some aspect of macroscopic quantumness. We do not only aim to give a technical summary of the measures, but also to facilitate a discussion of the motivation and intuition behind them, as well as relations between the different proposals. We then discuss fundamental difficulties and limitations to prepare, maintain and certify macroscopic quantum states, and briefly mention potential applications in foundations of quantum mechanics as well as quantum metrology and quantum computation.
 Finally, we evaluate the current status of experimental progress by reviewing experiments with different systems, and applying different measures and proposals to assess their level of macroscopic quantumness.
 In the remainder of the introduction, we specify more precisely the scope of the review, clarify the motivation and the terminology and mention the structure of this paper.

\subsection{Defining macroscopic quantumness: not an easy task}
\label{sec:not-an-easy}

It is often claimed that quantum mechanics is one of the most successful theories in physics. The basis of this assertion is its passing of all experimental tests so far. This is certainly true in the microscopic realm. Here, we are interested in large systems, that is, experiments involving many atoms, photons or electrons. There, the experimental evidence and its interpretation are less clear. As mentioned by several physicists such as \textcite{Leggett_Macroscopic_1980}, many well-established large-scale experiments can be seen as a macroscopic accumulation of \textit{microscopic quantum effects}. As an example, the genuine quantum effect of Cooper pair formation in the BCS theory of superconductivity is a two-electron problem and many-body correlations are not necessary to observe superconductivity on human scales \cite{Leggett_Macroscopic_1980}.
Hence, as opposed to microscopic quantum effect, one could define a \textit{macroscopic quantum effect} as a situation in which the experimental evidence excludes a model based on an accumulated microscopic quantum effect (regarding terminology, see Sec.~\ref{sec:preliminary-remarks}).
However, this definition does not seem to be restrictive enough, as it can be fulfilled in drastically different situations, some of which defy our intuition. Let us illustrate this point with the two following examples.

Consider a large atomic ensemble that was prepared in the ground state and coherently absorbed a single photon. The resulting atomic state is in a so-called Dicke state or W state \cite{Dicke_Coherence_1954,Dur_Three_2000}, that is, a coherent superposition of all states where exactly one atom has absorbed the photon. This state is genuinely multipartite entangled, that is, nonseparable for any bipartition of the ensemble.
According to the previous definition, genuine multipartite entanglement between all atoms constitutes a genuine macroscopic quantum effect. In particular, this has direct observable consequences if we consider the spontaneous reemission of the photon. The coherent  phase relation between the atoms leads to a temporally and directionally well-defined emission of the photon \cite{Dicke_Coherence_1954,Duan_Long-distance_2001,Scully_Directed_2006}. On the other side, if one divides the ensemble into several groups without a shared phase relation (and thus losing the entanglement between the groups), the spatial emission pattern progressively becomes isotropic. Yet this quantum effect has a completely different character than the following example.

In his seminal paper, \citet{Schrodinger_Present_1935} sketched the famous cat paradox, where the total system consists of an atom with two levels (ground $\ket{g}$ and excited $\ket{e}$) and a cat, which is coupled to the atom via a mechanism $U$ that kills the cat whenever the excited atom decays to the ground state. By applying the linearity of quantum mechanics even to macroscopic scales, Schrödinger argued that the coherent superposition of the atom being excited and already decayed results in a \textit{micro-macro entanglement} between atom and cat
\begin{align}
\label{eqI:1}
 U \ket{\text{Alive}}\otimes \left( \ket{e} + \ket{g}\right)\to\\
\ket{e}\otimes \ket{\mathrm{Alive} }+\ket{g}\otimes \ket{\mathrm{Dead}}.
\end{align}
This paradox challenges our world view much more than the example of the absorbed single photon, even though the size of the atomic ensemble can be truly macroscopic. But what are the essential features of this example that cause the unease of Schrödinger and many others? We now list some aspects frequently appearing in the literature that is discussed in this review.

\label{sec:first-classification}

(1) The superposition principle is one of the most straightforward illustrations of the drastic difference between classical and quantum physics. For any two possible quantum states $\ket{\mathcal{A}}$ and $\ket{\mathcal{D}}$, the superposition $\ket{\mathcal{A}} \pm \ket{\mathcal{D}}$ is also a valid quantum state.
This is a well accepted fact for microscopic systems or even for macroscopic systems if $\ket{\mathcal{A}}$ and $\ket{\mathcal{D}}$ are ``hardly'' distinguishable. However, the superposition principle appears paradoxical when $\ket{\mathcal{A}}$ and $\ket{\mathcal{D}}$ represent states that are drastically different or irreconcilable in classical physics, like an object being here and there or a cat being dead and alive. This is a key feature in Schrödinger's example, as the two superposed states $\ket{\t{Alive}}$ and $\ket{\t{Dead}}$ are not only orthogonal but also \emph{macroscopically distinct}, using a term coined by \citet{Leggett_Macroscopic_1980}.

(2) The number of biological cells in the cat is in the order of trillions; the number of atoms in the order of $10^{26}$. But it is not only the bare number of constituents. In the biological cat many degrees of freedom (or, modes) are ``acitve'', that is, are accessible via interactions between them. This leads to an enormous complexity within an unthinkably large Hilbert space. Some physicists see the presence of this large number of degrees of freedom (and not only a large number of particles) as a necessary perquisite for quantum systems to be called macroscopically quantum \cite{Shimizu_Stability_2002}. Others tend to drop this aspect when going to realistic systems like Bose-Einstein condensates or superconducting devices (see Sec.~\ref{sec:implementations}), where only one or few modes effectively exist.

(3) Finally, an aspect that is often emphasized is the micro-macro entanglement between the atom and the cat. It is a priori not clear what happens if --in a bipartite scenario-- the system size of one party is increased. Several experiments in quantum optics and optomechanics aim to realize this aspect of Schrödinger's cat (see discussion later). Also some of the proposals presented in this review reflect this idea.

In this review, we will focus on recent contributions that aim for a more systematic approach to the topic of macroscopic quantum mechanics. In many papers, the Schrödinger-cat paradox serves as a starting point to formalize concepts such as macroscopic quantum effects or macroscopic quantum states. There is a broad agreement in the literature that coherence between macroscopically distinct states (1) has to be necessarily present in experiments that aim to mimic the Schrödinger's cat example. In contrast, high complexity in terms of many accessible degrees of freedom (2) is, for many but not for all authors, neither necessary nor sufficient. Likewise, the micro-macro entanglement (3) plays only a minor role in the papers discussed here.

\subsection{Motivation}
\label{sec:motivation}

The theoretical and experimental study of macroscopic quantum systems is motivated by a wide range of interests. Many open questions in the foundations of quantum mechanics touch on quantumness on large scales, the quantum-to-classical transition and the measurement problem. This includes comparisons of standard quantum mechanics against potential modifications relevant on large scales but also a better understanding of quantum mechanics itself. The latter is expected to have a cross fertilization with applications of quantum mechanics that are particularly interesting when performed with large quantum systems (e.g., for quantum computation and metrology). Let us discuss some of these points in more depth.

Historically, Schr\"odinger's motivation for his paradox was to demonstrate the interpretational problems of what he calls the ``blurriness'' of the wave function \cite{Schrodinger_Present_1935}. He argued that at a scale of a radioactive nucleus one might be able to accept that the state of a quantum system cannot be described by a well-defined collection of properties like position, momentum, excitation level etc. The speculative reason is that we anyway cannot directly access these small scales and everyday intuition breaks down. However --assuming full validity of quantum mechanics also at large scales-- one can easily construct examples where the initial microscopic blurriness is translated to human scales. In Schr\"odinger's example, it is the coherent superposition of a cat that is dead and alive correlated with a radioactive substance being decayed and excited. We can easily determine the basic vital function of a cat, which prevents us to accept a ``blurred model'' as an accurate picture of reality.

Very generally, there is a natural motivation to formalize this intuition and essential aspects of Schrödinger's example into a mathematically solid and abstract tool. This would allow us to benchmark experimental progress, tell us at which scales we did verify quantum laws, and maybe get closer to the original question: Are quantum laws, such as the superposition principle, valid or at least observable at all scales? In fact, one might hear different answers to this question, which all provide motivation to study macroscopic quantumness.

The common intuition is that, while nature might allow for quantum effects on macroscopic scale, it makes them practically impossible to observe. This is due to technical limitations that forbid one to perfectly isolate a system from its environment and to perform measurements with unlimited precision. This leads to an \emph{effective} quantum-to-classical transition, which can be ideally derived from the quantum laws themselves \cite{Zurek_Decoherence_2003,Joos_Decoherence_2003}.

One might also take a more radical attitude on this question, saying that nature prohibits even the existence of Schrödinger cat states, such that quantum laws have to be supplemented with an explicit collapse mechanism. This is commonly done by introducing a stochastic extension of the Schrödinger equation. This extension basically does not affect any microscopic quantum system composed of a few atoms. Large masses and distances, however, lead to an efficient collapse whenever the wave function is widely spread over a characteristic amount of time (see \citet{Bassi_Models_2013,Arndt_Testing_2014} and references therein). Collapse models are a way to stay within an extended quantum theory and at the same time avoiding paradoxes à la Schrödinger. Hence, the experimental verification of superposing two macroscopically distinct states (in position space) is a typical way to falsify such modifications (however, see the discussion in Sec.~\ref{sec:cat-regard-purp} about ``more suited'' states to test collapse models under realistic conditions).

Along the same lines an additional motivation to study macroscopically distinct states stems from the measurement problem, that is, the appearance of a single measurement outcome irreversibly chosen among all possible outcomes \cite{Schlosshauer_Decoherence_2005}. In particular, in standard quantum mechanics (Copenhagen interpretation) the problem is explicitly solved by introducing the measurement postulate -- the ability of the measurement apparatus to collapse the wave function. However, this postulate also has provoked controversial discussions, as it is not a priori clear what qualifies to be a measurement apparatus. It is supposed to be a macroscopic device obeying the laws of classical physics, but down the line any such device is made of atoms -- quantum systems with reversible unitary dynamics. Where should this boundary between microscopic and macroscopic (quantum and classical), postulated by Copenhagen, be found? On the opposite side, the many-world interpretation deprives the measurement process from such a special role \cite{Everett_"Relative_1957,Wheeler_Assessment_1957}. In this theory, a measurement is just a particular case of unitary dynamics in which the measurement apparatus entangles with the system. Hence, all possible outcomes, in fact, occur, as the state of the system and the apparatus after the coupling is a superposition of all possible outcomes. Nevertheless, we, conscious beings, have the impression to live in a world where only a single outcome occurs. This discrepancy also provoked many discussions, summarized in the famous Wigner's friend paradox \cite{Wigner_Remarks_1961}. Schrödinger's thought experiment is of course at the heart of this debate \cite{Leggett_Testing_2002,Schlosshauer_Decoherence_2005}.

Besides the interest in the transition from quantum mechanics to our classical world, effective or not, it is important to understand the structure of quantum mechanics itself when applied to large systems. In entanglement theory, the structure of multipartite entanglement becomes richer and more complex as the number of parties increases. A similar behavior is expected for other aspects of quantum mechanics when brought to large scales. Although we here review papers that are often in the vicinity of Schrödinger's cat example, this can be seen as just one out of many interesting aspects of macroscopic quantumness. In general, the systematic study leads to new insight on quantum effects, new proposals for experiments and constraints on the experimental requirements to prepare, maintain and observe macroscopic quantumness.

Such a broad account might give general insight also useful for applications of quantum mechanics. In quantum computation and quantum metrology, for example, the performance of algorithms and protocols is often measured as a function of the system size \nocite{Nielsen_Quantum_2000,Giovannetti_Advances_2011}. Quantum advantage is particularly striking for large system sizes. Connections made between foundations and applications can lead to a new point of view and additional understanding of the mechanisms (e.g., see Sec.~\ref{sec:stat-occurr-quant}).

\subsection{Terminology}
\label{sec:preliminary-remarks}

A primary obstacle when discussing the present topic is the used terminology. First, as already mentioned before, the topic is strongly filled with emotions and preconceptions. Second, different authors use words like macroscopic or large in different ways. Hence, on the one hand we prefer to avoid loaded terminology such as Schrödinger-cat states, but on the other we cannot completely ignore the common terminology. Let us clarify how we here use some frequently used words.

In this review, \textit{macroscopic}\footnote{There are even papers on the term macroscopic in the foundations of quantum mechanics \cite{Jaeger_What_2014}.} is a synonym for large. All physical systems considered here necessarily consist of a large number of microscopic constituents (e.g., atoms, electrons or photons). This is referred to as the \textit{system size} or \textit{number of constituents}. However, the system size is not necessarily in the order of $10^{23}$ or even close to it. While it could be misleading to use the word macroscopic even for ``mesoscopic'' system sizes, it frequently appears in many of the discussed papers. Note that macroscopic can refer to a scaling or to a single number, depending on the context and the intention of the reviewed papers. On an abstract level, one might prefer to discuss the properties of a state family and the behavior as a function of the system size. Given a specific situation like real data from an experiment, one is probably more interested in extracting the bare numbers than the hypothetical scaling.

As stated before, we are interested in \textit{macroscopic quantumness}, which is more than a quantum state of a macroscopic system. Determining the precise meaning of macroscopic quantumness is the goal of most of the contributions discussed in this review. This cannot be captured by a single characteristic trait but the problem is expected to be multidimensional. Thus, a peaceful coexistence of different ideas is likely to be possible. Very generally, different concepts could be referred to as macroscopic quantumness. In this review however, macroscopic quantumness means quantum coherence between \textit{macroscopically distinct} states inspired by point (1) in Sec.~\ref{sec:not-an-easy}. We emphasize that the two aspects --macroscopicity and quantumness-- are sometimes separately studied, sometimes they are combined to a single concept.
Being one aspect of macroscopic quantumness, quantum coherence between far distant parts in the spectrum of a given observable is called \textit{macroscopic coherence}.

In many papers, the authors are interested in finding a function $f$ that assigns a nonnegative number $f(\rho) \geq 0$ to a quantum state $\rho$. This number should ideally reflect the degree of macroscopic quantumness of $\rho$. In this review, we call $f$ a \textit{measure (of macroscopic quantumness)} and $f(\rho)$ is called \textit{effective size} (or simply \textit{size}) of $\rho$. For a unified notation, we name the measures by the authors who first proposed them, which is also used to title the subsections in Sec.~\ref{sec:preliminary-measures}. We generally keep the mathematical symbols for the measures as introduced in the original papers (see table \ref{tab:overview2}).

\subsection{Physical systems}
\label{sec:physical-systems}

Macroscopic quantumness is not restricted to a single physical realization but can be studied for many different degrees of freedom (see \citet{Leggett_Testing_2002,Chou_Macroscopic_2011,Caldeira_Introduction_2014} for an introduction to the physics of macroscopic quantum phenomena). In this review, we treat few ``canonical'' systems for which we introduce the terminology and the notation in this section. The only ``global'' conventions we state here are that $\hbar = 1$  and $(\Delta A)^2 = \langle A^2 \rangle - \langle A \rangle^2$ for the variance of an operator $A$. Furthermore, the components \alive and \dead of a superposition \da are referred to as \textit{branches} or \textit{components}. Sometimes, a pure state as an argument of a function $f$ is abbreviated $f(\left| \psi \right\rangle\!\left\langle \psi\right| ) \equiv f(\psi)$.

\subsubsection{Spin ensemble}
\label{sec:spin-ensemble}

When we consider many microscopic constituents each carrying an identical, discrete and finite degree of freedom we speak of a \textit{spin ensemble}. Whether this degree of freedom is a physical spin or a pseudo-spin is of little relevance here. Alternatively, some authors call this system \textit{atomic ensemble}. The constituents are called \textit{particles} in the following. If not stated otherwise, each particle is considered to be a two-level system (spin 1/2 particles); hence the Hilbert space is $\mathcal{H} = \mathbbm{C}^{2 \otimes N}$. In this case, one might call the system a \textit{qubit ensemble}. The system size is the number of constituents, $N$.

An important class of operators are \textit{local operators}
\begin{equation}
\label{eq:18}
A = \sum_{l = 1}^N a_l^{(l)},
\end{equation}
where $a_l^{(l)} \equiv \mathbbm{1}^{\otimes l-1} \otimes a_l \otimes \mathbbm{1}^{\otimes N-l}$ is a single-particle operator $a_l$ acting the $l$th spin. For convenience and without loss of generality, we set $\lVert a_l^{(l)} \rVert = 1$ and $\mathrm{Tr} a_l^{(l)} = 0$. For two-level systems, $a_l$ can be uniquely decomposed into Pauli operators $\sigma_x, \sigma_y$ and $\sigma_z$. Local operators are sometimes called \textit{linear operators}. Nonlinear operators describe interactions between the particles and are typically written as polynomials of local operators. As an extension of local operators, we also consider \textit{quasi-local} operators. These observables are sums of operators where each addend $ a_l^{(l)}$ has a nontrivial support on group $l$ consisting of $O(1)$ particles. Here, we limit the $ a_l^{(l)}$ to act on nonoverlapping groups only and again we set $\lVert a_l^{(l)} \rVert = 1$. In this way, one can see quasi-local operators as local operators of quasi-particles (``molecules composed of atoms'') each living in a high-dimensional space.

Local operators for which all single-particle terms are identical (i.e., $a^{(l)}_l \equiv a^{(l)}$) are called \textit{collective operators} or \textit{extensive observables}. For spin-1/2 particles, every collective operator is hence a weighted sum of $S_x = \sum_l \sigma_x^{(l)}$,  $S_y = \sum_l \sigma_y^{(l)}$ and  $S_z = \sum_l \sigma_z^{(l)}$, which fulfill the $SU(2)$ commutation relation $[S_x, S_y ] = 2i S_z$ (and permutations). The unusual factor 2 comes from the choice of normalization. The ladder operators are defined as $S_{\pm} = \frac{1}{2} (S_x \pm i S_y)$.

An important class of operations are local operations and classical communication (LOCC). This implies access to single particles, arbitrary operations on them and processing possible measurement results for future operations.

Note that here we are not much concerned of whether the particles are distinguishable (by an additional degree of freedom such as position) or whether they are in a bosonic mode and hence symmetrized. The difference is that we only deal with collective operators in the latter case.

There are several important state families for spin 1/2 particles. The computational basis is denoted by $\left\{  | 0 \rangle , \left| 1 \right\rangle \right\}$, that is, the eigenbasis of $\sigma_z$. We denote $| \pm \rangle $ as the eigenstates of $\sigma_x$.

Among the symmetric states, the spin-coherent states $| \phi \rangle ^{\otimes N}$ are typically seen as the pure states with the ``most classical'' properties\footnote{Spin-coherent states have a vanishing relative uncertainty $\Delta S/\lVert S \rVert \leq O(N^{-1/2})$ for all collective operators $S$. In addition, the Heisenberg uncertainty relation $\Delta S_x \Delta S_y \geq \left| \langle S_z \rangle \right|$ is tight in case $| \phi \rangle ^{\otimes N}$ is polarized somewhere in the $x-z$ or in the $y-z$ plane. \label{fn:1}}. The single-qubit state is conveniently parametrized by the Bloch angles $| \phi \rangle = \cos \vartheta/2 \left| 0 \right\rangle + e^{i \varphi} \sin \vartheta/2 \left| 1 \right\rangle $.

A quantum state that is often discussed in the present context is the multipartite Greenberger-Horne-Zeilinger (GHZ) state \cite{Greenberger_Going_1989a}
\begin{equation}
\label{eq:39}
\left| \mathrm{GHZ_N^{\pm}} \right\rangle = \frac{1}{\sqrt{2}} \left( \left| 0 \right\rangle ^{\otimes N} \pm \left| 1 \right\rangle ^{\otimes N}\right),
\end{equation}
which is considered to be macroscopically quantum by many physicists. This state is a limiting case of the ``generalized GHZ state''
\begin{equation}
\label{eq:40}
\left| \Phi_{\epsilon} \right\rangle \propto \left| 0 \right\rangle ^{\otimes N} + \left| \epsilon \right\rangle ^{\otimes N},
\end{equation}
where $| \epsilon \rangle = \cos \epsilon \left| 0 \right\rangle + \sin \epsilon \left| 1 \right\rangle $.

We are also interested in the symmetric superposition states
\begin{equation}
\label{eq:41}
\left| N,k \right\rangle \propto \sum_{\pi: \mathrm{perm}} \pi \left| 0 \right\rangle ^{\otimes N-k} \otimes \left| 1 \right\rangle ^{\otimes k},
\end{equation}
where $\pi$ are particle permutations. These quantum states are often called Dicke states \cite{Dicke_Coherence_1954}. For $k = 0,N$, the states are product states, while all other Dicke states are genuinely multipartite entangled. An important instance is $| N,1 \rangle $, which was called W state in Sec.~\ref{sec:not-an-easy} \cite{Dur_Three_2000}. It is typically cited as a counterexample to a macroscopic quantum state, despite its widespread entanglement.

Another important state class are spin-squeezed states as defined by \citet{Kitagawa_Squeezed_1993}. There are various ways to generate spin-squeezing. A well-known method is the one-axis twisting,
\begin{equation}
\label{eq:9}
\left| S_{\mu} \right\rangle = e^{-i\nu S_{x}} e^{-i \mu S_z^2} \left| + \right\rangle^{\otimes N},
\end{equation}
where the rotation generated by $S_x$ is just to align the squeezed axis to $z$ (thus, $\nu = \nu(N,\mu)$). The optimal squeezing is achieved for $\mu = 24^{1/6}(N/2)^{-2/3}$ \cite{Kitagawa_Squeezed_1993}. A common way to characterize squeezed states without reference to its generation is the squeezing parameter
\begin{equation}
\label{eq:42}
\xi^2 \equiv \frac{N (\Delta S_{n_1})^2}{\langle S_{n_2} \rangle^2 + \langle S_{n_3} \rangle^2 },
\end{equation}
which is strictly smaller one, $\xi <1$, for squeezed states \cite{Sorensen_Many-particle_2001}. Here, $(n_1, n_2, n_3)$ are three orthogonal orientations of collective spin operators. This means that the states has to exhibit a large polarization in the $n_2 - n_3$ plane and simultaneously a small (i.e., squeezed) variance in $n_1$.

The last class of spin states introduced here are the cluster states \cite{Briegel_Persistent_2001}. Cluster states are special instances of graph states \cite{Hein_Entanglement_2006}, which are constructed by applying controlled phase gates $U_{C}^{(i,j)} = \left| 0 \right\rangle\!\left\langle 0\right|^{(i)}  + \left| 1 \right\rangle\!\left\langle 1\right|^{(i)} \sigma_z^{(j)}$ on a set of spin pairs $(i,j)$. If the $U_C$ are applied to nearest neighbors in a certain geometry, the graph state is called cluster state $| \mathrm{Cl} \rangle = \mathcal{U}_C \left| + \right\rangle ^{\otimes N}$; for example, a one-dimensional arrangement gives
\begin{equation}
\label{eq:51}
\mathcal{U}_C = \prod_{i = 1}^{N-1} U_C^{(i,i+1)}.
\end{equation}

\subsubsection{Photonic systems}
\label{sec:photonic-modes}

We consider well-defined temporal, spatial, frequency and polarization modes. For every mode, we define the usual \textit{quadrature operators} $X$ and $P$ with the canonical commutation relation $[X,P] = i$ and the decomposition into dimensionless creation and annihilation operators $X = 1/\sqrt{2}(a^{\dagger} + a)$ and $P = i/\sqrt{2}(a^{\dagger} - a)$. The notion of a linear operator refers to operators that are linear in $a$ and $a^{\dagger}$. The system size is the mean photon number $\langle a^{\dagger} a \rangle$.

Sometimes we discuss multi-mode scenarios, which have some parallels with spin ensembles in case of many modes (except that there are infinitely many levels per mode). The equivalence of a local operator is a sum of single-mode operators linear in $X^{(l)}$ and $P^{(l)}$. The system size is the total mean photon number.

Several quantum states are repeatedly discussed in the remainder of the paper. A useful countable basis in a single mode consists of Fock states (or photon number states) $ a^{\dagger} a \left| n \right\rangle = n \left| n \right\rangle$ for integer numbers including the vacuum $| 0 \rangle $.

The pure state that behaves ``most classically''\footnote{Regarding classicality, a similar comment as footnote \ref{fn:1} applies, where collective operators are replaced by quadrature operators.} is the coherent state $| \alpha \rangle $ defined via $a \left|\alpha  \right\rangle = \alpha \left| \alpha \right\rangle $ for all $\alpha \in \mathbbm{C}$. It can equally be seen as a displacement of the vacuum $\left| \alpha \right\rangle = D(\alpha)\left| 0 \right\rangle \equiv \exp(\alpha a^{\dagger} - \alpha^{*}a) \left| 0 \right\rangle $. The mean number of photons is $|\alpha|^2$ with a spread of $|\alpha|$ in the photon number spectrum. The variance of the quadratures is independent of $\alpha$.

Among other representations, the Wigner function is a well-established way of representing quantum states in phase space. The Wigner function is a quasi-probability distribution, whose marginals gives the statistics of the quadrature operators \cite{Wigner_Quantum_1932,Scully_Quantum_1997}.

Well-studied states in the present context are superpositions of coherent states (SCS).\footnote{In the literature, SCS often stands for Schrödinger-cat state, a frequent name for state Eq.~(\ref{eq:43}).} A typical instance is
\begin{equation}
\label{eq:43}
\left| \mathrm{SCS} \right\rangle \propto \left| \alpha \right\rangle + \left| -\alpha \right\rangle,
\end{equation}
whenever $\alpha$ is large. The overlap between the two coherent states, $\left\langle \alpha \right| -\alpha\left. \right\rangle  = \exp(-2|\alpha|^2)$, vanishes for $\alpha \gg 1$.
The equivalent state in photon number space is $| 0 \rangle + \left| N \right\rangle $, with $N \gg 1$. However, since $| N \rangle $ is considered to be highly nonclassical, $| \mathrm{SCS} \rangle $ and the superposition of Fock states have a different characteristics.

Another important state class are squeezed states. The most important instance of squeezing is the one of squeezed coherent states
\begin{equation}
\label{eq:1}
| \zeta,\alpha \rangle = S_{\zeta}^{(1)} \left| \alpha \right\rangle,
\end{equation}
where the squeezing operator $S_{\zeta}^{(1)} = \exp(-\frac{1}{2}(\zeta a^{\dagger 2} - \zeta^{*}a^2))$ reduces the variance of the quadrature $1/\sqrt{2}(e^{i \arg(\zeta)} a^{\dagger} + e^{-i \arg(\zeta)} a)$ by a factor $e^{-2|\zeta|}$ and increases the variance of $i/\sqrt{2}(e^{i \arg(\zeta)} a^{\dagger} - e^{-i \arg(\zeta)} a)$ by $e^{2|\zeta|}$.

All these examples have extensions to multi-mode systems. For SCS, this includes states like $| \alpha,\beta \rangle + \left| -\alpha, -\beta \right\rangle $. A famous two-mode superposition of Fock states is the NOON state $\left| N,0 \right\rangle + \left| 0,N \right\rangle $. Finally, two-mode squeezing of vacuum with the operator
\begin{equation}
\label{eq:twomodesqueezedstate}
S_{\zeta}^{(2)} = \exp(-\zeta a^{\dagger}b^{\dagger} + \zeta^{*}a b)
\end{equation}
plays an important role in spontaneous parametric down-conversion.

\subsubsection{Massive systems}
\label{sec:other-phys-real}

Massive systems are clearly of high importance for the topic of macroscopic quantumness. Mathematically, they are similarly treated as a photonic mode, except that $X$ and $P$ now become position and momentum. In particular for a massive particle in a harmonic trap, the creation and annihilation operators play the same role as for photons. In this case, state classes like coherent states, SCS and squeezed states are also important here.
Unlike photons, however, the role of parameters, in particular, the mass $m$, plays a crucial role in massive systems and is typically considered to be the system size. In addition, also in the case of ``free'' particles, the role of the distance in a spatial superposition is not obvious.

\subsubsection{Superconducting systems}
\label{sec:other-systems}

There are many different degrees of freedom that potentially show macroscopic quantum behavior. Notably, superconducting circuits such as superconducting quantum interference devices (SQUIDs) play an important role in the present context. Often, one works with collective degrees of freedom such as the total flux of the system, $\Phi$. Mathematically, this is equivalent to a massive particle moving in one-dimensional potential. The system size in this case might be defined as the total number of electrons that are involved in the experiment.

\subsection{Structure of the review and reading guide}
\label{sec:scope-revi-prel}

The purpose of the review is two-fold. On the one hand, we provide an introduction to different aspects of macroscopic quantum states for interested non-expert readers. On the other hand, we also give a comprehensive overview and detailed discussion of various approaches to quantum macroscopicity. This involves detailed discussions and comparison between approaches, some of which might only be relevant to experts in the field as technicalities, subtilities and particular aspects are discussed. In order to make the article accessible for both, experts and non-experts, we provide here a readers guide.

Section \ref{sec:introduction} provides the basis of this review: the precise scope and motivation of the review, the most important terminology, the mathematical notation and the discussed physical systems. In Sec.~\ref{sec:meas-macr-superp} we give an overview of different measures, and provide detailed discussions on their relations and differences. Some of these discussions are rather technical and might be hard to understand for a non-expert reader. Such a reader has to keep this in mind, and do not get stack or intimidated with the technicalities. To facilitate this, we provide --for each of the measures reviewed in section~\ref{sec:meas-macr-superp}-- a boxed text that conveys the general idea behind each measure in non-technical terms. We suggest to only read the encapsulated text first, and continue with sections \ref{sec:measurediscussion} and \ref{sec:macr-superp-vs}. At that point, the reader is now equipped with the necessary background information to proceed reading the rest of the article following his or her interest.

Instructive examples and an in depth discussion of the measures are provided in Secs.~\ref{sec:examples}--\ref{sec:summaryCht2}. Section \ref{sec:limits-observ-quant} contains a detailed discussion on limitations to prepare, maintain and measure macroscopic quantum states, and might be of particular relevance for readers interested in fundamental questions regarding macroscopic quantum states. In turn, Sec.~\ref{sec:potent-macr-quant} deals with potentials of macroscopic quantum states and contains a brief discussion of possible applications in the context of probing the limits of quantum theory, in quantum metrology and in quantum computation.
Section \ref{sec:implementations} provides an overview of implementations and previous attempts to generate macroscopic quantum states using different setups. We summarize and conclude in Sec.~\ref{sec:discussions-outlooks}.

%%% Local Variables:
%%% mode: latex
%%% TeX-master: "master"
%%% End:

\section{Measures for macroscopic superpositions and quantum states}
\label{sec:meas-macr-superp}

After motivating and specifying macroscopic quantumness in Sec.~\ref{sec:not-an-easy}, we now have a closer look at theoretical proposals that aim at formalizing intuitive ideas in a mathematical framework. The terminology and notation used in this section is defined in Secs.~\ref{sec:preliminary-remarks} and \ref{sec:physical-systems}. We start with a detailed summary of several measures in Sec.~\ref{sec:preliminary-measures} focusing on the motivation, the mathematical formulation and some basic properties. Inside this subsection, we stay close to the original language and notation of the reviewed papers. In addition every entry is opened with an encapsulated text, where we try to convey the idea behind each measure as simply as possible. This is meant to help readers novice to the topic, but comes at the price of giving our own interpretation which might not exactly match the authors' original motivation. A reader familiar with the literature might skip this part. In Sec.~\ref{sec:examples}, we apply the measures to several examples: standard situations discussed in many papers as well as specialized examples to see similarities and differences between the proposals. The discussion is continued in Secs.~\ref{sec:measurediscussion}--\ref{sec:meas-exper}, in which we elaborate on several details and comparisons (see table \ref{tab:overview2} for an overview). We summarize in Sec.~\ref{sec:summaryCht2}.
This section is meant to complement and extend previous contributions\footnote{For example, \citet{Frowis_Measures_2012,frowis15,Jeong_Characterizations_2015,Farrow_Classification_2015}.}.

\begin{table*}
  \centering
  \begin{tabular}{l| l | a c a c a c a c a c a c a c a c }
    \multicolumn{2}{r}{}
    &      \begin{turn}{86}  \citet{Leggett_Macroscopic_1980,Leggett_Testing_2002} \end{turn}
    &      \begin{turn}{86}  \citet{Dur_Effective_2002} \end{turn}
    &      \begin{turn}{86}  \citet{Shimizu_Stability_2002} \end{turn}
    &      \begin{turn}{86}  \citet{Bjork_Size_2004} \end{turn}
    &      \begin{turn}{86}  \citet{Cavalcanti_Signatures_2006,Cavalcanti_Criteria_2008} \end{turn}
    &      \begin{turn}{86}  \citet{Korsbakken_Measurement-based_2007} \end{turn}
    &      \begin{turn}{86}  \citet{Marquardt_Measuring_2008} \end{turn}
    &      \begin{turn}{86}  \citet{Lee_Quantification_2011} \end{turn}
    &      \begin{turn}{86}  \citet{Park_Quantum_2016} \end{turn}
    &      \begin{turn}{86}  \citet{Frowis_Measures_2012} \end{turn}
    &      \begin{turn}{86}  \citet{Nimmrichter_Macroscopicity_2013} \end{turn}
    &      \begin{turn}{86}  \citet{Sekatski_Size_2014} \end{turn}
    &      \begin{turn}{86}  \citet{Sekatski_General_2017} \end{turn}
    &      \begin{turn}{86}  \citet{Laghaout_Assessments_2015} \end{turn}
    &      \begin{turn}{86}  \citet{Yadin_Quantum_2015} \end{turn}
    &      \begin{turn}{86}  \citet{Kwon_Disturbance-Based_2016} \end{turn} \\
    \hline \hline
    \multirow{4}{*}{Basic approach}
   & Preferred observable         & x&  & o& x& x&  &  &  & o& o&  & x& x& x&  & x\\
   & Partition in subsystems      & x& x& x&  &  & x& x&  & x& x&  &  &  &  & x&  \\
   & Preferred representation     &  &  &  &  &  &  &  & x&  &  &  &  &  & x&  &  \\
   & Preferred decoherence        &  & o&  &  &  &  &  & x&  &  & x&  &  &  &  &  \\
    \hline
    \multirow{6}{*}{Motivation/goal}
   & Macroscopic coherence        & x&  & x& x& x& o& o& x& x& x&  & o& o& x&  & x\\
   & Effective particle number    & x& x&  &  &  & x& x&  &  & o&  &  &  &  & x&  \\
   & Ease to distinguish          &  &  &  &  &  & x& o&  &  &  &  & x& x& x&  &  \\
   & Relative improvement         &  &  &  & x&  &  &  &  &  & o&  &  &  &  &  &  \\
   & Collapse models/ fragility   &  & o& x&  &  &  &  & o& o&  & x&  & o&  &  & x\\
   & Nonclassicality              &  & x&  &  &  &  &  & x& x&  &  &  &  & x& x&  \\
    \hline
    \multirow{2}{*}{State structure}
   & \da                          &  & x&  & x&  & x& x&  &  &  &  & x&  &  &  &  \\
   & General state                & x&  & x&  & x&  &  & x& x& x& x&  & x& x& x& x\\
    \hline
    \multirow{4}{*}{Physical system}
   & Spin ensemble                & x& x& x& x& x& x& x&  & x& x&  & x& x&  & x& x\\
   & Photons                      &  &  &  & x& x& o& o& x&  & x&  & x& o& x&  & o\\
   & Mass                         &  &  &  & o&  &  &  &  &  &  & x&  &  &  &  &  \\
   & SQUID                        & x&  &  & x&  & x& x&  &  &  & x&  &  &  &  &  \\
    \hline
\multicolumn{2}{l|}{Mathematical symbol for measure} 
& $D,\Lambda$&  & $p,q$ & $M$& $S$ & $C_{\delta}$& $\bar{D}$& $\mathcal{I}$ & $\mathcal{I}_S$& $N_{\mathrm{eff}}$& $\mu$ & Size & MIC & $\mathcal{N,D}$ & $N^{*}$& $M_{\sigma}$\\
    \hline  \hline
  \end{tabular}
  \caption{ Properties of measures as discussed in Secs.~\ref{sec:measurediscussion}--\ref{sec:physical-setups}. Each column corresponds to a measure that is reviewed in Sec.~\ref{sec:preliminary-measures} in the order of appearance. Four different categories are compared:
    the basic approach of the measure to break the unitary equivalence (Sec.~\ref{sec:observ-vs.-part});
    the motivation or goal of the measure (Sec.~\ref{sec:cat-regard-purp});
    the state structure for which the measure is formalized (Sec.~\ref{sec:macr-superp-vs});
    and the physical system to which the measure can be applied (Sec.~\ref{sec:physical-setups}).
The symbol ``x'' means general applicability; the symbol ``o'' stands for partial applicability (e.g., ``macroscopic coherence'' is not the primary motivation but a consequence; or some measures are applied to photonic states only for specific instances). The last line summarizes the mathematical symbols used by the authors for their measures (if any). As some papers leave room for slightly different view points, we understand this table not as an ultimate judgment but as a starting point for further discussions.}
  \label{tab:overview2}
\end{table*}

\subsection{Summary of measures}
\label{sec:preliminary-measures}
\label{sec:macr-quant-stat}
\label{sec:Superposition size}
\label{sec:superpositions partition based}
\label{sec:superpositions observable based}

Although the goal of this section is to give a summary of the relevant literature as neutral and objective as possible, we would like to draw the reader's attention to some basic observations. While it is true that different authors partially use different concepts to define macroscopic quantumness, it is obvious that a common goal is to distinguish ``interesting'' from ``uninteresting'' states and to find an ordering between states. However, since all pure quantum states are connected via unitary operations, one has to find a way to break this unitary equivalence. This is similar to entanglement theory, where the partition of a large Hilbert space into subspaces is the structure necessary to define separable states (i.e., ``uninteresting'' states; \citet{Horodecki_Quantum_2009}). For the present topic, this prestructuring is far less obvious. We invite the reader to observe which mechanisms for breaking the unitary equivalence has been chosen in the following summary (and refer to table \ref{tab:overview2} and Secs.~\ref{sec:measurediscussion} and \ref{sec:macr-superp-vs} for further discussion). Some authors partition the system into subsystems like in entanglement theory. Others choose an observable to specify a basis and a spectrum (i.e., a set of eigenvalues), which is close to attempts in coherence theory. Alternatively, one can focus on a specific state representation like the Wigner function in phase space.

To give another example for a difference in the structure of the proposals, the measures vary in their range of applicability. Some discuss specific examples, others work with pure states decomposed into a structure of ``dead and alive'', $| \mathcal{A} \rangle + \left| \mathcal{D} \right\rangle $,  while some proposals are defined for general mixed states.

Let us finally summarize the typical way a measure is constructed. It starts with an intuition or examples, which is then formalized. This normally consists of the basic framework and a sort of fine-tuning, for example, by choosing the right observable or fixing a parameter. When discussing the measures in details, it is worth keeping the implications of each step (intuition, basic framework and fine-tuning) in mind.

\subsubsection{\citet{Leggett_Macroscopic_1980,Leggett_Testing_2002}}
\label{sec:legg-contr}

\fbox{\parbox{0.9\columnwidth}{For a superposition state \da of a system composed of a large number of particles, the extensive difference $\Lambda$ is the difference between the expectation values of some extensive observable of the states \alive and $| \mathcal{D} \rangle $. The disconnectivity $D$ measures the quantumness (nonseparability and purity) of the total superposition state. A macroscopic quantum state is required to have both $D$ and $\Lambda$ large.}}\\

Motivated by the question ``What experimental evidence do we have that quantum mechanics is valid at the macroscopic level?'', \citet{Leggett_Macroscopic_1980} was the first to point out a qualitative difference between quantum effects on microscopic scales amplified to large scales and genuine large-scale quantum signatures.
As Leggett argues, the common feature of macroscopic quantum states is their long-range coherence. This cannot be revealed with single local measurements. Only simultaneous measurements of a large number of particles allow to distinguish this from an incoherent mixture.

In order to quantify this insight, \citet{Leggett_Macroscopic_1980,Leggett_Testing_2002} introduces two concepts for a large system consisting of $N$ of particles. First, consider a superposition of \da and calculate the difference in the expectation value for a particular extensive variable (e.g., total charge, total magnetic moment, or total momentum). This number divided by a characteristic microscopic unit (e.g., the Bohr magneton for the magnetic moment) is called \textit{extensive difference} $\Lambda$. Second, to characterize the quantumness of the state (here, the entanglement between the particles), Leggett defines the \textit{disconnectivity} $D$ as follows. For simplicity, we consider symmetric states $\rho$ of spin ensembles. The von Neumann entropy $S_M = - \mathrm{Tr} \rho_M \log \rho_M$ of the reduced density matrix $\rho_M = \mathrm{Tr}_{M+1,\dots,N} \rho $ is a measure of the correlation between two parts with $M$ and $N-M$ particles, respectively. The quantity
\begin{equation}
\label{eqIb:2}
\delta_M = \frac{S_M}{\min_n (S_{n} + S_{M-n})}
\end{equation}
compares $S_M$ with the minimal correlations from smaller partitions to the rest of the system. If numerator and denominator are zero, $\delta_M$ is set to one and $\delta_1 \mathrel{\mathop:}= 0$. Leggett now defines $D$ to be the largest integer $M$ for which $\delta_M$ is smaller than some predetermined small fraction.
In words, a large disconnectivity is found for states that are rather pure when considering a large part of the system (i.e., $S_M \ll 1$ for $M = O(N)$) but all smaller parts of size $n$ still exhibit large entropy indicating entanglement to the rest of the system.\footnote{Note, however, that Leggett himself does not attach much importance to the precise mathematical formulation of this idea, which ``could almost certainly be substantially improved'' \citet{Leggett_Testing_2002}.}

How to combine the two measures $D,\Lambda$ is ultimately an open question. In the examples discussed by Leggett, both should be somehow large in order to call a state macroscopically quantum.

\subsubsection{\citet{Dur_Effective_2002}}
\label{sec:effect-size-comp}

\fbox{\parbox{0.9\columnwidth}{Two ways to quantify the macroscopic quantumness of a many-body state $\ket{\Phi_\epsilon}$ of type Eq.~(\ref{eq:40}) are introduced, both via a comparison with GHZ states. First, the effective size of $\ket{\Phi_\epsilon}$ is identified with the size of GHZ that has the same decoherence rate under local noise. Second, it is identified with the largest GHZ state, that can be obtained from $\ket{\Psi_\epsilon}$ with local operations and classical communication. Both lead to the same result.}}\\

Motivated by several proposals and experiments, \citet{Dur_Effective_2002} study the generalized GHZ state $| \Phi_{\epsilon} \rangle $, Eq.~(\ref{eq:40}), with $\epsilon \ll 1$, that is, $ |\left\langle 0 \right| \epsilon\left. \right\rangle|^2 \approx 1-\epsilon^2$.
While the two branches are orthogonal even for $\epsilon$ close to zero (as long as $N \epsilon^2 = O(1)$), \citeauthor{Dur_Effective_2002} argue that Eq.~(\ref{eq:40}) effectively corresponds to the ideal state $\left|\mathrm{GHZ}_n  \right\rangle $, Eq.~(\ref{eq:39}), with $n \ll N $.
To show this, one generally chooses a key property (or a set of key properties) for which $| \Phi_{\epsilon} \rangle $ and $\left|   \mathrm{GHZ}_n\right\rangle$ coincide. As argued by the authors, properties connected to the nonclassicality of the states are clearly of importance. They choose two methods: (i) the rate with which the coherence decays is compared; (ii) the average size $n$ one can achieve by going from $| \Phi_{\epsilon} \rangle $ to $|  \mathrm{GHZ}_n \rangle $ with LOCC. For both methods, they find that $n \approx N \epsilon^2$ in the limit $N\gg 1$ and $\epsilon \ll 1$.
This example, which was not directly expanded to a general measure, was later used as a test bed for proposals to measure arbitrary macroscopic quantum states.

\subsubsection{\citet{Shimizu_Stability_2002} and followups}
\label{sec:index-p}

\fbox{\parbox{0.9\columnwidth}{The macroscopic quantumness of pure states -- the index $p$ is identified via the maximal variance of the state with respect to all extensive observables $A$. The extension to mixed state is formally introduced via an additional maximization to uncover widespread coherence.}}\\

\citet{Shimizu_Stability_2002} study the stability of finite macroscopic quantum states under weak noise and local measurements. The systems they consider can be decomposed into subsystems like the spin ensembles introduced in Sec.~\ref{sec:spin-ensemble}. They find a difference for the decay rate $\Gamma$ of the purity depending on whether or not a quantum state has the so-called cluster property, whose absence implies long-range correlations in a many-body system and wide-spread entanglement for pure states. The noise and decoherence is supposed to be generated by some operator $A$. From a physical viewpoint, it is clear that $A$ has to be a local operator. While the detailed results regarding fragility are discussed in Sec.~\ref{sec:decay-purity}, we emphasize here that the authors are able to connect $\Gamma$ with the variance $(\Delta A)^2$.

A series of papers was devoted to formalize this basic insight \cite{Ukena_Appearance_2004,Shimizu_Detection_2005,Morimae_Macroscopic_2005}.
\citet{Shimizu_Detection_2005} remark that two pure states of $N$ subsystems can be reasonably called macroscopically distinct if there exist some local operator $A$ such that the difference of its expectation value between the two states is $O(N)$ (with the normalization of Sec.~\ref{sec:spin-ensemble}). A general state $\rho$ is more macroscopically quantum the more coherence between macroscopically distinct states it contains. Defining the eigenstates $A\ket{a_k,\nu}= a_k\ket{a_k,\nu}$ ($\nu$ accounts for possible degeneracy) of a given local observable $A$, the amount of macroscopic coherence is quantified by the total weight of all terms $|\bra{a_k, \nu} \rho \ket{a_l, \nu'}|$  with $|a_k-a_l|=O(N)$. The authors propose to measure this coherence in the following way.

For a pure state $\rho =\prjct{\psi}$, the spread of coherence is quantified by the index $p$ (with $1\leq p\leq 2$) of the state \cite{Ukena_Appearance_2004}, defined as the best scaling of the variance of the state with respect to all normalized local observables
\begin{equation}
\max_{A: \mathrm{local}}  (\Delta A)_{\psi}^2  = O(N^p).\label{eq:13}
\end{equation}
Expressing $A$ that maximizes Eq.~(\ref{eq:13}) with single-spin operators $a^{(l)}_l$, one finds that the variance reads
\begin{equation}
\label{eq:14}
(\Delta A)^2 = \sum_{l = 1}^{N} (\Delta a^{(l)}_l)^2 + \sum_{l \neq l^{\prime} = 1}^N \langle a^{(l)}_l a^{(l^{\prime})}_{l^{\prime}} \rangle - \langle a^{(l)}_l \rangle \langle a^{(l^{\prime})}_{l^{\prime}} \rangle.
\end{equation}
The scaling $p = 1$ hence implies that the quantum correlations expressed in the second term in Eq.~(\ref{eq:14}) do not play a significant role. For every spin $l$, there exist only $O(1)$ ``neighbors'' with which the spin shares correlations. \citet{Shimizu_Stability_2002} call this the cluster property. In contrast, $p > 1$ and in particular $p = 2$ means that as growing number of pairs share nonvanishing quantum correlations.\footnote{Even though the name ``index $p$'' was first introduced by \citet{Ukena_Appearance_2004}, we refer to \citet{Shimizu_Stability_2002} for measure (\ref{eq:13}).}

However, for mixed states this is not sufficient since the variance of a state with respect to $A$ only depends on the diagonal terms $|\bra{a_k, \nu} \rho \ket{a_k, \nu'}|$. To overcome this problem, \citet{Shimizu_Detection_2005} introduce the so-called index $q$ defined as (in the formulation of \citet{Morimae_Superposition_2010})
\begin{equation}
\label{eq:2}
\max \left\{ N, \max_{A: \mathrm{local}} \lVert [A,[A,\rho]] \rVert_1 \right\} = O(N^q).
\end{equation}
While the ``outer'' maximization is just to guarantee $1 \leq q \leq 2$, a large trace norm $\lVert [A,[A,\rho]] \rVert_1 = O(N^2)$ for local operators $A$ is only possible with significant contributions from elements $(a_k - a_l)^2 \bra{a_k, \nu} \rho \ket{a_l, \nu'} = O(N^2)$, which reflects the initial motivation as discussed in the beginning of this subsection. For pure states, it is shown that $q=1\Rightarrow p=1$, $p=1\Rightarrow q \leq 1.5$ and $q=2\Leftrightarrow p=2$ (which are the corrected statements from \citet{Shimizu_Erratum:_2016,Tatsuta_Conversion_2017}).

\subsubsection{\citet{Bjork_Size_2004}}
\label{sec:bjork}

\fbox{\parbox{0.9\columnwidth}{The macroscopic quantumness $M$ of a superposition state is identified with the advantage it offers for interferometry as compared to the individual components \alive or $| \mathcal{D} \rangle $. The interforometric usefulness is defined by how fast the state becomes orthogonal to itself when subject to a unitary evolution $e^{i\theta A}$ for some fixed observable $A$.}}\\

\citet{Bjork_Size_2004} motivate their contribution by the need for an operational meaning of a measure rather than focusing on particle and mode numbers. In their opinion, a genuinely macroscopic quantum state should give some advantage over states that lack this feature in some practical application. \citeauthor{Bjork_Size_2004} consider the interferometry as such a task. This general idea is formalized for a state of the form \da in the following way. (i) The starting point is to identify a preferred observable $A$ which results from the experimental context as the most useful one for a particular interferometric application. (ii) One identifies some semi-classical states $\ket{c_a}$, that are pure and have a smooth but rather narrow distribution $c(A-a)$ in the eigenvalues of $A$ centered at $a$. (iii) One imagines the superposition state to undergo an evolution $e^{\ii\,\theta A}$ generated by the observable $A$. In such an interferometric scenario, large oscillation frequencies between the initial state and the finial state indicate a large separation between coherent components of a superposition state, which do not occur for mixtures. This last observation is formalized as follows.
Consider a superposition state
\be
\ket{\psi} \propto \ket{c_{a_1}} + \ket{c_{a_2}},
\ee
where the two components have a negligible overlap. The overlap of the evolved state with the original one is given by
  \be
  |\bra{\psi}e^{\ii \theta A}\ket{\psi}|=2\biggl|\cos\frac{\theta(a_1-a_2)}{2}
   \underbrace{\int e^{\ii \theta A} c(A) dA}_{\tilde c(\theta)} \biggr|,
\ee
with two contributions to the oscillation of this overlap. The first one comes from the shape of underlying classical states $\tilde c(\theta)$, while the second and more rapid effect is due to the superposition $\cos\left(\frac{\theta(a_1-a_2)}{2}\right)$. Hence, one compares the minimal $\theta$ for which the superposition evolves to an orthogonal state, $\theta_{\mathrm{sup}} \approx \pi/(a_1-a_2)$, to the corresponding $\theta$ for a single classical state, $\theta_\text{c}\approx \pi/(\Delta A)_c$ (with $(\Delta A)_c$ the width of the distribution $c(A)$). Hence the dimensionless ratio between the two ``orthogonalization times''
\begin{equation}
M=\frac{\theta_\text{c}}{\theta_\text{sup}}\label{eq:11}
\end{equation}
quantifies the ``interferometric macroscopicity'' of the superposition state.

\subsubsection{\citet{Cavalcanti_Signatures_2006,Cavalcanti_Criteria_2008}}
\label{sec:calvalcanti-reid}

\fbox{\parbox{0.9\columnwidth}{For a given observable $A$, the macroscopic quantumness of a state is identified with the maximum spectral range $S$ on which the state exibits quantum coherence. That is, the density matrix contains some coherence terms between eigenstates of $A$ with eigenvalues different by at least $S$.}}\\

\begin{figure}[htbp]
\centerline{\includegraphics[width=\columnwidth]{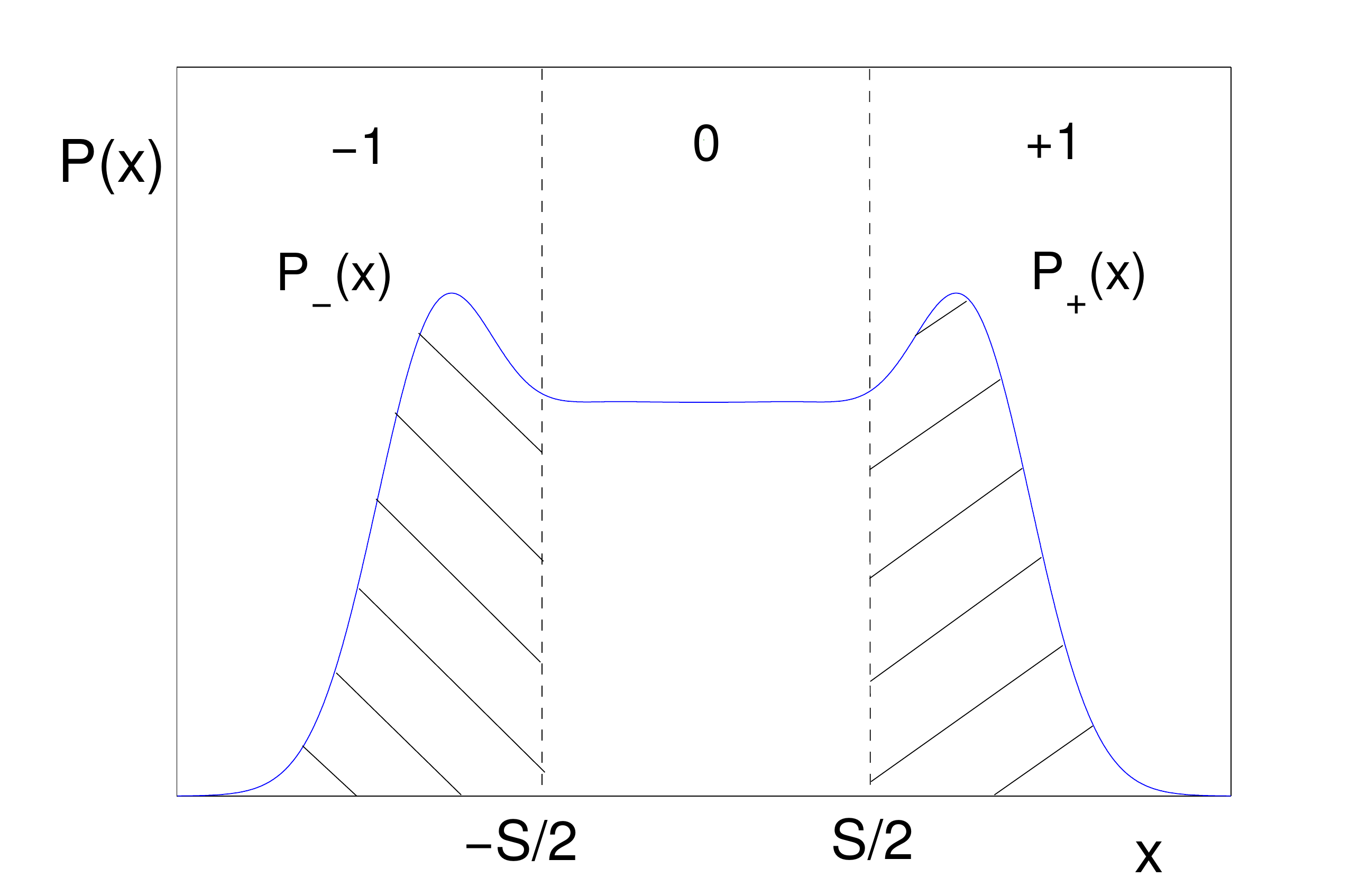}}
\caption[]{\label{fig:CavalcantiReid2006} Division of the spectrum of the observable $X$ into three intervals. The curve is a schematic example of the probability distribution of the considered state $\rho$. \citet{Cavalcanti_Signatures_2006} aim to verify coherence between the left (-1) and the right interval (+1). From \citet{Cavalcanti_Signatures_2006}.}
\end{figure}

\citet{Cavalcanti_Signatures_2006} are interested in witnessing coherent superpositions on the macroscopic scale. Following their argument, macroscopic coherence contradicts a general notion of macroscopic reality, in which an object is necessarily in one out of several states. To be more specific, let us consider a preselected observable $X$ on the real line which is divided into three intervals called left $I_{-}$, central $I_0$ and right $I_{+}$ (see Fig.~\ref{fig:CavalcantiReid2006}). The interval $I_0$ has length $S$. A general pure state $| \psi \rangle $ can be written as a superposition of states located in the respective intervals
\begin{equation}
\label{eq:15}
\left| \psi \right\rangle = c_- \left| \psi_{-} \right\rangle + c_0\left| \psi_0 \right\rangle + c_{+}\left| \psi_{+} \right\rangle.
\end{equation}
The goal is now to guarantee that a state prepared in an experiment exhibits coherence between the left and the right interval. Working with general mixed states $\rho$, this means that states of the form Eq.~(\ref{eq:15}) with contributions from the left \textit{and} the right interval are present in every pure state decomposition of $\rho$.

To derive a witness, \citeauthor{Cavalcanti_Signatures_2006} point out that a state with nonvanishing $c_{-}$ and $c_{+}$ must have a minimal spread in the spectrum of $X$. From the Heisenberg uncertainty relation it follows that states with lower variance in $X$ necessarily have a larger variance in a conjugate observable $P$. This argument can be extended to mixed states. The authors derive such general bounds for the canonical commutation relation\footnote{In this subsection, we adapt to the convention of \citeauthor{Cavalcanti_Signatures_2006}. $X$ and $P$, as introduced in Sec.~\ref{sec:photonic-modes}, would give $[X,P] = i$. This only changes some constants in Eqs.~(\ref{eq:17}) and (\ref{eq:52}) and later in example \ref{ex:squeezed}.} $[X,P] = 2i$. If $\rho$ does not contain a state like Eq.~(\ref{eq:15}), it can only be written in the form
\begin{equation}
\label{eq:16}
\rho = P_L \rho_L + P_R \rho_R,
\end{equation}
where $\rho_L$ completely lies in the intervals $I_{-}$ and $I_0$ and $\rho_R$ in the intervals $I_0$ and $I_{+}$. Let us define the normalized distributions $p_{\pm}(x) = p(x|x \in I_{\pm})$ and the weights $\pi_{\pm} = \int_{I_{\pm}} p_{\pm}(x) dx$ and $\pi_0 = 1 - \pi_{+} - \pi_{-}$. From this, we define the mean $\mu_{\pm}$ and the variance $(\Delta X)^2_{\pm}$ of $p_{\pm}(x)$,  $(\Delta X)^2_{\mathrm{ave}} = \pi_{+} (\Delta X)^2_{+} + \pi_{-} (\Delta X)^2_{-}$ and $\delta = (\mu_{+} + S/2)^2 + (\Delta X)^2_{+} + (\mu_{-} - S/2)^2 + (\Delta X)^2_{-} + S/2$. Then, \citet{Cavalcanti_Signatures_2006} show that for all states (\ref{eq:16}) it holds that
\begin{equation}
\label{eq:17}
[(\Delta X)^2_{\mathrm{ave}} + \pi_0 \delta] (\Delta P) ^2\geq 1.
\end{equation}
Violating Eq.~(\ref{eq:17}) proves coherence in $X$ between $I_{+}$ and $I_{-}$.

This inequality can be generalized or modified in several ways \cite{Cavalcanti_Criteria_2008}. First, one can use other uncertainty relations (e.g., involving sums of variances rather than products); one can derive them for bipartite scenarios; and one can drop the predetermined binning into three intervals and only verify coherence with a minimal distance $S$. The latter is gives a simpler expression: The violation of the bound
\begin{equation}
\label{eq:52}
(\Delta P) ^2 \geq \frac{2}{S}
\end{equation}
implies that the corresponding states exhibits coherence with a spread of at least $S$.

\subsubsection{\citet{Korsbakken_Measurement-based_2007}}
\label{sec:Korsbakken}

\fbox{\parbox{0.9\columnwidth}{The macroscopic quantumness $C_\delta$ of a many-body superposition \da is related to the minimum number of subsytems that have to be measured in order to learn the ``which-path'' information (i.e., \alive or $| \mathcal{D} \rangle $).}}\\

Let us consider a system that can be divided into $N$ subsystems, called particles. The quantum state is supposed to be of the form $| \mathcal{A} \rangle + \left| \mathcal{D} \right\rangle $. The basic intuition of \citet{Korsbakken_Size_2010} is that the (macroscopic) distinctness of the components \alive and \dead can be quantified by asking how many particles have to be measured in order to distinguish the two states, or equivalently to collapse the superposition \da to just one branch. This is formalized in the following way. Given the desired guessing probability $P_g =1-\delta$, \citeauthor{Korsbakken_Measurement-based_2007} consider the minimal number $n_{\min}$ of particles one has to measure on average in order to distinguish \alive from \dead with probability $P_g$ (subsystems are drawn at random). Then, the size of the superposition $\left| \Psi \right\rangle \propto \left| \mathcal{A} \right\rangle + \left| \mathcal{D} \right\rangle $ is defined by
\begin{equation}
C_\delta(\Psi) \equiv \frac{N}{n_\text{min}}.\label{eq:37}
\end{equation}

The optimal guessing probability between two states $\rho$ and $\sigma$ is given by
\be
P[\rho,\sigma] = \frac{1}{2}+\frac{1}{4}\text{tr}|\rho-\sigma|.
\ee
Hence, one can compute the average guessing probability with $n$ subsystems given all the $\binom{N}{n}$ reduced density matrices $\rho_{\mathcal{A}}^{(n)}= \text{Tr}_{N-n}\prjct{\mathcal{A}}$  and $\rho_{\mathcal{D}}^{(n)}= \text{Tr}_{N-n}\prjct{\mathcal{D}}$. The complexity of the minimization is significantly reduced if the initial state is permutationally invariant.

The authors discuss several aspects of their measures, including the dependency on the choice of splitting into \alive and \dead, cases for which the two branches do not have equal amplitudes and slight variations of the definition of $C_\delta(\Psi)$. In particular, the author emphasize that $P_g$ should be chosen to be close to one for a reasonable notion of macroscopic distinctness.
\citet{Volkoff_Measurement-_2014} discuss a potential formulation for photonic systems (see also Sec.~\ref{sec:link-meas-phot}).

\subsubsection{\citet{Marquardt_Measuring_2008}}
\label{sec:Marquardt}

\fbox{\parbox{0.9\columnwidth}{The macroscopic quantumness $\bar{D}$ of a superposition \da is defined by the number of elementary operations that have to be applied in order to map \alive to $| \mathcal{D} \rangle $.}}\\

If the system admits a partition into $N$ subsystems or consists of $N$ particles, one can also define a set of all elementary operations. For example, such a set can include all operations that only affect one subsystem, or only modify the state of one particle. \citet{Marquardt_Measuring_2008} focus on the example of $N$ fermions, where an elementary operation is naturally given by the exchange of one fermion $c_j^\dag c_k$. Their idea is to quantify the distinctness of the states \alive and \dead by counting how many elementary operations one has to apply to go from one state to the other. More precisely, they introduce a hierarchy of Hilbert spaces $\cH_d$ for $d\geq 0$, where $\cH_0=\text{span}\{\ket{\mathcal{A}}\}$ and each new subspace $\cH_{d+1}$ is constructed from the previous spaces in the following way.  One takes the span $\tilde{\cH}_{d+1}$ of all states that can be obtained by applying a single elementary operation to all states in $\cH_{d}$. Then, $\tilde{\cH}_{d+1}$ is made orthogonal to all subspaces $\cH_0,\dots,\cH_d$ by subtracting them, which is called $\cH_{d+1}$. In this way all subspaces of the hierarchy are orthogonal (hence the direct sum notation is justified) and, for a properly chosen set of elementary operations, one finds $\cH = \bigoplus_{d=0}^\infty\cH_d$. Consequently, the state \dead admits a unique decomposition
\be
\ket{\mathcal{D}}=\sum_{d=0}^\infty \lambda_d \ket{\nu_d},
\ee
where each $\ket{v_d}\in \cH_d$ is the projection of the state on the corresponding subspace. The distinctness between the states \alive and \dead is then defined as
\begin{equation}
\bar D = \sum_{d=0}^\infty |\lambda_d|^2 d\label{eq:38}
\end{equation}
quantifying the average number of elementary operations that are needed to go from \alive to $| \mathcal{D} \rangle $. The authors note that the measure is in general not symmetric under exchange of \alive and $| \mathcal{D} \rangle $. \citet{Volkoff_Measurement-_2014} propose an extension of this measure to multi-mode photonic systems by discussing a specific example.

\subsubsection{\citet{Lee_Quantification_2011} and \citet{Park_Quantum_2016}}
\label{sec:LeeJeong}

\fbox{\parbox{0.9\columnwidth}{The macroscopic quantumness $\mathcal{I}$ of the state is definied by quantifying the non-classical features of its phase-space representation (i.e., the frequency and amplitude of Wigner function oscillations).}}\\

Several measures discussed so far have been defined with respect to an observable $A$, which gives rise to a preferred basis. Alternatively, one can also consider representations of quantum states that are not just expansions in a given basis. In particular, the Wigner function representation is commonly used in phase space to visualize quantum states. \citet{Lee_Quantification_2011} observe that the Wigner function of states that are intuitively considered to be macroscopically quantum (e.g., $\ket{\alpha}+\ket{-\alpha}$) exhibit two or more distinct peaks with some oscillating pattern between the peaks.
In contrast, classical states are known to have a positive and smooth Wigner function.
Following this intuition \citeauthor{Lee_Quantification_2011} propose to quantify the size of a state by the ``frequency'' of oscillations of its Wigner function. Formally, the size of a quantum state $\rho$ for $M$ bosonic modes is defined as
\begin{equation}
\mathcal{I}(\rho) = \frac{\pi^M}{2} \int d{\bm \alpha}^2 W({\bm \alpha})\sum_{m=1}^M\left(-\frac{\partial^2}{\partial\alpha_m\partial\alpha_m^*}-1 \right)W({\bm \alpha}),\label{eq:5}
\end{equation}
 which takes a simpler form when expressed in terms of the characteristic function $\chi({\bm \xi})=\text{tr}\, \rho\, \exp(\sum_m \xi_m a_m +\xi_m^* a_m^\dag)$. As the authors show, the quantity $\mathcal{I}(\rho)$ can equally be expressed as the susceptibility of the state to lose purity when all the modes are subject to photon loss, that is,
\begin{equation}
\mathcal{I}(\rho)=-\frac{1}{2} \left.\frac{d\,\mathcal{P}(\rho_t)}{dt}\right|_{t=0} ,   \label{eq:6}
\end{equation}
with the purity of a state $\mathcal{P}(\rho)\equiv \text{tr}\rho^2$. The decoherence process is specified by
\begin{equation}
\dot \rho_t = \sum_m \left(a_m \rho_t a_m^\dag - \frac{1}{2}\{\rho_t,a^\dag_m a_m \}\right).\label{eq:7}
\end{equation}

\citeauthor{Lee_Quantification_2011} emphasize that $\mathcal{I}(\rho)$ simultaneously quantifies  the quantumness and the macroscopicity of a state. In particular there is no need to assume that the state is pure. One simply expects that the oscillations of the Wigner function are smoothed with noise, as it happens for the examples discussed in the paper. For later comparison, we note that $\mathcal{I}(\Psi) = \frac{1}{2}[(\Delta X)^2 + (\Delta P)^2 - 1]$ for pure states. \citet{Gong_Comment_2011} remarks that $\mathcal{I}(\rho)$ can become negative, which can be simply fixed by adding $1/2$ to the definition of $\mathcal{I}(\rho)$ \cite{Jeong_Reply_2011}.

This measure for the Wigner function was later generalized to spin ensembles with $N$ particles with each spin $S$ \cite{Park_Quantum_2016}. A similar reasoning as before leads to measure the frequency components of the so-called Stratonovich-Weyl distribution, which is the spin-equivalent of the Wigner function \cite{Klimov_Group-Theoretical_2009,Agarwal_State_1998}. In contrast to the phase space treatment, \citeauthor{Park_Quantum_2016} choose to maximize the frequency measure over all quantization axes (using collective spin operators $A$). In addition, they add the purity in the denominator and define
\begin{equation}
\label{eq:27}
\mathcal{I}_S(\rho) = \frac{1}{N S \mathrm{Tr}\rho^2} \max_{A: \mathrm{collective}} \mathrm{Tr} \left[ A^2 \rho^2 - \rho A \rho A \right].
\end{equation}
The measure is particularly compared to the quantum Fisher information (see Sec.~\ref{sec:QFI}). In addition, its role in quantum phase transitions is discussed (see Sec.~\ref{sec:quant-phase-trans}).

\subsubsection{\citet{Frowis_Measures_2012} and followups}
\label{sec:QFI}

\fbox{\parbox{0.9\columnwidth}{The macroscopic quantumness $N_\text{eff}$ of a many-body state is related to its maximal quantum Fisher information with respect to all extensive observables $A$. The QFI can be read as a signature of interferometric improvement offered by the state  over product states, or as an extension of the variance to mixed states.}}\\

We consider a spin ensemble with $N$ particles. \citet{Frowis_Measures_2012} argue that a genuinely macroscopic quantum state of the joint system should exhibit some quantum effect that does not reduce to an accumulation of microscopic quantum effects displayed by its individual constituents. In this sense the effective size of a state can be intuitively thought of as the minimal irreducible number of constituents. To formalize this guideline, \citeauthor{Frowis_Measures_2012} use the quantum Fisher information  \cite{Helstrom_Quantum_1976,Holevo_Probabilistic_2011,Braunstein_Statistical_1994}
\begin{equation}
\label{eq:28}
\mathcal{F}(\rho,A) = 2 \sum_{k,l} \frac{(\pi_k - \pi_l)^2}{\pi_k + \pi_l} \left| \left\langle \psi_k \right| A \left| \psi_l \right\rangle  \right|^2,
\end{equation}
with the spectral decomposition of the state, $\rho = \sum_k \pi_k \left| \psi_k \right\rangle\!\left\langle \psi_k\right|$. The quantum Fisher information is a measure of the susceptibility of $\rho$ to small influences generated by $A$ (see Sec.~\ref{sec:witn-macr-quant}).

The effective size of the state is defined as
\begin{equation}
N_{\text{eff}}(\rho) = \frac{1}{4N}\max_{A: \mathrm{local}} \mathcal{F}(\rho,A).\label{eq:10}
\end{equation}
Note that the quantum Fisher information reduces to four times the variance for pure states. Furthermore, it is the convex roof of the variance (see \citet{Yu_Quantum_2013} and Sec.~\ref{sec:conn-betw-meas-1}). The normalization factor $1/(4N)$ is chosen such that all pure separable states have size 1, while the maximal possible size is $N$ (attained by the GHZ state).
This measure has an operational aspect as, on the one hand, a large quantum Fisher information can be witnessed via fast unitary time evolution generated by $A$ \cite{Frowis_Kind_2012}. On the other hand, it has an applied aspect as $N_{\mathrm{eff}}(\rho)$ tells us how much better $\rho$ can be in a potential parameter estimation scenario compared to the best separable states (see Sec.~\ref{sec:quantum-metrology}). The definition Eq.~(\ref{eq:10}) might be extended to quasi-local observables $A$ (i.e., sums of few-particle operators; see also example \ref{ex:cluster} in Sec.~\ref{sec:ex:spin-ensemble}).

In order to explicitly deal with a ``dead and alive'' structure, the authors define the so-called relative Fisher information for $| \Psi \rangle \propto \left|   \mathcal{A}\right\rangle +  \left|\mathcal{D}\right\rangle$,
\begin{equation}
\label{eq:29}
N_{\mathrm{eff}}^{\mathrm{rel}}(\Psi) = \frac{N_{\mathrm{eff}}(\Psi)}{\frac{1}{2}N_{\mathrm{eff}}(\mathcal{A}) + \frac{1}{2}N_{\mathrm{eff}}(\mathcal{D})}.
\end{equation}

Later, the proposal Eq.~(\ref{eq:10}) was extended to photonic systems. As argued by \citet{frowis15,Oudot_Two-mode_2015}, the equivalent of local operators in spin ensembles are quadrature operators in phase space. The factor $1/(4N)$ is generally replaced by the quantum Fisher information of the ``most classical'' state $N_{\mathrm{eff}}(\Psi_{\mathrm{classical}})$ (see also \citet{Frowis_Lower_2017}). For spin ensemble, this is chosen to be product states. In phase space, coherent states are selected as the most-classical pure states. To handle multi-mode situations, one considers sums of quadrature operators $X_{\bm{\theta}} = \sum_{m = 1}^M X_{\theta_m}^{(m)}$ with $X_{\theta}^{(m)} = \cos(\theta) X^{(m)} + \sin(\theta) P^{(m)}$ and maximizes over the angles $\bm{\theta} = (\theta_1,\dots,\theta_M)$, that is,
\begin{equation}
\label{eq:12}
N_{\mathrm{eff}}(\rho) = \frac{1}{2M}\max_{\bm{\theta}} \mathcal{F}(\rho,X_{\bm{\theta}}).
\end{equation}

\subsubsection{\citet{Nimmrichter_Macroscopicity_2013}}
\label{sec:NimmrichterHornberger}

\fbox{\parbox{0.9\columnwidth}{The macroscopic quantumness $\mu$ of an experimental setup is defined via the range of unconventional mass-induced decoherence models (as potential modifications of standard quantum mechnics) of some type that are ruled out by the experiment.}}\\

The proposal presented in this paragraph differs from most other ideas, as \citet{Nimmrichter_Macroscopicity_2013} do not consider the macroscopic quantumness of an isolated state (including the structure of additional operators or partitions). They attribute a size to a whole experiment, from the preparation step to the time evolution and to the observation of measurement results. This is motivated by their goal to evaluate how well experiments exclude slightest variations of standard quantum mechanics. In particular, \citeauthor{Nimmrichter_Macroscopicity_2013} focus on dynamical modifications also known as collapse models for spatial superpositions of massive systems \cite{Bassi_Models_2013}.

More precisely, the authors consider a model in which the master equation of the system composed of $N$ particles $\dot \rho = [H, \rho]/(\ii \hbar) + \mathcal{L}_N(\rho)$ is modified by the addition of a dissipative term
\begin{align}\label{eq:ME}
\mathcal{L}_N(\rho)= \frac{1}{\tau_e}\int {\bf d}({\bf s},{\bf q}) \left( W_N({\bf s},{\bf q})\rho W_N^\dag({\bf s},{\bf q})\right. \nonumber \\ - \left.\frac{1}{2}\{W_N^\dag({\bf s},{\bf q})W_N({\bf s},{\bf q}),\rho\} \right)
\end{align}
with
\begin{equation}
W_N({\bf s},{\bf q})= \sum_{n=1}^N \frac{m_n}{m_e}\exp\left(\frac{\ii}{\hbar}(\frac{m_e}{m_n} {\bf s}\cdot {\bf P}_n - {\bf q}\cdot{\bf X}_n) \right),\label{eq:8}
\end{equation}
where ${\bf X}_n$, ${\bf P}_n$ and $m_n$ are the position operator, the momentum operator and the mass of the $n$-th particle, respectively, while ${\bf d}({\bf s},{\bf q}) = g_e({\bf s},{\bf q}) d^3 {\bf s} d^3{\bf q}$ is a measure with an isotropic phase-space distribution $g_e({\bf s},{\bf q})$. The particular form of the modification term is motivated by several physical requirements imposed on the model, like the invariance under Galilean transformation, symmetry under particle exchange and others \cite{Nimmrichter_Macroscopicity_2013}. At this stage, it is fully specified by the coherence time parameter $\tau_e$ and the distribution $g_e({\bf s},{\bf q})$. (The reference mass parameter $m_e$ can be absorbed in the previous two.) The authors further assume the distribution to be a product of Gaussians for ${\bf s}$ and ${\bf q}$, with the corresponding width $\sigma_s$ and $\sigma_q$. Hence, the dynamical modification of quantum mechanics is just described by three parameters $\tau_e$, $\sigma_s$ and $\sigma_q$. From this, the macroscopicity measure $\mu$ of an experiment is defined as a recalibration
\be
\mu \equiv \log_{10}\left(\frac{\tau_e}{1 \text{s}}\right)
\ee
of the greatest time parameter $\tau_e$ excluded by the experiment. For this, one optimizes over the other two parameters with $\sigma_s\leq 10$ pm and $\hbar/\sigma_q\geq 10$ fm, which are argued to be the limiting value for which a non-relativistic treatment is still valid. Concretely, the modified master equation Eq.~\eqref{eq:ME} predicts an evolution where branches that correspond to different phase space configurations of massive particles progressively decohere. This effect constrains the results of any measurement that register the interference between different branches (e.g., interference visibility in a matter-wave interferometer). Reciprocally any experimental data obtained from such a measurement puts a limit on the modification term. The stringency of this limit is quantified by the measure $\mu$.

\subsubsection{\citet{Sekatski_Size_2014,Sekatski_General_2017}}
\label{sec:Sekatski}
\label{sec:sekatski2017}

\fbox{\parbox{0.9\columnwidth}{The macroscopic quantumness --Size for \da superpositons and MIC for multicomponent superposition states-- is identified by how much one can learn about the state with a classical detector that lacks microscopic resolution.}}\\

\citet{Sekatski_Size_2014} define macroscopic distinctness for the two components of a pure state \da by asking how well the two states can be distinguished with a coarse-grained measurement of some observable $A$. Similarly as in other proposals, the observable is chosen depending on the experimental context. This measurement is said to be performed with classical detectors. If the distributions of the states \alive  and \dead with respect to the eigenbasis of the observable $A$ are given by $p^0_{\mathcal{A}}(a)$ and $p^0_{\mathcal{D}}(a)$ respectively, the distributions observed with a coarse-grained measurement are given by their convolution
\begin{equation}
p_{\mathcal{A}}^\sigma(\lambda)=\sum_a n_\sigma(a|\lambda) p^0_{\mathcal{A}}(a),\label{eq:31}
\end{equation}
with the coarse-graining $n_\sigma(a|\lambda) = n_\sigma(a - \lambda)$ typically chosen as a Gaussian distribution of width $\sigma$ and mean $a$ (and similar for $| \mathcal{D} \rangle $). Hence, the probability to correctly distinguish the two states with such a measurement is given by
\begin{equation}
P^\sigma[\,\ket{\mathcal{A}},\ket{\mathcal{D}}] = \frac{1}{2}+\frac{1}{4}\int |p_{\mathcal{A}}^\sigma(\lambda)-p_{\mathcal{D}}^\sigma(\lambda)| d\lambda.\label{eq:32}
\end{equation}

A measure of macroscopicity is constructed from this probability provided a choice of a target guessing probability $P_g$ and a calibration set of states. This set is a range of superposition states $\ket{\Psi_N}$ with a naturally defined size $N$. To do so, one first computes the maximal allowed coarse-graining $\sigma_{\max} = \sup \{\sigma| P^\sigma[\,\ket{\mathcal{A}},\ket{\mathcal{D}}] \geq P_g \}$ that still allows one to distinguish \alive and \dead with a probability of at least $P_g$. Then, one identifies the size of the superposition $| \Psi \rangle = \left| \mathcal{A} \right\rangle + \left| \mathcal{D} \right\rangle $ with the smallest state from the calibration set that attains the same guessing probability $P_g$ under the same amount of coarse-raining $\sigma_\text{max}$
\begin{equation}
\text{Size}(\Psi) = \inf \{ N| P^{\sigma_\text{max}}[\ket{\Psi_N}]\geq P_g \}.\label{eq:34}
\end{equation}
A natural choice of the calibration states $\ket{\Psi_N}$ are the superpositions of two eigenstates of $A$ separated by $N$. With such a choice one has $P^{\sigma_\text{max}}[\ket{\Psi_N}] =(1+\text{Erf}(N/(2\sqrt{2}\sigma_\text{max})))$ and the size can be directly expressed as a function of $\sigma_\text{max}$ and $P_g$.

The idea of using classical detectors with limited resolution to define macroscopic distinctness \cite{Sekatski_Size_2014} was extended to general states without a predetermined \da structure. \citet{Oudot_Two-mode_2015} adapt the work of \citet{Sekatski_Size_2014} to argue that two-mode squeezed states can be considered as macroscopically quantum  (see Sec.~\ref{sec:examples}, example~\ref{ex:squeezed}).

Later, \citet{Sekatski_General_2017} rephrased the question of how well \alive and \dead can be distinguished by asking how much information can be learned from a state by measuring it with a classical device. Similar as before, consider a pure state $| \Psi \rangle $ with the probability distribution $p^0_{\Psi}(a)$ and the convolution $p^\sigma_{\Psi}(\lambda)$.  The information the classical detector can learn about $| \Psi \rangle $ can be quantified by the mutual information
\begin{equation}
\label{eq:30}
I_{\sigma} = H(p^0_{\Psi}(a)) - \int p^\sigma_{\Psi}(\lambda) H(n_\sigma(a|\lambda)),
\end{equation}
where $H(p(x)) = - \sum_x p(x) \log p(x)$ is the Shannon entropy. \citet{Sekatski_General_2017} define the size of $| \Psi \rangle $ as the largest $\sigma$ for which $I_{\sigma}$ gives at least $b$ bits of information
\begin{equation}
\label{eq:33}
\mathrm{MIC}_{b}(\Psi) = \max \left\{ \sigma | I_{\sigma} \geq b  \right\}.
\end{equation}
To be more precise, Eq.~(\ref{eq:33}) defines a family of measures parametrized by $b$. A natural way of extending Eq.~(\ref{eq:33}) to mixed states, $\widehat{\mathrm{MIC}}_b(\rho)$, is done via the convex roof construction (see Sec.~\ref{sec:extens-vari-mixed2}).

This idea can be rephrased by using von Neumann's pointer model in which the system is coupled to an auxiliary system $E$ in state $| \xi_{\Delta} \rangle $ with initial spread $\Delta$ via $U = \exp(-i A \otimes p)$. The disturbance of the post-measured state $\rho^{\prime} = \mathrm{Tr}_E U \rho \otimes \left| \xi_{\Delta} \right\rangle\!\left\langle \xi_{\Delta} \right| U^{\dagger}$ is measured via
\begin{equation}
\label{eq:50}
C_{\Delta}(\rho) = S(\rho^{\prime}) - S(\rho),
\end{equation}
where $S$ is the von Neumann entropy. \citet{Sekatski_General_2017} show that this is a sensitive measure for macroscopic coherence similar to Eq.~(\ref{eq:30}). Like before, inversion of Eq.~(\ref{eq:50}) leads to a measure that scales (at most) with the spectral radius and that is applicable to general quantum states, but without the need of a convex-roof construction.

\subsubsection{\citet{Laghaout_Assessments_2015}}
\label{sec:laghaout}

\fbox{\parbox{0.9\columnwidth}{The measure consists of two parts. The objective part $\mathcal{N}$ quantifies the quantum fluctuations of the state in phases space. The subjective part $\mathcal{D}$ quantifies the average distinguishability of each component of in the superposition from all the others. The product $\mathcal{N}\times \mathcal{D}$ is proposed as a measure for macroscopic quantumness.}}\\

\citet{Laghaout_Assessments_2015} are interested in characterizing systems that are large, quantum and are composed of macroscopically distinct branches in at least some of its subsystems. The authors exclusively treat pure states and divide their attempt into two parts, which they call ``objective'' and ``subjective'' macroscopicity.

For the objective macroscopicity, they note that the spread of phase space distribution accounts for the quantum fluctuations (i.e., coherence). They hence define
\begin{equation}
\label{eq:19}
\mathcal{N} = \frac{1}{2}[ (\Delta X)^2 + (\Delta P)^2 - 1]
\end{equation}
as the quantum fluctuations that go beyond the spread of a coherent state (cf.~\citet{Lee_Quantification_2011} in Sec.~\ref{sec:LeeJeong}).

The subjective part of the measure builds on previous work on characterizing the macroscopic distinctness using classical detectors with limited resolution (in particular,  \citet{Sekatski_Size_2014}). The attribute ``subjective'' comes from the choice that has to be done for the measurement. This reduces the problem to a distinction of probability distributions $P(\lambda)$ with measurement outcomes $\lambda$. Unlike previous works, \citeauthor{Laghaout_Assessments_2015} develop a formalism that allows to define macroscopic distinctness for more than two states and for different weights. Suppose that a pure state $| \psi \rangle $ is written as a superposition of preselected states $| b_k \rangle $ with probability amplitudes $c_k$. Then, one calculates the distance between any $| b_k \rangle $ and the mixture of all other $\left\{ | b_l \rangle  \right\}_{l\neq k}$
\begin{equation}
\label{eq:20}
\tilde{\rho}_k = \frac{\sum_{l\neq k} |c_l|^2 \left| b_l \right\rangle\!\left\langle b_l\right|  }{\sum_{l\neq k} |c_l|^2}.
\end{equation}
The specific formulation depends on the chosen distance measure. The authors discuss one based on the Bhattacharrya coefficient \cite{Bhattacharyya_Measure_1946}
\begin{equation}
\label{eq:21}
\mathcal{D}^{\mathrm{BC}} = 1 - \sum_k |c_k|^2 \int \sqrt{P(\lambda| b_k  ) P(\lambda| \tilde{\rho}_k) } d\lambda
\end{equation}
and another one based on the Kolmogorov distance \cite{Fuchs_Cryptographic_1999}
\begin{equation}
\label{eq:22}
\mathcal{D}^{\mathrm{KD}} = \frac{1}{2} \sum_k |c_k|^2 \int |P(\lambda| b_k  ) -P(\lambda| \tilde{\rho}_k)| d\lambda.
\end{equation}
The subjectivity of $\mathcal{D}$ becomes clear by noting that, for orthogonal $| b_k \rangle $ and optimizing over \textit{all} measurements, the distinguishability can always be maximal, that is, $\mathcal{D} = 1$.

The combined proposal is a product of both measures of macroscopicity
\begin{equation}
\label{eq:23}
\mathcal{M} = \mathcal{N} \times \mathcal{D}.
\end{equation}

\subsubsection{\citet{Yadin_Quantum_2015}}
\label{sec:yadin}

\fbox{\parbox{0.9\columnwidth}{The macroscopic quantumness $N^*$ of a many-body state $\ket{\psi}$ is defined as the maximal average size of the GHZ state that can be prepared from $\ket{\psi}$ with local operation and classical communication.}}\\

The work of \citet{Yadin_Quantum_2015} is devoted to view macroscopic quantumness as a statement about long-range quantum correlations. In the doctrine of multipartite entanglement, LOCC cannot create nonclassicality. Using LOCC protocols one can only reveal entanglement properties that have been present before. From this point of view, it is natural not only to consider, for example, the GHZ state as macroscopically quantum but also all other quantum states that can be brought to a GHZ state by means of LOCC.

To formalize this insight, \citeauthor{Yadin_Quantum_2015} propose to consider generalized local measurements $M_a$ that map $| \psi \rangle $ to $| \mathrm{GHZ}_{n_a} \rangle $, $n_a \leq N$,  with a probability $p_a$ (cf.~\citet{Dur_Effective_2002}; the remaining $N-n_a$ qubits are no longer of interest). Then, the effective size of $| \psi \rangle $, $N^{*}(\psi)$, is defined as
\begin{equation}
\label{eq:24}
N^{*}(\psi) = \max_{M_a} \sum_a p_a n_a.
\end{equation}

There exist quantum states detected as macroscopically quantum by \citet{Yadin_Quantum_2015} that are not identified by any other proposal (see example \ref{ex:cluster} in Sec.~\ref{sec:ex:spin-ensemble}). The reason is that the LOCC paradigm --which forms the basis of this proposal-- stems from entanglement theory and is not considered in other proposals.

\subsubsection{\citet{Kwon_Disturbance-Based_2016}}
\label{sec:kwon}

\fbox{\parbox{0.9\columnwidth}{The macroscopic quantumness $M_\sigma$ of a state is quantified by the sensitivity of the state to dephasing noise generated by an observable $A$. A large sensitivity is argued to reveal the presence of coherence between spectrally distant eigenstates of $A$ in the state, i.e. eigenstates with very different eigenvalues.}}\\

The approach of \citet{Kwon_Disturbance-Based_2016} is motivated by finding an operational meaning of quantum macroscopicity in terms of coherence (in the sense of \citet{Yadin_General_2016}, see Sec.~\ref{sec:conn-reso-theory}). After fixing a spectrum and a basis by choosing an observable $A = \sum_k a_k \left| k \right\rangle\!\left\langle k\right| $, the authors define a dephasing channel with Kraus operators $Q^{\sigma}_x = \sum_k \sqrt{q^{\sigma}_k(x)} \left| k \right\rangle\!\left\langle k\right| $, where $q^{\sigma}_k(x)$ is a Gaussian function with mean value $a_k$ and spread $\sigma$. The physical realization of this dephasing channel can come from environmentally induced decoherence or from a low-resolution measurement for which $\sigma$ is the resolved scale. Then, a state of interest $\rho$ is compared with the same state after dephasing, $\Phi_{\sigma}(\rho) = \sum_x Q^{\sigma}_x \rho Q^{\sigma \dagger}_x$. The measure is defined as
\begin{equation}
\label{eq:25}
M_{\sigma}(\rho) = D(\rho, \Phi_{\sigma}(\rho)),
\end{equation}
where $D$ is a distance-like function. The authors mention the Bures distance and the quantum relative entropy but characterize the general properties of $D$ such that Eq.~(\ref{eq:25}) fulfill the criteria of \citet{Yadin_General_2016} (see Sec.~\ref{sec:conn-reso-theory}) for all $\sigma > 0$.

As noted by \citeauthor{Kwon_Disturbance-Based_2016}, $M_{\sigma}(\rho)$ may lead to surprising results for small $\sigma$. For example, the spin-coherent state gives higher values than the GHZ state contradicting all other proposed measures and our intuition. To fix this issue, the authors argue that $\sigma$ has be sufficiently large in order to faithfully represent macroscopic distinctness. In particular with the interpretation as the scale of measurement resolution, $\sigma$ should be in the ``classical'' regime. Several heuristic arguments show that a measurement with resolution $\sigma \gtrsim O(\sqrt{N})$ for collective spin operators and $\sigma \gtrsim O(1)$ for quadrature operators in phase space can be reasonably called classical. This is further supported by the inequality
\begin{equation}
\label{eq:26}
M_{\sigma}(\rho) \leq 2 \left( 1 - e^{-\frac{I_W(\rho,A)}{4 \sigma^2}} \right),
\end{equation}
where $I_W(\rho,A) = -\frac{1}{2} \mathrm{Tr}[\sqrt{\rho},A]^2$ is the Wigner-Yanase-Dyson skew information, which reduces to $I_W(\Psi,A) =(\Delta A)^2/2$ in the case of pure state $\left| \Psi \right\rangle$. Hence, large variance as a measure for macroscopic quantumness proposed by \citet{Shimizu_Stability_2002,Lee_Quantification_2011,Frowis_Measures_2012} is necessary for large values of $M_{\sigma}(\Psi)$ in the case of $\sigma = O(\sqrt{N})$. For mixed states, the relation to the quantum Fisher information $4 I_W(\rho,A) \leq \mathcal{F}(\rho,A)\leq 8 I_W(\rho,A)$ \cite{Kwon_Disturbance-Based_2016} makes further connections to the measure of \citet{Frowis_Measures_2012}.

\subsection{Examples}
\label{sec:examples}

We now present a selection of quantum states for spin ensembles and photonic systems, to which we apply the measures presented in the previous section. The states and operators appearing in the examples are defined in Sec.~\ref{sec:physical-systems} and the measures are defined in Sec.~\ref{sec:preliminary-measures}. Note that massive systems and superconducting states are discussed in Sec.~\ref{sec:superc-quant-interf}, as they are more specific examples of real experiments.

Not all measures are originally conceived to be applicable to both spins and photons. Measures not mentioned in an example implies that it is not clear how to apply it to this specific case. The reason might be the measure is not applicable to the physical system (spins or photons) or that it is difficult to apply it to the specific instance. In the following, approximate expressions are valid in the large $N$ regime. We refer to further examples discussed by \citet{Volkoff_Measurement-_2014,Volkoff_Macroscopicity_2014} and by the papers reviewed in Sec.~\ref{sec:preliminary-measures}.

\subsubsection{Spin ensemble}
\label{sec:ex:spin-ensemble}

While connections and differences between the measures are discussed later in Secs.~\ref{sec:measurediscussion}--\ref{sec:physical-setups}, we would like to point out similarities between some measures when applied to pure spin states $| \Psi \rangle $. \citet{Shimizu_Stability_2002,Frowis_Measures_2012,Park_Quantum_2016} optimize the variance of the state over all local operators\footnote{Strictly speaking, \citet{Park_Quantum_2016} optimize over collective operators. Since in all examples the states are symmetric, this leads to the same results.}. This means that \citet{Frowis_Measures_2012,Park_Quantum_2016} will result in the same values $\mathcal{I}(\Psi) = N_{\mathrm{eff}}(\Psi)$. \citet{Shimizu_Stability_2002} are merely interested in the scaling of the variance in $N$, that is, $\mathcal{I} = O(N^{p-1})$. Note that there is an efficient method to calculate the maximal variance \cite{Ukena_Appearance_2004,Guhne_Covariance_2007}. For the measure of \citet{Bjork_Size_2004} there is some room for variations, but following some of their spin examples, one way of defining their measure $M$ is to look at the maximal standard deviation over all collective operators divided by the maximal spread of a spin-coherent state (i.e., $\sqrt{N}$). It follows that this exactly corresponds the square root of the variance-based measures \cite{Frowis_Measures_2012,Park_Quantum_2016}. In summary, for pure spin ensemble states $| \Psi \rangle $ we find that\footnote{See table \ref{tab:overview2} for an overview of the mathematical symbols used for the measures.}
\begin{equation}
\label{eq:54}
M(\Psi)^2 = \mathcal{I}(\Psi) = N_{\mathrm{eff}}(\Psi) = O(N^{p-1}).
\end{equation}
In addition, note that the measure of \citet{Kwon_Disturbance-Based_2016} is connected to $N_{\mathrm{eff}}$ via $M_{\sigma}(\rho) \lesssim \frac{N}{2\sigma^2} N_{\mathrm{eff}}(\rho)$ in the limit $\sigma \gg \sqrt{N}$ (see paragraph around Eq.~(\ref{eq:26})). Similarly, the measure of \citet{Sekatski_General_2017} has a relation to $N_{\mathrm{eff}}$ in the limit of small $b$ via $\widehat{\mathrm{MIC}}_b(\rho) \approx \sqrt{\frac{N_{\mathrm{eff}}(\rho) N}{2 b \log 2}}$  \cite{Sekatski_General_2017}.

\begin{Exp}
  \label{ex:GHZ}[Generalized GHZ state]

\begin{table}
  \centering
  \begin{tabular}{p{1.2cm} p{1.7cm} p{5cm}}
    
   Scaling & Connection to width & Measures \\
    
    \hline \hline

$N \epsilon^2$ & $(\Delta A)^2/N$ & \citet{Dur_Effective_2002}, \citet{Korsbakken_Measurement-based_2007}, \citet{Marquardt_Measuring_2008}, \citet{Frowis_Measures_2012}, \citet{Park_Quantum_2016}, \citet{Yadin_Quantum_2015}, \citet{Kwon_Disturbance-Based_2016}\footnote{In case of fixed $\sigma/\sqrt{N}$.}\\
    
    \hline

    $N \epsilon$ & $\Delta A$ & \citet{Leggett_Macroscopic_1980} ($\Lambda$), \citet{Kwon_Disturbance-Based_2016}\footnote{In case of fixed $\sigma$.}, \citet{Sekatski_General_2017}\\
    
    \hline

    $\sqrt{N} \epsilon$ & $(\Delta A)/\sqrt{N}$ & \citet{Bjork_Size_2004} \\

    \hline

    $N$ & Independent & \citet{Leggett_Testing_2002} ($D$)\footnote{For a large range of $\epsilon$, see text.}, \citet{Shimizu_Stability_2002}\footnote{For all $\epsilon = O(1)$.}\\
    
    \hline  \hline
  \end{tabular}
\caption{Evaluation of generalized GHZ state, example \ref{ex:GHZ}, with several measures. Disregarding prefactors, we find four different scaling classes for the effective size, which can be connected to the width of the state maximized over all collective observables $A$. }
  \label{tab:ghz}
\end{table}

  This state, Eq.~(\ref{eq:40}), is discussed in two different regimes. For $\epsilon = \pi/2$, one recovers the GHZ state, Eq.~(\ref{eq:39}), which is considered to be macroscopically quantum by all measures applicable to spin ensembles. In the other regime, $\epsilon \ll 1 $, \citet{Dur_Effective_2002} show that the state behaves like a GHZ state with reduced system size $N \epsilon^2$ (see Sec.~\ref{sec:effect-size-comp} and Fig.~\ref{fig:TwoGaussians}). This example was used by many subsequent proposals to test and scale the measures.
In particular, as shown in the respective papers, many measures\footnote{\citet{Korsbakken_Measurement-based_2007,Marquardt_Measuring_2008,Frowis_Measures_2012,Park_Quantum_2016,Yadin_Quantum_2015}; the measure of \citet{Korsbakken_Measurement-based_2007} assigns the value $C_{\delta}\approx N \epsilon^2/\log(1/\delta)$ for $\epsilon,\delta \ll 1$.} have the same effective size in this parameter regime. The framework of \citet{Shimizu_Stability_2002} leads to $p = 2$ whenever $\epsilon = O(1) > 0$. The measure of \citet{Bjork_Size_2004} gives $M \approx \sqrt{N} \epsilon$. Measures for which one optimizes over local operators are maximal for $A_{\mathrm{opt}} = \cos \epsilon S_x + \sin \epsilon S_z$.

\begin{figure}[htbp]
\centerline{\includegraphics[width=.9\columnwidth]{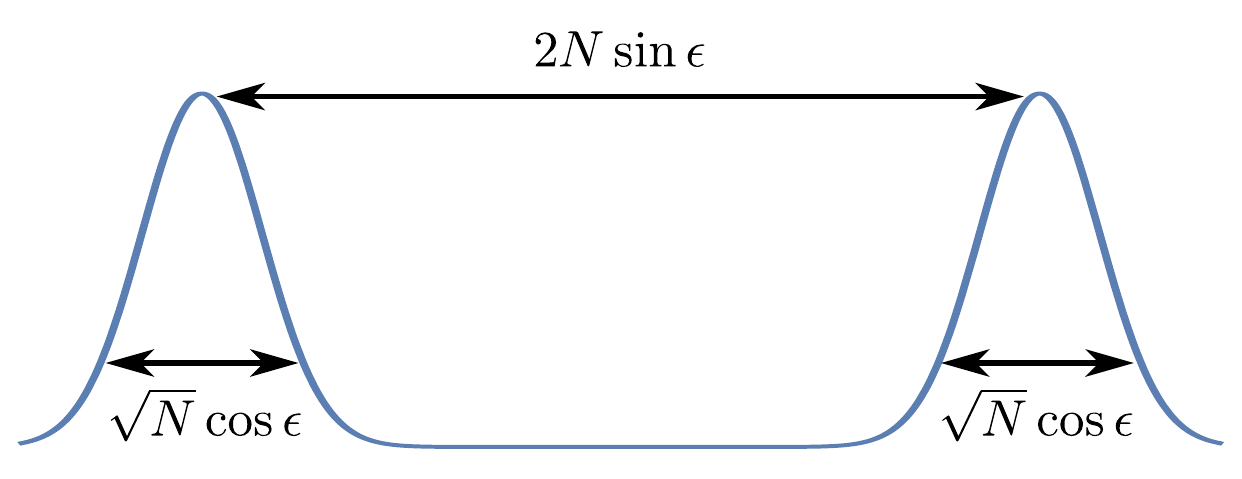}}
\caption[]{\label{fig:TwoGaussians} The probability distribution of the generalized GHZ state, example \ref{ex:GHZ}, in the basis of $A_{\mathrm{opt}}$, which is the local operator maximizing the difference between the expectation values. In many papers, one considers lowest-order approximations of $\epsilon$.}
\end{figure}

  The behavior of the disconnectivity from \citet{Leggett_Macroscopic_1980} is rather different for this example. For instance in the case $1 \ll N \epsilon^2 \ll N$, $\delta_M$ drops to zero only for $M_{\max} \gtrsim N - c/\epsilon^2$ with some characteristic constant $c$. Hence, the disconnectivity can be much larger than $N \epsilon^2$. The difference becomes clear when noting that $\delta_M$ is large when the bipartite splitting $M : (N-M)$ is entangled (by roughly one ebit), which is the case up to $M_{\max}$. This suggests to see Leggett's extensive difference as a measure for macroscopic distinctness (here, $\Lambda = N \sin \epsilon$) and interpret the disconnectivity as a measure of quantumness (see also the discussion about separate consideration of macroscopic distinctness and quantumness in Sec.~\ref{sec:macr-superp}).

  In order to calculate the measure of \citet{Sekatski_Size_2014}, one chooses the optimal measurement basis $A_{\mathrm{opt}}$. In the regime in which $N \epsilon \gg \sqrt{N}$, the actual width of the two components does not play a significant role. The effective size of the generalized GHZ state simply scales as the distance between the two peaks, that is, $\mathrm{Size} \propto 2 N \epsilon$. The same result in scaling is found by applying the framework of \citet{Sekatski_General_2017} in the regime $b < 1$, which is the relevant regime when discussing the macroscopic distinctness between two peaks.

  Even though the inter-peak distance and the standard deviation of the total state are basically the same (up to a factor of two), the measures of \citet{Bjork_Size_2004} and \citet{Sekatski_Size_2014,Sekatski_General_2017} differ in scaling of $\sqrt{N}$. The reason is that \citet{Bjork_Size_2004} introduce a normalization with ``classical states'' exhibiting a width $\sqrt{N}$. A similar renormalization is --implicitly or explicitly-- done by all measures that find an effective size of $N \epsilon^2$.

  The issue of renormalization also appears when applying the measure of \citet{Kwon_Disturbance-Based_2016} to the generalized GHZ state. Any measure should provide an answer to, for example, how much $N$ has to be increased when $\epsilon$ decreases in order to keep the macroscopic quantumness constant. For this example, we find that, in the ``classical regime'' of the measurement resolution $\sigma$ (i.e., $\sigma \gtrsim \sqrt{N}$), the measure $M_{\sigma}$ gives the same values for $(N,\epsilon)$ if $2 N \sin \epsilon = \mathrm{const}$ (see Fig.~\ref{fig:KwonGenGHZ} (a)) and is hence comparable with \citet{Sekatski_Size_2014,Sekatski_General_2017}. If we want to compare $M_{\sigma}$ with measures that find an effective size of $\approx N \epsilon^2$, one has to rescale the measurement resolution to $\sigma/\sqrt{N}$ (see Fig.~\ref{fig:KwonGenGHZ} (b)).

\begin{figure}[htbp]
\centerline{\includegraphics[width=\columnwidth]{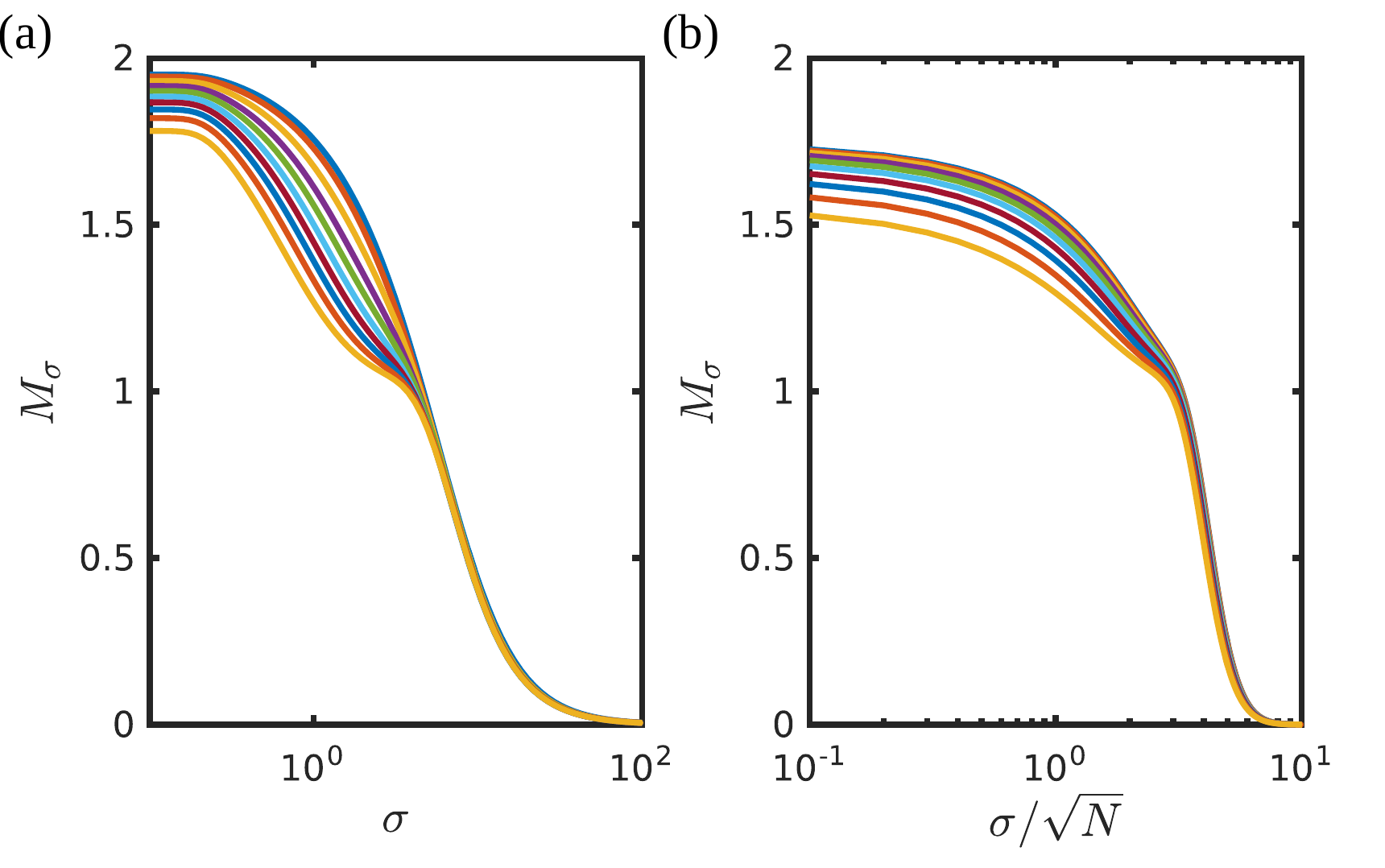}}
\caption[]{\label{fig:KwonGenGHZ} The measure of \citet{Kwon_Disturbance-Based_2016} applied to the generalized GHZ state, example \ref{ex:GHZ}, for different $N$ and $\epsilon \in \left\{0.01,0.02,\dots,0.1  \right\}$ (from top to bottom). (a) $N$ is chosen such that $2 N \sin \epsilon = 20$. For large $\sigma$, $M_{\sigma}$ is the same for all pairs $(N,\epsilon)$, implying that these states have the same macroscopic quantumness according to \citet{Kwon_Disturbance-Based_2016}. (b) $N$ is chosen such that $N \sin^2 \epsilon + \cos^2 \epsilon= 20$ (which is the exact value of the variance divided by $N$). In order to find similar values of $M_{\sigma}$, one has to rescale $\sigma$ to $\sigma/\sqrt{N}$.}
\end{figure}

Finally, we discuss the simplest example of a mixed state. We consider a damping of the coherence terms of the GHZ state (i.e., $\epsilon = \pi/2$),
\begin{equation}
\label{eq:46}
\left| 0 \right\rangle\!\left\langle 1\right| ^{\otimes N} \rightarrow p \left| 0 \right\rangle\!\left\langle 1\right| ^{\otimes N}
\end{equation}
and similar for the conjugate term. The quantum state is hence a mixture of $| \mathrm{GHZ}^{\pm}_{N} \rangle $ with weights $\mu_{\pm} = \frac{1}{2}(1\pm p)$, respectively. In the following, all measures are evaluated for $A = S_z$, which is optimal for $p \gtrsim 1/N$. We give the effective sizes relative to the pure state $p = 1$. Regarding the index $q$ of \citet{Shimizu_Detection_2005}, we calculate $Q(p) = \lVert [S_z,[S_z,\rho]] \rVert_1$.  Simple calculations lead to
\begin{equation}
\label{eq:36}
\begin{split}
  Q(p)/Q(1) &= p\\
  \mathcal{I}_S(p)/\mathcal{I}_S(1)& = \frac{2p^2}{1+p^2}\\
  N_{\mathrm{eff}}(p)/N_{\mathrm{eff}}(1) &= p^2\\
C_{\Delta}(p)/C_{\Delta}(1)&\rightarrow 1+\mu_{+} \log \mu_{+} +\mu_{-} \log \mu_{-}  \\
M_{\sigma}(p)/M_{\sigma}(1) &\rightarrow 1-\sqrt{1-p^2}
\end{split}
\end{equation}
for the measures of \citet{Shimizu_Detection_2005}, \citet{Park_Quantum_2016}, \citet{Frowis_Measures_2012}, \citet{Sekatski_General_2017} and \citet{Kwon_Disturbance-Based_2016}, respectively (see Fig.~\ref{fig:NoisyGHZ}). The last two expressions are valid in the limit $\Delta/N \rightarrow 0$ and $\sigma/N\rightarrow 0$, respectively. Note that $p$ can be $N$-dependent in common dephasing models (e.g., $p = e^{-\lambda N}$ or $p = e^{-\lambda N^2}$); in this case, $Q(p) \neq O(N^2)$ in general.

\begin{figure}[htbp]
\centerline{\includegraphics[width=\columnwidth]{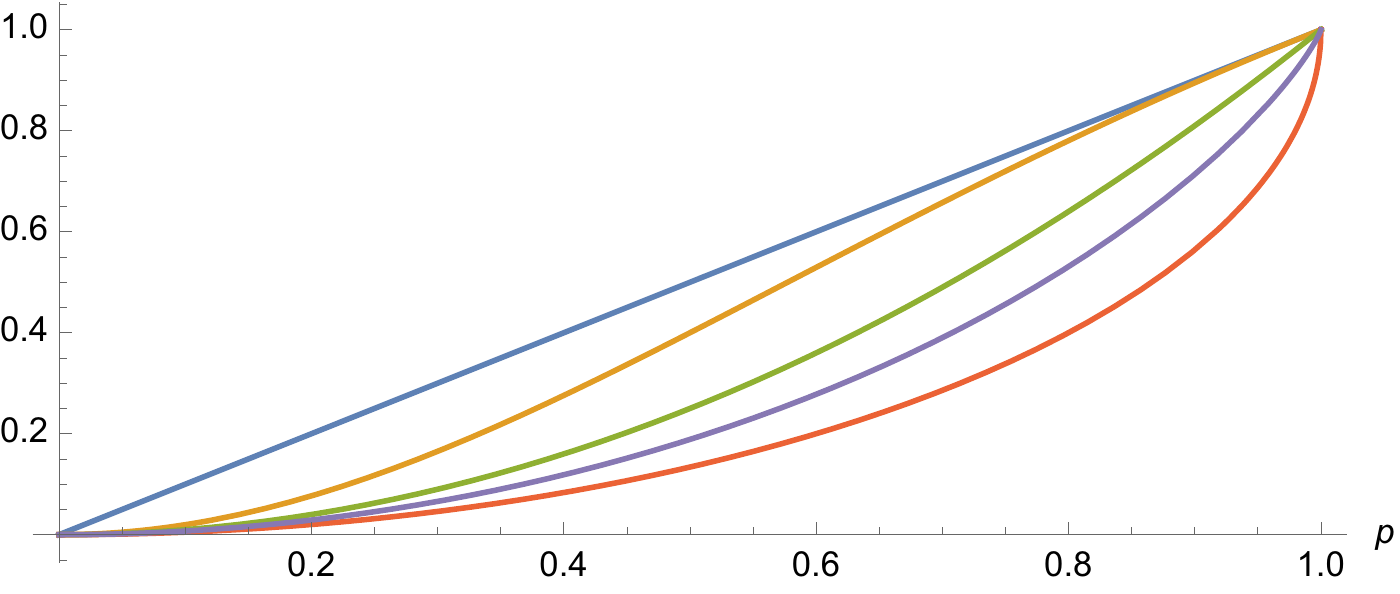}}
\caption[]{\label{fig:NoisyGHZ} Relative effective size for the noisy GHZ state, Eq.~(\ref{eq:46}), for measures listed in Eq.~(\ref{eq:36}) (from top to bottom in the order of Eq.~(\ref{eq:36})). All measures except the index $q$ scale with $p^2$ for small $p$. For $p$ close to one (i.e., close to purity), the first derivative lies between 1 and infinity. The only nonconvex function is $\mathcal{I}_S(p)/\mathcal{I}_S(1)$.}
\end{figure}
\end{Exp}

\begin{Exp}[Dicke states]\label{ex:Dicke}\label{ex:Korsbakken}
  Dicke states $| N,k \rangle $, Eq.~(\ref{eq:41}), do not offer an obvious splitting \da and are hence primarily addressed by measures that do not impose such a structure (however, see discussion later in this example). Since $| N,k \rangle $ have the same properties as $\left| N,N-k \right\rangle$ we restrict the following discussion to $0 \leq k \leq N/2$. \citet{Morimae_Macroscopic_2005} evaluated in index $p$ for this state class in detail. They find that $p = 2$ for Dicke state with $k = O(N)$. If $k = O(1)$, one has $p = 1$ and hence these states are not considered to be macroscopically quantum despite being genuinely multipartite entangled. More quantitatively, \citet{Frowis_Measures_2012,Park_Quantum_2016} find that $N_{\mathrm{eff}} = \mathcal{I} \approx 2 k(N-k) + 1$ which confirms the findings of \citet{Morimae_Macroscopic_2005}. The optimal observable is any collective operator on the equator. We work with $S_x$ in the following without loss of generality.

  \citet{Kwon_Disturbance-Based_2016} agree with the previous measures qualitatively. Similar conclusions can also drawn applying the measure of \citet{Sekatski_General_2017}.

The disconnectivity of \citet{Leggett_Macroscopic_1980} again shows a different behavior. Except for $k = 0$ (which are product states), all Dicke states exhibits large $\delta_{\max} = O(N)$. The reason is that one can divide the ensemble in $\approx k+1$ groups such that each group is entangled with the rest of the spins with about one ebit. This means that $\delta_M$ is far from being zero up to $M \approx N k/(k+1)$. The extensive difference, $\Lambda$, is not rigorously enough defined by \citet{Leggett_Testing_2002} to be directly applicable to Dicke states. If it is understood to be the spread of the probability distribution, then only for $k = O(N)$ the spread is broad enough to have a large $\Lambda$.

\begin{figure}[htbp]
\centerline{\includegraphics[width=\columnwidth]{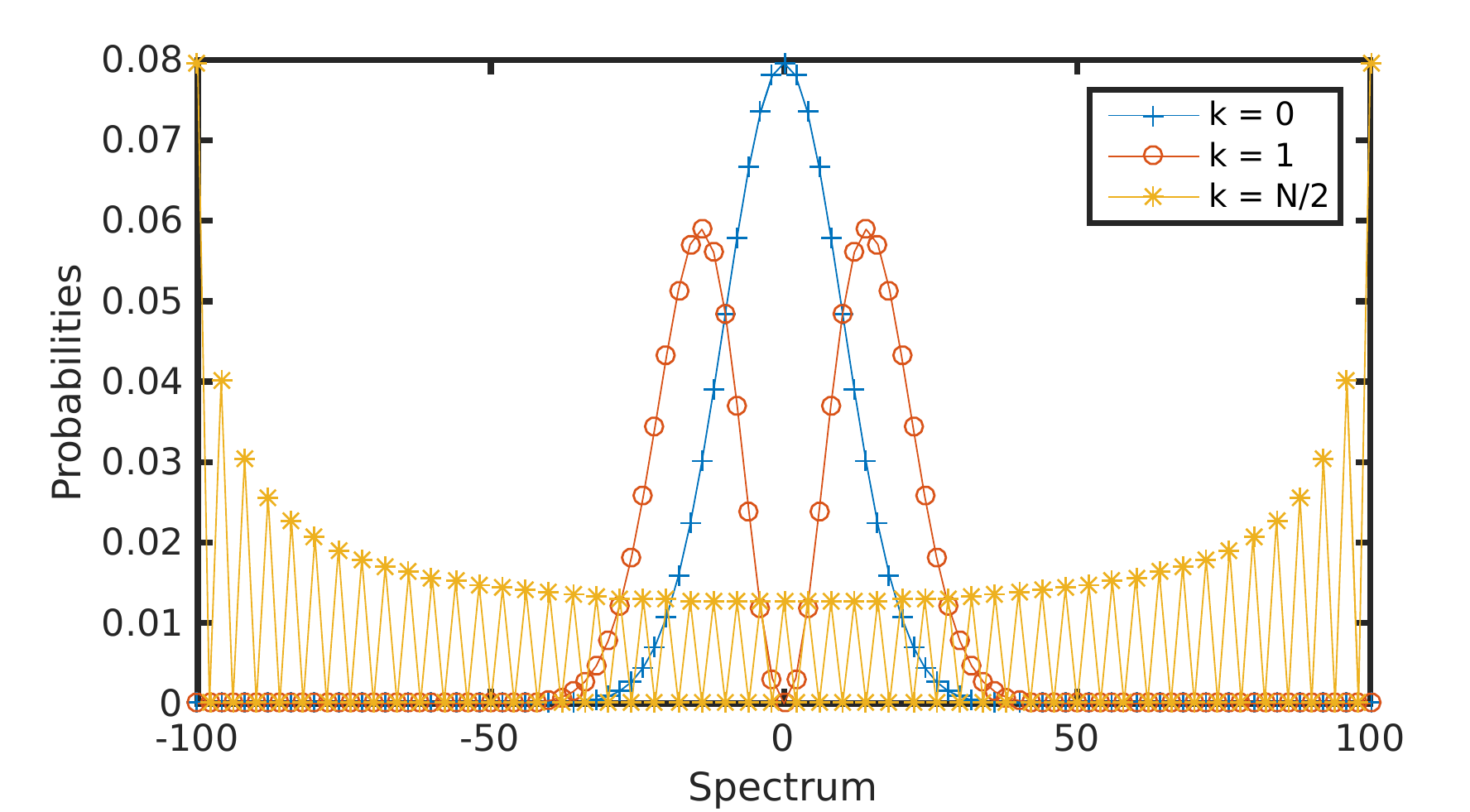}}
\caption[]{\label{fig:Dicke} Three examples of probability distributions for $S_x$ and Dicke states $| N ,k \rangle $ with $N = 100$ and $k \in \{0,1,N/2\}$. The spread is roughly $\sqrt{N}, \sqrt{3N}$ and $N/\sqrt{2}$, respectively. }
\end{figure}

The example of $| N,N/2 \rangle $ is interesting because it also highlights the impact of the choice of the splitting in cases where \alive and \dead are not two well-separated wave functions but \alive and \dead themselves are widely spread over the spectrum\footnote{The following analysis can be repeated for the example discussed by \citet{Korsbakken_Measurement-based_2007}, Sec.~III A, or the SQUID example by \citet{Marquardt_Measuring_2008}}. We now discuss two possible splittings. First, the appearance of high-$k$ Dicke states in so-called optimal covariant cloning \cite{Bruss_Phase-covariant_2000,DAriano_Optimal_2003} might be used to split the state into $| \mathcal{A} \rangle + \left| \mathcal{D} \right\rangle $. Indeed, one could ask whether an optimal cloning device could amplify the components of $\left| 0 \right\rangle \propto | + \rangle + \left| - \right\rangle $ to create a superposition of two macroscopically distinct states. Considering $1\rightarrow N$ cloning, one finds $| + \rangle \rightarrow \left| \mathcal{A} \right\rangle \propto \left| N,(N-1)/2 \right\rangle + \left| N, (N+1)/2 \right\rangle $ and $| - \rangle \rightarrow \left| \mathcal{D} \right\rangle \propto \left| N,(N-1)/2 \right\rangle - \left| N, (N+1)/2 \right\rangle $ for odd $N$ and similarly for even $N$ \cite{Bruss_Phase-covariant_2000,DAriano_Optimal_2003}. While the total state $| \mathcal{A} \rangle + \left| \mathcal{D} \right\rangle \propto \left| \Psi \right\rangle = \left| N, (N-1)/2 \right\rangle $ is macroscopically quantum for all previously mentioned measures, the measures of \citet{Korsbakken_Measurement-based_2007,Bjork_Size_2004,Marquardt_Measuring_2008} were shown to not classify \da as a superposition of macroscopically distinct states \cite{Frowis_Cloned_2012}.\footnote{For the measure of \citet{Bjork_Size_2004}, \citet{Frowis_Cloned_2012} choose the reference state to be \alive, which differs from the ``classical reference'' mentioned in the beginning of the section.}

A similar analysis using the measure of \citet{Sekatski_Size_2014} gives comparable results (see Fig.~\ref{fig:sekatski_Dicke}). One finds that $\mathrm{Size}(\Psi ) = O(N)$, whenever $P_g < P(\Delta = 1) \approx 0.8183$; however, $\mathrm{Size}(\Psi ) \lesssim 1$ for larger $P_g$. This example clearly illustrates the influence of the success probability $P_g$ that also plays a role in the approach of \citet{Korsbakken_Measurement-based_2007}. The authors of the latter argue that $P_g$ should be ``very close'' to one and hence the total state is not considered to be a superposition of macroscopically distinct states \cite{Frowis_Cloned_2012}. This is in contrast to \citet{Sekatski_Size_2014} who consider success probabilities in the order of 2/3.

Second, we look for a splitting that increases the effective size for the measure at hand. An obvious choice for $\left| N, (N-1)/2 \right\rangle $ is to define \alive as the superposition of all eigenstates of $S_x$ with positive eigenvalues (with weights coming from $\left| N, (N-1)/2 \right\rangle $) and likewise \dead for the negative part of the spectrum. Then, the distinguishability in the approach of \citet{Sekatski_Size_2014} is close to one also for $\Delta \gg 1$ (see Fig.~\ref{fig:sekatski_Dicke}). Even for $P_g = 0.99$, we numerically find a linear scaling of the size. The same splitting leads to similar results for \citet{Korsbakken_Measurement-based_2007}, while the framework of \citet{Marquardt_Measuring_2008} is more difficult to apply for this splitting.

\begin{figure}[b]
\centerline{\includegraphics[width=\columnwidth]{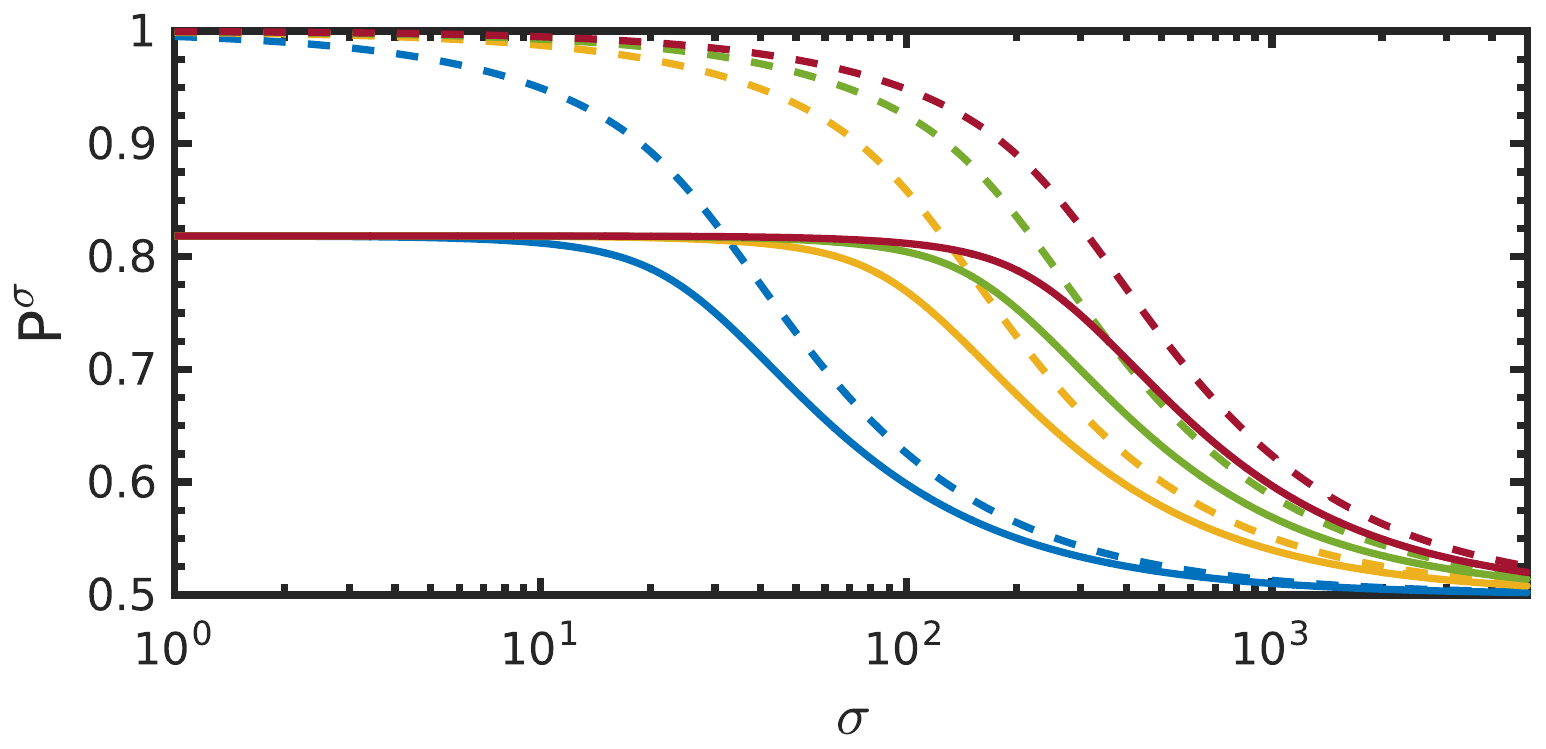}}
\caption[]{\label{fig:sekatski_Dicke} Success probability $P^{\sigma}$ as a function of the coarse-graining $\sigma$ for $| N, (N-1)/2 \rangle $ with $N \in \left\{ 101,401,701,1001 \right\}$ (from left to right). Inverting this functions lead to effective size in the proposal of \citet{Sekatski_Size_2014}. The success probability depends on the splitting $| N, (N-1)/2 \rangle \propto | \mathcal{A} \rangle + \left| \mathcal{D} \right\rangle $. The ``natural'' splitting based the cloner scenario (solid lines, see text) leads to a reduced $P^{\sigma}$ and hence the effective size is $O(N)$ only if $P^{\sigma} \lesssim 0.8183$. The splitting based on a cut of the spectrum into two parts (dashed lines, see text) allows for high distinguishability also for $\Delta \gg 1$. As a consequence, $\mathrm{Size}(\Psi) = O(N)$ even for $P_g$ close to one.}
\end{figure}

\end{Exp}

\begin{Exp}[Spin-squeezing] \label{ex:spinsqueezing}
  The discussion for spin-squeezed states is similar as for Dicke states, as it is a general quantum state with a broad distribution in some collective observable without obvious splitting into $| \mathcal{A} \rangle + \left| \mathcal{D} \right\rangle $. Hence we only mention some specific points. Spin-squeezing means that the spread of the state in one quantization axis, say $S_1$, is reduced while keeping the polarization in axis $S_2$ large. From the Heisenberg uncertainty relation, it follows that the state is widely spread over the axis $S_3$.

  One-axis twisted spin-squeezed states, Eq.~(\ref{eq:9}), exhibit a scaling of $(\Delta S_3)^2 = O(N^{5/3})$ for the optimal squeezing \cite{Kitagawa_Squeezed_1993}; hence $\mathcal{I}(\Psi) = N_{\mathrm{eff}} = O(N^{2/3})$. This means that large effective sizes can be achieved, even if the scaling is less than for some Dicke states or the GHZ state. Thus, one has $p = 5/3$ and the state is not macroscopically quantum according to \citet{Shimizu_Stability_2002}. The two-axis twisting leads to a maximal variance in $S_3$ \cite{Kitagawa_Squeezed_1993} and hence $\mathcal{I}(\Psi) = N_{\mathrm{eff}} = O(N)$ and $p = 2$.
\end{Exp}

\begin{Exp}
  [Cluster states]\label{ex:cluster}
  Despite being multi-partite entangled states, cluster states are seldom considered by most of the discussed papers. An exception are \citet{Yadin_Quantum_2015}, who propose cluster states as macroscopically quantum because one can distill GHZ states $| \mathrm{GHZ}_n \rangle $ with $n = O(N)$ out of cluster states. However, all other presented measures do not assess this state as macroscopically quantum (essentially because it does not have a large spread in any local observable, see also Sec.~\ref{sec:mbqc-entangl-macr}).

However, a superposition of cluster states is an interesting case study to understand the difference between some measures. Consider the operator $\mathcal{U}_C$, Eq.~(\ref{eq:51}), that maps a product state $| + \rangle ^{\otimes N}$ to a one-dimensional cluster state.  Then, the state $\left| \operatorname{Cl-GHZ} \right\rangle \propto \mathcal{U}_C(\left| + \right\rangle ^{\otimes N} + \left| - \right\rangle ^{\otimes N})$ is a state that has a similar structure as a GHZ state. The measures of \citet{Korsbakken_Measurement-based_2007,Marquardt_Measuring_2008} indeed assign an effective size of $N/3$ \cite{Frowis_Measures_2012}. On the other side, every measure that is build upon local operators will still qualify this state as not being macroscopically quantum because it does not exhibit a large spread in any local operator. For this reason, \citet{Frowis_Measures_2012} extended their measure to quasi-local operators. The state $| \operatorname{Cl-GHZ} \rangle $ has a large variance for the operator $\mathcal{U}_C S_x \mathcal{U}_C^{\dagger}$, which is a sum of three-body interactions $\sigma_z^{(i-1)} \sigma_z^{(i)} \sigma_z^{(i+1)}$ for $i = 2,\dots,N-1$ plus two-body boundary terms. In the spectrum of this operator, $| \operatorname{Cl-GHZ} \rangle $ is maximally spread and every observable-based measure using this operator will find a large effective size.
\end{Exp}

\subsubsection{Photonic systems}
\label{sec:photonic-systems}

Similar as for spin examples, we first note a number of mathematical connections between some measures. \citet{Laghaout_Assessments_2015} have two elements that they combine to single measure. For pure, single-mode states, the element $\mathcal{N} = \frac{1}{2}[(\Delta X)^2 + (\Delta P)^2 - 1] = \langle a^{\dagger} a \rangle_{\Psi} - |\langle a \rangle_{\Psi}|^2$ is identical to the measure of \citet{Lee_Quantification_2011}, $\mathcal{I}$, if applied to pure states. In the photonic extension of \citet{Frowis_Measures_2012}, one optimizes the quantum Fisher information over all quadrature operators, which reduces to four times the variance for pure states \cite{Oudot_Two-mode_2015}. In summary, one finds for all single-mode pure states $| \Psi \rangle $ that
\begin{equation}\label{eq:55}
  \begin{split}
    \mathcal{N}(\Psi) &=
    \mathcal{I}(\Psi)\\
    \frac{1}{4}N_{\mathrm{eff}}(\Psi) &<
    \mathcal{I}(\Psi) + \frac{1}{2} \leq \frac{1}{2}
    N_{\mathrm{eff}}(\Psi).
  \end{split}
\end{equation}
\citet{Bjork_Size_2004} use either quadrature or number operators to calculate $M$ in their example. The authors calculate the standard deviation and rescale it with the standard deviation of a coherent state. When using quadrature operators, this implies that $M(\Psi)^{2} = N_{\mathrm{eff}}(\Psi)$. \citet{Sekatski_Size_2014} generally work with photon number measurements to calculate $\mathrm{Size}(\Psi)$.

\begin{Exp}
  [Superposition of coherent states]\label{ex:SCS}
  An archetypal instance of a photonic state with macroscopic quantumness is the superposition of coherent states $| \mathrm{SCS} \rangle $ (see Eq.~(\ref{eq:43}) and Fig.~\ref{fig:SCS} (a) for a plot of the Wigner function for a specific $\alpha$). Without loss of generality, we assume $\alpha \in \mathbbm{R}$ in the following. One easily finds\footnote{\cite{Lee_Quantification_2011,Frowis_Measures_2012,Laghaout_Assessments_2015,Sekatski_Size_2014,Bjork_Size_2004}}
  \begin{equation}
    \begin{split}
      \mathcal{N}(\mathrm{SCS}) &= \mathcal{I}(\mathrm{SCS}) = \langle
      a^{\dagger} a \rangle_{\mathrm{SCS}} = \alpha^2 \tanh
      \alpha^2\\
      N_{\mathrm{eff}}(\mathrm{SCS}) &\approx 4 \alpha^2 +
      1\\
      M(\mathrm{SCS}) &\approx 2
      \alpha\\
      \mathrm{Size}(\mathrm{SCS}) &\approx 4 \alpha^2
    \end{split}
\label{eq:56}
\end{equation}
 The latter two expressions are valid for large $\alpha$. Note that $\mathrm{Size}(\mathrm{SCS})$ is evaluated for $\left| 0 \right\rangle + \left| 2\alpha \right\rangle $. The second element of \citet{Laghaout_Assessments_2015}, the distinguishability between $| \alpha \rangle $ and $| -\alpha \rangle $, quickly goes to one for $\alpha \gtrsim 1$ using homodyne detection.

  The SCS has macroscopic coherence also in the spirit of \citet{Cavalcanti_Signatures_2006,Cavalcanti_Criteria_2008}, in the order of $S = 4\alpha$. However, the witness Eq.~(\ref{eq:17}) is more suited for squeezed states (see example \ref{ex:squeezed}) and less to detect large coherence in SCS. Consequently, the achievable lower bound is only $S_{\max} \approx 2.5$ for $\alpha = 0.5$. This value is just above $S = 2$ which is found for coherent states \cite{Cavalcanti_Criteria_2008}.

\begin{figure}[htbp]
\centerline{\includegraphics[width=\columnwidth]{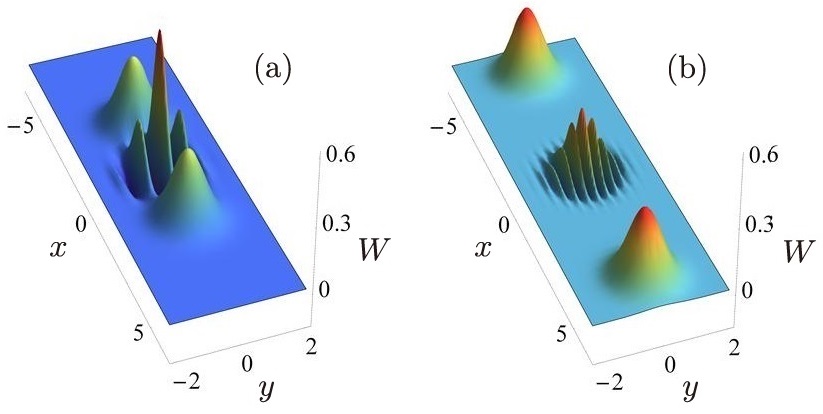}}
\caption[]{\label{fig:SCS} Plot of the Wigner function of $| \alpha \rangle + \left| -\alpha \right\rangle $ (a) for $\alpha = 2.3$ and (b) for $\alpha = 4.96$ with reduced visibility $\Gamma \approx 0.464$ (see text). The two states exhibit the same effective size $\mathcal{I} \approx 5.29$ and $N_{\mathrm{eff}}\approx 22.2$ for the measures of \citet{Lee_Quantification_2011} and \citet{Frowis_Measures_2012}, respectively. From \citet{Jeong_Detecting_2014}.}
\end{figure}

Consider the case in which the visibility is nonunit, that is, $\left| \alpha \right\rangle\!\left\langle -\alpha\right| \rightarrow \Gamma \left| \alpha \right\rangle\!\left\langle -\alpha\right|$ (see Fig.~\ref{fig:SCS} (b)). The measures of \citet{Lee_Quantification_2011,Frowis_Measures_2012} can be straightforwardly applied to this mixed states. In particular, one finds $\mathcal{I}(\Gamma)\approx \Gamma^2  \mathcal{I}(\mathrm{SCS})$ and $N_{\mathrm{eff}}(\Gamma)\approx   4 \alpha^2 \Gamma^2 + 1$. One has to go to more complex examples to see a difference between $\mathcal{I}$ and $N_{\mathrm{eff}}$ \cite{Yadin_General_2016}.

There exist two interesting multimode extension of $| \mathrm{SCS} \rangle $: $M$ identical copies $\left| \mathrm{SCS}^{M}_1  \right\rangle =| \mathrm{SCS} \rangle^{\otimes M} $ or the entangled version $| \mathrm{SCS}^M_2 \rangle \propto \left| \alpha \right\rangle ^{\otimes M} + \left| -\alpha \right\rangle ^{\otimes M}$ (see also \citet{Volkoff_Measurement-_2014}, who studied the second example in depth for several measures and \citet{Volkoff_Nonclassical_2015} for so-called hierarchical photonic superposition states). The additivity of the measure of \citet{Lee_Quantification_2011} results in a similar effective size for both states $\mathcal{I}(\mathrm{SCS}^M_1) = M \alpha^2 \tanh (\alpha^2)$ and $\mathcal{I}(\mathrm{SCS}^M_2) = M \alpha^2 \tanh (M \alpha^2)$, while the approach of \citet{Frowis_Measures_2012} yields $N_{\mathrm{eff}}(\mathrm{SCS}^M_1) = N_{\mathrm{eff}}(\mathrm{SCS})$ but $N_{\mathrm{eff}}(\mathrm{SCS}^M_2) \approx 4 M \alpha^2 + 1$.
\end{Exp}

\begin{Exp}[Fock states] \label{ex:Fock}
Photon number states defined via $a^{\dagger} a \left| n \right\rangle = n \left| n \right\rangle $, also called Fock states, are not often discussed in the context of macroscopic quantumness. One reason might be the lack of a natural decomposition into $| \mathcal{A} \rangle + \left| \mathcal{D} \right\rangle $. Note, however, that $| n \rangle $ can be written as a superposition of coherent states with large (i.e., $O(\sqrt{n})$) difference between the amplitudes $\alpha$. For the measures of \citet{Lee_Quantification_2011,Frowis_Measures_2012}, the Fock state has an effective size of $N_{\mathrm{eff}}(n) = 2n+1$ and $\mathcal{I}(n) = n$, respectively. Hence, Fock states become macroscopically quantum with increasing $n$.
\end{Exp}

\begin{Exp}[Squeezed states] \label{ex:squeezed}
  Similar to Fock states, squeezed states, Eq.~(\ref{eq:1}), do not exhibit a clear \da structure. Measures that are not focusing on a binary division typically recognize these states as macroscopically quantum for large squeezing despite having a positive Wigner function. This is traced back to the large spread in phase space in the ``anti-squeezed'' direction. In the following, we consider $\zeta \in \mathbbm{R}$ without loss of generality. This implies squeezing of the $X$ quadrature and anti-squeezing of $P$. It directly follows that $\mathcal{I}(\zeta) = \sinh^2(\zeta)$ \cite{Lee_Quantification_2011} and $N_{\mathrm{eff}}(\zeta) = e^{2\zeta}$ \cite{Frowis_Measures_2012}, which scales linearly with the mean photon number $\langle a^{\dagger}a \rangle = \sinh^2(\zeta)$.

Squeezed states are the ideal states to verify macroscopic coherence in the framework of \citet{Cavalcanti_Signatures_2006,Cavalcanti_Criteria_2008}. By measuring $\Delta X$ and by using Eq.~(\ref{eq:17}), one can show coherence up to $S = \frac{1}{2} \Delta P$. However, under experimentally realistic conditions, a finite $S$ can only be verified if $\Delta X \Delta P \lesssim 1.6$ \cite{Cavalcanti_Signatures_2006}. The witness Eq.~(\ref{eq:52}), in which the binning of $I_{-}$, $I_0$ and $I_{+}$ is dropped \cite{Cavalcanti_Criteria_2008}, is much more robust. More specifically, coherence up to $S = 2\Delta P$ independent of the value $\Delta X \Delta P$ can be shown. Note the similarity between Eq.~(\ref{eq:52}) and using the Heisenberg uncertainty relation, Eq.~(\ref{eq:48}), to witness large quantum Fisher information to lower bound $N_{\mathrm{eff}}$ (\citet{Frowis_Measures_2012}; see Sec.~\ref{sec:meas-exper} for details).

  Similar considerations can be done for two-mode squeezed states, Eq.~(\ref{eq:twomodesqueezedstate}). \citet{Lee_Quantification_2011,Frowis_Measures_2012,Cavalcanti_Criteria_2008,Cavalcanti_Signatures_2006} give basically the same results as for single-mode squeezed states. In addition, the framework of \citet{Sekatski_Size_2014} can be modified to be applicable to two-mode states \cite{Oudot_Two-mode_2015}. The idea is to rephrase the original formulation by having a bipartite, entangled system, where one party ``Alice'' prepares the state of the other party ``Bob'' by means of local measurement (i.e., a kind of steering scenario). Then, Bob has to guess which state Alice measured using only measurements with finite resolution. The quantum state used for this protocol is said to be a superposition of macroscopically distinct states if Bob is able to distinguish different preparation with very low resolution. Using coarse-grained quadrature operators, this results in $\mathrm{Size} = O(\sqrt{N})$, where $N$ is the mean photon number of the states. Note that this is the optimal scaling in this scenario.
\end{Exp}

%%% Local Variables:
%%% mode: latex
%%% TeX-master: "master"
%%% End:

\subsection{Classifications of measures}
\label{sec:measurediscussion}

In the previous sections we reviewed several proposals to quantify macroscopic quantumness and discussed relevant examples. In this and the following sections, we discuss and compare different contributions (see table \ref{tab:overview2}).

However, the measures do not all share the same motivation or goals, neither they have the same range of applicability. Hence, one observes a family resemblance situation, where objects in a set are not connected by a unique common feature, but rather they are connected by a series of features, each shared by some but not all of the objects. The goal of the following subsections is therefore to discuss several recurring aspects of the measures and to identify common features and differences within subsets of measures. Some original proposals leave room for interpretation or small modifications that influence the final measure. Hence we understand this comparison not as an ultimate judgment but as a starting point for further discussions.

\subsubsection{Mechanisms to break unitary equivalence}
\label{sec:observ-vs.-part}

In the introduction of Sec.~\ref{sec:preliminary-measures}, we drew the reader's attention to the structure that breaks the unitary equivalence between quantum states. This is necessary in order to define a measure of macroscopic quantumness and to introduce a hierarchy of macroscopic quantumness. On a formal level this requires to identify some additional structure over the Hilbert space $\cH$ associated with the physical system. After reviewing the literature, we are now in the position to comment more on the mechanisms to break the unitary equivalence between all pure states. We identify at least three basic approaches.

(i) One can choose a preferred hermitian operator $A = \sum_k a_k \prjct{k}$ (or $\int dk$) as a starting point\footnote{\citet{Leggett_Testing_2002,Cavalcanti_Signatures_2006,Cavalcanti_Criteria_2008,Sekatski_Size_2014,Laghaout_Assessments_2015,Kwon_Disturbance-Based_2016,Bjork_Size_2004,Sekatski_General_2017}; the ``extensive difference'' part of \citet{Leggett_Testing_2002} and ``subjective'' part of \citet{Laghaout_Assessments_2015}.}. This operator can be an observable that is measured or a generator of unitary transformations. In any case, this choice defines a preferred basis $\ket{k}$ and a notion of spectral distance $a_k-a_{k'}$ between any two eigenstates $\ket{k}$ and $\ket{k'}$. Thus, a measure based on a preferred observable is a function of the spectral distribution and the coherence properties of the considered state.

(ii) In some cases the authors start by identifying a specific partition of the system into subsystems\footnote{ \citet{Leggett_Macroscopic_1980,Frowis_Measures_2012,Dur_Effective_2002,Yadin_Quantum_2015,Korsbakken_Measurement-based_2007,Marquardt_Measuring_2008,Shimizu_Stability_2002,Park_Quantum_2016}}, for example, $\cH=\cH_1\otimes \cH_2\dots$. Often, like for spin systems, the subsystems correspond to physical particles\footnote{ \citet{Leggett_Macroscopic_1980,Frowis_Measures_2012,Dur_Effective_2002,Yadin_Quantum_2015,Korsbakken_Measurement-based_2007}}, but can be also identified with a locality argument \cite{ Shimizu_Stability_2002} for systems with a spatial extent.

Though the starting point here is different from (i), in some cases  the final measure can be equally well understood from the perspective of a preferred operator \cite{Shimizu_Stability_2002,Frowis_Measures_2012,Park_Quantum_2016}. This is because the measures are computed via optimizing some property of the state over a set of local operators. This corresponds to the identification of a preferred observable $A$.

(iii) The third approach is to look at a preferred representation of the considered state; here, it is the phase space for position and momentum of a particle or conjugated quadratures of a bosonic mode. This is the starting point of \citet{Lee_Quantification_2011,Laghaout_Assessments_2015}, who focus on the the Wigner function of a given state.
The quantification of high-frequency components leads to an equivalent formulation of their measure based on a decoherence model.
Despite the different motivation, \citet{Nimmrichter_Testing_2011} is directly related to \citet{Lee_Quantification_2011} \cite{Yadin_General_2016}\footnote{The connection is valid if only the ideal quantum state plus the canonically introduced modification is considered.}.
This is because the modification of quantum mechanics considered in \citet{Nimmrichter_Testing_2011} also corresponds to a continuous spontaneous localization in phase space of the particles. Alternatively, one may say that the starting point here is the choice of a preferred Lindbladian superoperation $\mathcal{L}(\varrho)= \sum_j(L_j \varrho L_j^\dag - \frac{1}{2}\{L_j^\dag L_j, \varrho\})$, which is used to model dissipative evolution in terms of a master equation. Interestingly, the observables that appear as Lindblad operators in the master equation of the decoherence are linear in the quadrature operators and frequently appear in the proposals of group (i).

\subsubsection{Goals of the measures}
\label{sec:cat-regard-purp}

All proposals aim for measuring a kind of macroscopic quantumness, but one can identify different variations in formulating the precise goal.

(i) Macroscopic coherence. For a fixed operator $A=\sum_k a_k \prjct{k}$, the macroscopic quantumness of a state is related to its coherence spread\footnote{\citet{Leggett_Testing_2002,Shimizu_Detection_2005,Cavalcanti_Signatures_2006,Cavalcanti_Criteria_2008,Lee_Quantification_2011,Laghaout_Assessments_2015,Kwon_Disturbance-Based_2016,Bjork_Size_2004,Park_Quantum_2016,Frowis_Measures_2012}; the ``extensive difference'' of \citet{Leggett_Testing_2002}, the ``objective'' part of \citet{Laghaout_Assessments_2015} and the large-$\sigma$ regime of \citet{Kwon_Disturbance-Based_2016}.}, that is, the amount of coherence between basis state $| k \rangle , \left| k^{\prime} \right\rangle $ that are far apart in the spectrum $| a_k -a_k'| \gg 1$. The superposition of far-distant parts of the spectrum is seen as a macroscopic quantum effect if the system size is large. In the case of a pure state $\ket{\Psi}=\sum \psi_k \ket{k}$, one often considers the variance of the state $(\Delta A)^2_{\ket{\Psi}}$, or the total amount of coherence with distance $\geq S$, that is, $\sum_{|a_k-a_k'|>S} |\psi_k \psi_{k'}|$.

(ii) Effective particle number. Some measures \footnote{\citet{Leggett_Macroscopic_1980,Dur_Effective_2002, Korsbakken_Measurement-based_2007,Marquardt_Measuring_2008,Yadin_Quantum_2015}} explicitly aim to quantify how many particles effectively participate to a macroscopic superposition. As an example \cite{Korsbakken_Electronic_2009}, consider a system of $n$ electrons in a pure quantum state building up the magnetic flux in a superconducting device (see Sec.~\ref{sec:superc-quant-interf}). Suppose that the induced magnetic moment from a superposition of right \alive and left \dead circulating currents has a spread $\Delta\mu = n \mu_B$, where $\mu_B$ denotes the Bohr magneton. For \citet{Korsbakken_Electronic_2009}, it matters whether a single electron makes the difference between the \alive and \dead or whether all electrons behave differently in the two branches. In contrast, representatives from (i) might care less because a single particle in a superposition of macroscopically different momenta could be seen as one with a very large effective mass. From this example, the concept of effective particle number seems to be more restrictive than (i) as the large spread for some extensive quantity necessarily appears whenever the effective particle number is large (see Sec.~\ref{sec:macr-superp-vs}).

(iii) Ease to distinguish. Macroscopic quantumness is identified with the easy to distinguish between the superposed components in the state \cite{Korsbakken_Measurement-based_2007, Sekatski_Size_2014,Laghaout_Assessments_2015, Sekatski_General_2017}. The ease can be quantified by measuring how invasive a measurement apparatus should be in order to extract the desired information about the superposition state, e.g., distinguish \alive and \dead with the desired probability. Such a non-invasive measurement apparatus can be modeled as a weak measurement of a fixed operator $A$ \cite{Sekatski_Size_2014,Sekatski_General_2017}, or as a general measurement that only acts on a limited number of subsystems \cite{Korsbakken_Measurement-based_2007}. In the case of weak measurement of a fixed observable, easily distinguishable states clearly exhibit a large macroscopic coherence (i), as the superposed states have to be far enough in the spectrum of $A$.

(iv) Relative improvement. \citet{Bjork_Size_2004,Frowis_Measures_2012} motivated their proposals with situations where the superposition of (quasi-)classical states overcome the performance of its single branches. Although not explicitly mentioned, it seems that ``macroscopic quantum effects'' like the single excitation mentioned in Sec.~\ref{sec:not-an-easy} should be excluded. Instead, the authors focus on the speed of unitary evolution or on quantum-enhanced sensing. While the motivation might differ from (i), the choice of the figure of merit leads to measures based on the variance for pure states \cite{Frowis_Measures_2012} or at least is closely connected to it in relevant situations \cite{Bjork_Size_2004}.

(v) Falsification of collapse models. The proposal by \textcite{Nimmrichter_Macroscopicity_2013} is explicitly motivated by evaluating quantum states (more precisely, entire experiments) that potentially show small deviations from standard quantum mechanics. Surely there are connections to (i), since typical collapse models predict larger collapse rates for larger spatial spread of the massive particle. Likewise, heavier objects generally trigger the hypothetical collapse much faster. However, if one is tempted to analyze the entire experiment including ``unavoidable'' imperfections and ask how much it is capable to show modifications of quantum mechanics, larger quantum systems (e.g., larger masses) are not necessarily more useful (i.e., give a larger number) for such tests as environmentally induced decoherence is more disturbing \cite{Nimmrichter_Optomechanical_2014}.

(vi) Amount of nonclassicality. \textcite{Lee_Quantification_2011} identify the total amplitude of wiggles in the Wigner function for the state of one or several optical modes. As coherent states correspond to classical fields and their Wigner function is smooth, such wiggles are considered as a trace of nonclassicality of the state. Similarly, \citet{Park_Quantum_2016,Dur_Effective_2002,Laghaout_Assessments_2015} are concerned with the nonclassical aspects of macroscopic quantum states.

\subsection{Structure of applied states}
\label{sec:macr-superp-vs}

The presented literature is rather diverse regarding the application to quantum states. While \citet{Dur_Effective_2002} basically discuss a class of examples, others\footnote{\citet{Sekatski_Size_2014,Laghaout_Assessments_2015,Bjork_Size_2004,Korsbakken_Measurement-based_2007,Marquardt_Measuring_2008}; \citet{Bjork_Size_2004} phrase their idea for general states, but the mathematical formulation for Eq.~(\ref{eq:11}) is derived for the \da structure. \citet{Laghaout_Assessments_2015} extend the \da structure to more than two states, but still need to provide the substructure.} derive a measure assuming an explicit decomposition of the state into $| \mathcal{A} \rangle + \left| \mathcal{D} \right\rangle $. In the following, we call states that are macroscopically distinct for a fixed partition \da \textit{macroscopic superpositions}. The index $p$ from \citet{Shimizu_Stability_2002} is applicable to pure states without determining any substructure. Finally, there exist several proposals\footnote{\citet{Cavalcanti_Signatures_2006,Cavalcanti_Criteria_2008,Kwon_Disturbance-Based_2016,Leggett_Macroscopic_1980,Frowis_Measures_2012,Yadin_Quantum_2015,Shimizu_Detection_2005,Sekatski_General_2017}; the basic idea of \citet{Cavalcanti_Signatures_2006,Cavalcanti_Criteria_2008} is for general states, but the derived bounds for the detection basically work only for squeezed, not necessarily pure, states.} that are defined for arbitrary mixed states. Large-size states beyond the \da structure are here called \textit{macroscopic quantum states}. Let us further elaborate on these differences.

\subsubsection{Macroscopic superpositions}
\label{sec:macr-superp}

\paragraph{Separate treatment of macroscopic distinctness and quantumness.} In most of the cases, measures designed for a \da structure aim for
quantifying the macroscopic distinctness between the two components, but they do not measure the coherence between the components. A mixed state $\left| \mathcal{A} \right\rangle\!\left\langle \mathcal{A}\right| + \left| \mathcal{D} \right\rangle\!\left\langle \mathcal{D}\right| $ has equally distinct components, but it has no coherence and hence cannot be called a macroscopic quantum state. Therefore, the distinctness of a macroscopic superposition and its quantumness appear as two independent values. Consequently, showing that a state contains distant components only reveals its macroscopic distinctness. This is not sufficient if the state can not be assumed to be pure. In practice, an additional measurement is required to certify the quantumness of the superposition state, for example, through the purity of the state, some entanglement to an auxiliary system or some interferometric visibility between $| \mathcal{A} \rangle $ and $| \mathcal{D} \rangle $.
This contrasts measures that aim to give a single number quantifying macroscopic quantumness.

\paragraph{Freedom for choosing the branching.} Measures for macroscopic superpositions often provide an intuitive account for characterizing macroscopic distinctness. Having the Schrödinger-cat example in mind, it is easy to connect the concept of macroscopic distinctness with the thought experiment. Macroscopic distinctness can mean, for instance, that measuring only one biological cell is enough to determine the vitality of the cat \cite{Korsbakken_Measurement-based_2007}; that it is necessary to modify the state all $N$ cells from ``alive'' to ``dead'' \cite{Marquardt_Measuring_2008}; or that a quick (i.e. coarse-grained) look suffices to decide whether the cat is dead or alive \cite{Sekatski_Size_2014}.
On the other side, given just a pure state $\ket{ \Psi}$, there are infinitely many ways to split it in two branches $\ket{ \Psi}=\ket{\mathcal{A}}+\ket{\mathcal{D}}$. As in example~\ref{ex:Korsbakken} in Sec.~\ref{sec:examples}, the decomposition  can become problematic when the state does not show two well-isolated distributions in the spectrum of a local observable that is particularly suited to distinguish $| \mathcal{A} \rangle $ and $| \mathcal{D} \rangle $. In other words, the most obvious decomposition does not have to lead to the maximal size in these cases. This ambiguity is particularly relevant when real experiments are discussed (e.g., \citet{Marquardt_Measuring_2008} discussing the experiment of \citet{vanderWal_Quantum_2000}).

The ambiguity is lifted (though not fully) when the state is given in the micro-macro entangled form
\begin{equation}
\ket{\Psi}= \ket{\uparrow}\ket{\mathcal{A}} + \ket{\downarrow}\ket{\mathcal{D}},\label{eq:53}
\end{equation}
where the macroscopic system is entangled with an ``atom'' like in Schrödinger's thought experiment. The entanglement in Eq.~(\ref{eq:53}) reduces the complexity of the problem and it suffices to find the optimal distinguishability in a two-dimensional subspace spanned by $\text{span}\{\ket{\mathcal{A}},\ket{\mathcal{D}}\}$. This is a feasible task in many examples.

\subsubsection{Macroscopic quantum states}
\label{sec:conn-betw-meas-1}

Measures for macroscopic quantum states quantify the macroscopic quantumness of a general state $\rho$, which does not need to be pure or have a required structure. The macroscopic distinctness and the coherence are hence measured with a single number. For reasons of clarity, it is useful to first consider the application of the measures to pure states, for which the interpretation is more direct. After that, we discuss how they are lifted to mixed states.

\paragraph{Variance as recurring measure for pure macroscopic quantum states.} In most of the cases, the formalization of measures for macroscopic quantum states amounts to a study of
coherences of the state in the basis of a given operator $A$. For pure states, once $A$ is fixed, the variance of the state with respect to $A$ is a natural choice that allows to quantify the coherent spread of the state. Indeed, despite the diversity, several proposals\footnote{ \citet{Shimizu_Stability_2002,Lee_Quantification_2011,Frowis_Measures_2012,Park_Quantum_2016}} directly have the variance of the state (for given $A$) at the core of their measures when applied to pure state.
\citet{Leggett_Testing_2002} defines the extensive difference as the spectral distance of \alive and \dead for a well-chosen extensive observable, which is a special case of the variance. Moreover, while \citet{Bjork_Size_2004} do not explicitly use the variance in their approach, the connection between the variance and improved sensitivity is well established for pure states (see Secs.~\ref{sec:meas-exper} and \ref{sec:quantum-metrology}).

In addition, there are measures that do not reduce to the variance but still are closely related.  \citet{Cavalcanti_Signatures_2006,Cavalcanti_Criteria_2008} are interested in the coherence between distant parts of the spectrum $|\bra{k} \rho \ket{j}| > 0$, with $|a_k-a_j| \geq S$, exhibited by the state. For pure states it is easy to see that any state with variance $(\Delta A)^2 \geq V$ necessarily exhibits coherence between spectral parts with a distance of at least $S=4 \sqrt{V}$.\footnote{The same argument holds for mixed states if the variance is replaced by the quantum Fisher information, since the latter is the convex roof of the variance \cite{Yu_Quantum_2013}} Hence, a large variance is sufficient for the presence of large coherence $S$ in the state. However, it is not necessary, as a state with arbitrarily large $S$ might have an arbitrarily small $V$.

In addition, the measure of \citet{Sekatski_General_2017} gives a lower bound to an expression proportional to the variance (and in general to the quantum Fisher information). For general $b$, this bound can be arbitrarily loose but it becomes tight in the limit of low information gain $b$.

Finally, the variance is fully sufficient to capture the spread of a probability distribution in the limit of many copies of the system \cite{Yadin_General_2016}.
All this shows an important role for the variance as quantifier of a specific aspect of macroscopic quantumness.

\paragraph{Extensions of the variance for mixed states.}
\label{sec:extens-vari-mixed2}

Since the presence or absence of coherence in the basis of $A$ does not influence the variance $(\Delta A)^2$, it can only be used as a measure of macroscopic quantumness if the state is pure. The generalization of the variance for mixed states is not unique and the proposals of \citet{Shimizu_Detection_2005,Lee_Quantification_2011,Frowis_Measures_2012,Park_Quantum_2016} all go in different directions.

In principle, any measure $f$ for pure states can be extended to a measure $\hat{f}$ for mixed states via the convex roof construction (e.g., see Sec.~\ref{sec:Sekatski}). To this end, consider all possible pure state decompositions (PSD) of the state $\rho = \sum_k \pi_k \left| \psi_k \right\rangle\!\left\langle \psi_k\right| $ and minimize the average size
\begin{equation}
\label{eq:35}
\hat{f}(\rho) = \min_{\mathrm{PSD}} \sum_k \pi_k f(\psi_k).
\end{equation}
By definition, the convex roof is the ``worst-case'' average $f$ among all possible pure-state decompositions of $\rho$. At the same time, $\hat{f}$ is the largest convex function that reduces to $f$ when $\rho$ is pure \cite{Toth_Extremal_2013}. The quantum Fisher information, which appears in the measure of \citet{Frowis_Measures_2012}, is (four times) the convex roof of the variance \cite{Yu_Quantum_2013}.

The index $q$ was introduced as a generalization of the index $p$ for mixed states (see Sec.~\ref{sec:index-p}). \citet{Shimizu_Detection_2005} are mainly interested in identifying ``fully macroscopic'' quantum states, for which the two measures match $p = q = 2$. But this is no longer true for ``less macroscopic'' states. Specifically, there exist examples for pure states with $p = 1$ (i.e., the variance is $O(N)$), but $q = 1.5$ \cite{Shimizu_Erratum:_2016}.
This suggests that the intuition for the index $q$ does not entirely correspond to coherent spread of pure states as measured by the variance.  In particular, there is no general relation between the quantum Fisher information and the index $q$.

The measure $\mathcal{I}$ of \citet{Lee_Quantification_2011} is similar to the quantum Fisher information for low-rank instances (see example \ref{ex:SCS}). However, they differ in general \cite{Yadin_General_2016}.
Furthermore, \citet{Yadin_General_2016} pointed out a problematic relation of $\mathcal{I}$ with the two-norm (see Sec.~\ref{sec:conn-reso-theory}).

\subsubsection{Connections between some measures}
\label{sec:conn-betw-meas}

\begin{figure}[htbp]
\centerline{\includegraphics[width=.8\columnwidth]{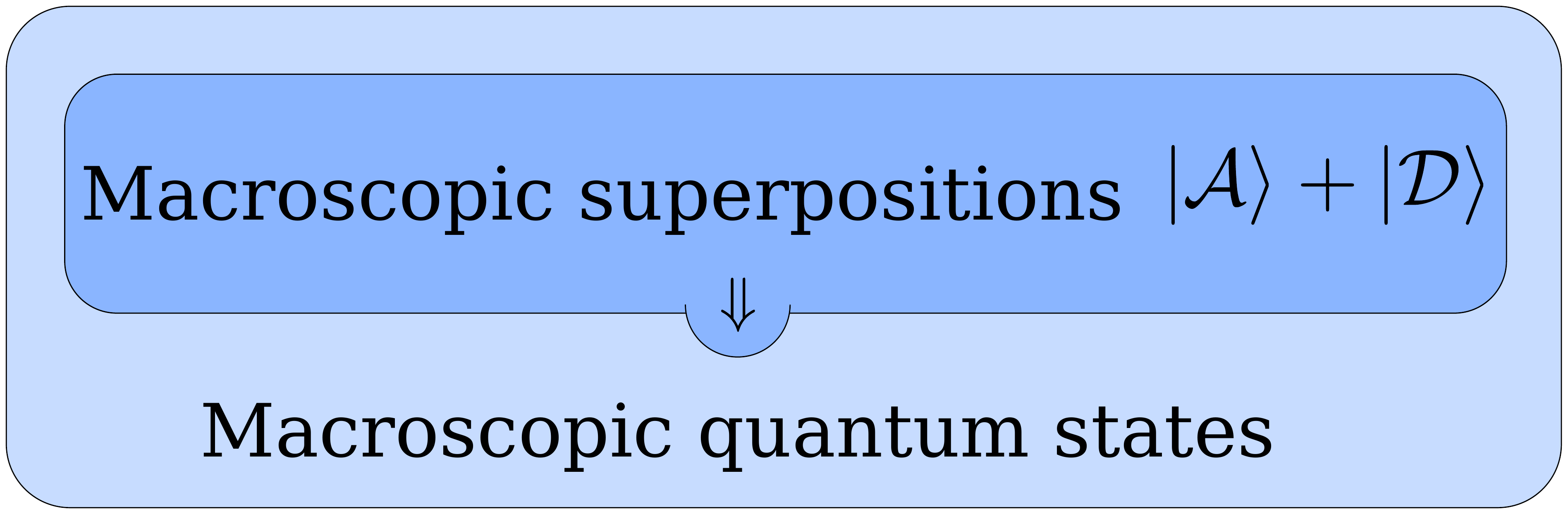}}
\caption[]{\label{fig:Hierarchy} Hierarchy between states that are macroscopically quantum according to some partition-based measures. Macroscopic superpositions according to \citet{Marquardt_Measuring_2008,Korsbakken_Measurement-based_2007} are also macroscopic quantum states according to variance-based measures if quasi-local operators are considered \cite{Frowis_Measures_2012}.}
\end{figure}

In this section, we review connections between some measures. In particular, one finds for some partition-based measures that macroscopic quantum states includes the concept of macroscopic superposition (\citet{Frowis_Measures_2012}; see Fig.~\ref{fig:Hierarchy}). Let us consider quantum states \da for spin ensembles. \citet{Frowis_Measures_2012} show that $C_{\delta} = O(N)$ (see Eq.~(\ref{eq:37})) if and only if $\bar{D} = O(N)$ (see Eq.~(\ref{eq:38})), which suggests a connection between the measures of \citet{Korsbakken_Measurement-based_2007} and \citet{Marquardt_Measuring_2008}. Next, if $C_{\delta} = O(N)$ then there exists a quasi-local operator $A$ (see Sec.~\ref{sec:spin-ensemble}) for which the variance scales quadratically $(\Delta A)^2 = O(N^2)$. Given that one accepts the step from local to quasi-local operators in the formulation of measures for macroscopic quantum states, this implies a large effective size for all measures based on the variance (see example \ref{ex:cluster}).

Further connections can be found between general measures\footnote{ \citet{Frowis_Measures_2012,Kwon_Disturbance-Based_2016,Sekatski_General_2017,Cavalcanti_Signatures_2006}} (see Secs.~\ref{sec:examples} and \ref{sec:link-meas-phot}).

Independent of these connections, the measure of \citet{Sekatski_General_2017} applies ideas for macroscopic superpositions (such as \citet{Korsbakken_Measurement-based_2007,Sekatski_Size_2014}) to general quantum states. This can be seen as a way to bridge measures for macroscopic superpositions and macroscopic quantum states.

Note that one can also spot significant differences between the measures, in particular, when the effective size is not $O(N)$. An interesting example are the SQUID experiments as discussed in Sec.~\ref{sec:superc-quant-interf}, for which diametrical results are obtained.

\subsection{Application to various physical setups}
\label{sec:physical-setups}
\label{sec:categ-regard-appl}
\label{sec:mathematical}

The proposed measures sometimes attempt to be applicable to all physical systems (such as spin ensembles, photons, superconducting devices and massive systems). Nevertheless, they are typically motivated with a specific setup in mind and one has to analyze the implications when applied to other systems. In the following, we focus on some general issues regarding the application of measures for macroscopic quantum states to various physical systems. This analysis is done for the variance. However, it can be partially repeated for other, more specific measures.

\subsubsection{Implications of large variance}
\label{sec:impl-large-vari}

Let us start with spin ensembles, for which the operator at hand is typically chosen to be a local operator $A =\sum_{l=1}^N a_l^{(l)}$. Since $||a_l^{(l)}|| = O(1)$, the spectral radius of the total operator is proportional to the number of spins, $N$. A state is said to be macroscopically quantum if the standard deviation (i.e., the square root of the variance) is large compared to the spectral radius.
A large number of spins is necessary for a large variance. Thus, this automatically leads to a connection between the large-scale character (i.e., the spread in the spectrum) with quantumness (through the coherence between the eigenstates). The connection to multipartite entanglement (in particular for the quantum Fisher information, see Sec.~\ref{sec:relat-mult-entangl}) further supports the use of the variance to measure macroscopic quantumness of pure spin-ensemble states.

For photons, one encounters a very similar situation. For simplicity, we discuss single-mode states in the following. By taking quadrature operators $A= X_\theta = \frac{1}{\sqrt{2}}\left(e^{\ii \theta} a  +e^{-\ii \theta} a^\dag \right)$  as the canonical choice for the variance, a large variance necessarily comes from many-photon states since $\langle a^{\dagger} a \rangle \geq \frac{1}{2}[(\Delta X)^2 + (\Delta P)^2 - 1]$. A large variance also implies more nonclassicality of the photonic state in the sense that $k$-partite mode-entanglement can be created by splitting the photonic mode into many spatial modes in a beamsplitter network. The same effect is achieved by letting the photons be absorbed by a sufficiently large atomic ensemble (see Sec.~\ref{sec:link-meas-phot}). Alternatively one can choose the photon number operator $A= a^\dag a$ for measuring the macroscopic spread. A large variance in photon number implies that large Fock states are involved in the superposition. However, it does not generally imply large mean photon number as examples with a fixed average photon number and arbitrarily large variance show.\footnote{For instance, $| \psi \rangle = \sqrt{1-1/k}\left| 0 \right\rangle + 1/\sqrt{k}\left| k \right\rangle $ for $k \gg 1$.} In addition, note that coherent states $\ket{\alpha}$ have a variance that increases linearly with the mean number of photons $|\alpha|^2$ for the photon number, but remains constant for any quadrature.

The simultaneous presence of quantumness and large system size for large-variance states in spin and photonic systems may encourage us to use it in other systems, such as superconducting devices or massive systems. After all, photonic states, superconducting circuits and motional degrees of freedom of massive systems are mathematically connected by using the same algebraic structure with the canonical commutation relation $[a^{\dagger}, a] = \mathbbm{1}$. However, the interpretation of the variance for photons is not trivially transferable to other systems. For example, the SQUID experiments from  \citet{vanderWal_Quantum_2000,Friedman_Quantum_2000,Hime_Solid-State_2006} lead to a controversial discussion of whether a macroscopic superposition of clockwise and anti-clockwise currents has been generated. While this is reviewed in more detail in Sec.~\ref{sec:superc-quant-interf}, we just summarize here the analysis of \textcite{Bjork_Size_2004}. Effectively, they calculate the variance of the ideal target state and find it to be roughly 1000 times larger than the variance of the ground state, which is assumed to be the most classical state.\footnote{Note that \textcite{Bjork_Size_2004} work with the standard deviation rather than the variance.} However, a clear interpretation of this number, for example, as an effective size of the electronic state is missing so far.

For massive systems, the situation is even more puzzling. For a fixed mass, it seems that the spread of the wave function in the spatial degree of freedom is a natural choice to measure the macroscopic quantumness of the system. However, the connection between large variance in position and large system size is lost. Moreover for partition based measures \cite{Leggett_Testing_2002,Korsbakken_Electronic_2009}  one does not change the macroscopic quantum character of, for example, a nucleus by increasing the distance of a superposition of being ``here and there'' from 1 cm to 1 m, as the states of individual protons and neutrons in the two branches are equally orthogonal in both cases. Similarly, it is an open question how the partition in subsystems should be made, for example it can be done equally well on the level of atoms or protons and neutrons. For the same reason the role of the mass of the individual constituents, and how it interacts with the spread in position, is open.

Note that if we go away from macroscopic distinctness and ask about quantum states/experiments that exclude small deviations from quantum mechanics, we indeed care about the distance between the two possible positions of the neutron. This is because typical collapse models are more effective for larger distances, and are have a precise mass dependence. From this point of view, the mathematical connection between \cite{Nimmrichter_Macroscopicity_2013} and \cite{Lee_Quantification_2011} (as mentioned in Sec.~\ref{sec:cat-regard-purp}) seems to be loose since the interpretation and meaning of the respective measures are different.

\subsubsection{Linking measures for photons and spins}
\label{sec:link-meas-phot}

Recently, \citet{frowis15} compared measures that are primarily defined for spin ensembles and photonic systems. The authors consider an ideal light-matter interaction between a spin ensemble in the ground state and an incoming photonic field. Under the assumption that the number of spins is much larger than the number of photons, a photon is linearly mapped to an atomic excitation. In technical words,
the first-order approximation of the Holstein-Primakoff transformation \cite{Holstein_Field_1940} allows one to identify $S_{-}/\sqrt{N} \leftrightarrow a$. This enables one to compare quantum states from the two physical systems and eventually entire measures defined for spins or photons. For example, it turns out that the measure of \citet{Korsbakken_Measurement-based_2007} could be formulated for a single photonic mode by asking how well \alive and \dead can be distinguished with highly inefficient detectors (i.e., ideal detectors with preceding photon loss). This resembles the ideas of \citet{Sekatski_Size_2014}, who considered coarse-graining instead of loss. Even though the different realization of the detector's imperfection changes the measure qualitatively, a strong conceptual connection of two or more measures leads to confidence in the principal idea. As another example, the connections between several works\footnote{\citet{Shimizu_Stability_2002,Shimizu_Detection_2005,Lee_Quantification_2011,Frowis_Measures_2012,Cavalcanti_Signatures_2006,Cavalcanti_Criteria_2008}} are further reinforced. Not only the measures are based on the variance (or related to it), the canonical choice of the operators (local operators for spin ensembles and quadratures in the photonic case) are intimately connected via this ideal light-matter interaction.

\subsection{Connection to multipartite entanglement}
\label{sec:relat-mult-entangl}

Many authors do not consider multipartite entanglement \textit{per se} when constructing a measure of macroscopic quantumness. An exception is the measure of \citet{Yadin_Quantum_2015}, who identify the macroscopic quantumness for the size of the maximally distillable GHZ state with LOCC. This yields a direct connection to entanglement theory, in which LOCC is the set of free operations. For other partition-based measures, LOCC are not free in general and might change the macroscopic quantum character of a state (e.g., the variance can increase under LOCC). Nevertheless, it is very natural to look for a relation between entanglement properties and macroscopic quantumness of a state, given that both have the partition of the system in subsystem  as a starting point.

The measure of \citet{Frowis_Measures_2012} based on the quantum Fisher information is related to multipartite entanglement in the following sense. An effective size of $N_{\mathrm{eff}}  \gtrsim k$ implies genuine entanglement within groups of at least $k$ spins (this statement can be made fully rigorous including $O(1)$ correction terms \cite{Hyllus_Fisher_2012,Toth_Multipartite_2012}). Furthermore, large $N_{\mathrm{eff}}$ is not only a witness for $k$-partite entanglement in the sense of \citet{Sorensen_Entanglement_2001}, but $N_{\mathrm{eff}} \gtrsim k$ also means that the two-body correlations within these entangled groups have a certain minimal strength \cite{Toth_Multipartite_2012}.
Note also that since a high value for the measures of \citet{Korsbakken_Measurement-based_2007} and \citet{Marquardt_Measuring_2008} implies a large quantum Fisher information, these measures are also sufficient of a large entanglement depth. Furthermore, \citet{Morimae_Superposition_2010} proves that a state with $q = 2$, Eq.~(\ref{eq:2}), contains multipartite entanglement measured with the distance to the set of separable states.

On the other hand there is no measure for which a large entanglement depth on its own is sufficient for macroscopic quantumness, as follows from the example of the state $\ket{W}=\ket{N,1}$ discussed in the sec. \ref{sec:not-an-easy}. This state is is fully non-separable (exhibits $N$-partite entanglement), but is not recognized as a macroscopic quantum state by any of measures listed above.

\subsection{Connection to resource theory of coherence}
\label{sec:conn-reso-theory}

In recent years, there has been a trend to formalize certain aspects of quantum mechanics such as entanglement \cite{Horodecki_Quantum_2009}, athermality \cite{Gour_resource_2015,Goold_role_2016}, asymmetry \cite{Gour_resource_2008} and coherence \cite{Baumgratz_Quantifying_2014,Marvian_How_2016,Yadin_Quantum_2016} in the framework of resource theories. The basic idea is to assume certain constraints on the generation and manipulation of quantum states. For this, one defines free states and free operations. Quantum states that are not producible from free states and free operations constitute a resource for a task with typically some quantum advantage. The most well-known resource theory is entanglement in multipartite settings, in which free operations are LOCC. Free states are states
that can be generated starting with product states and using LOCC.

\subsubsection{Resource theory for macroscopic coherence}
\label{sec:reso-theory-macr}

The concept of macroscopic distinctness is naturally connected to superposing states from different parts of a given spectrum. This suggests to work with a resource theory of coherence where not only a basis is chosen but a spectrum is associated to it. This is the framework of asymmetry or ``unspeakable coherence'' \cite{Marvian_How_2016}. There, quantum states that are not invariant under translations in the spectrum are a resource to detect the presence of processes that are generated by these translations. \textcite{Yadin_General_2016} realized the similarity between asymmetry and the attempts to define macroscopic quantum states.
Like in asymmetry, they define free operations as completely positive maps that cannot increase coherence between parts of a spectrum with a certain distance. This is in contrast to many variants of ``speakable coherence'' \cite{Marvian_How_2016}, in which coherence between neighboring basis states can be freely transferred to coherence between distant states. The motivation for this choice of free operations is obvious as it makes the creation of coherence a ``difficult'' task.
Note, however, that with this choice of free operations it is not more difficult to create far-distant coherence than coherence between nearby states. This is achieved by additionally requiring that any measure of quantum macroscopicity --in addition to be nonincreasing under free operations-- should assign larger values to superpositions of two basis states with increasing spectral distance.

It turns out that the variance is a valid measure for this resource theory of quantum macroscopicity for pure states. So the present resource theory agrees with all measures of general macroscopic quantum states that use the variance for pure states\footnote{\citet{Shimizu_Stability_2002,Lee_Quantification_2011,Frowis_Measures_2012,Park_Quantum_2016}}. The quantum Fisher information, which is used by \textcite{Frowis_Measures_2012} as an extension for mixed states, is the convex roof of variance \cite{Yu_Quantum_2013} and hence a valid measure in the framework of \textcite{Yadin_General_2016}. In contrast, the measure of \textcite{Lee_Quantification_2011} generally increase under the present free operations because of a problematic connection to the Hilbert-Schmidt norm, which is known to be not contractive under trace-preserving operations \cite{Ozawa_Entanglement_2000,Piani_Problem_2012,Yadin_General_2016}. Whether the index $q$ \cite{Shimizu_Detection_2005} fulfills the requirements is open (see Sec.~\ref{sec:conn-betw-meas}).

In this context, the work of \textcite{Kwon_Disturbance-Based_2016} gives further insight. Consider a distance function $D(\rho,\tau)$ that is (D1) positive semidefinite, (D2) invariant under joint unitary rotations of $\rho$ and $\tau$, (D3) contractive under physical maps and (D4) jointly convex. Using the dephasing map $\Phi_{\sigma}(\rho)$ defined in Sec.~\ref{sec:kwon}, the authors show that $D(\rho, \Phi(\rho))$ is a valid measure in the sense of \citet{Yadin_General_2016}. However, as pointed out by \citeauthor{Kwon_Disturbance-Based_2016}, the measure behaves in a surprising way if $\sigma$ is small (i.e., the distance between the original state and a strongly dephased stated is measured, see Sec.~\ref{sec:kwon}). If the dephasing is strong (i.e., $\sigma$ is small), the coherence between even nearby basis states is lost. In the case $\sigma = O(\sqrt{N})$, the measure coincides with the intuition for basic examples.

\subsubsection{Free operations in proposed measures}
\label{sec:free-oper-prop}

The arguments of \textcite{Kwon_Disturbance-Based_2016} suggest that while the axioms of \textcite{Yadin_General_2016} might be a good starting point for a resource theory of macroscopic quantumness, they seem to be not sufficient to unambiguously identify relevant measures for macroscopic quantum states. A further noteworthy point is that fully incoherent states in one basis generally exhibit large coherence in an other basis. Since in many examples for photonic systems or spin ensembles it is not possible to determine a single preferred observable but we have a set of observables to choose from, further improvements for the choice of free operations could lead to a resource theory that captures our intuition of macroscopic distinctness even better. Finally, the right choice of free operations is nontrivial in general \cite{Marvian_How_2016}. For example, LOCC in entanglement theory is not the most general class of operations that do not increase entanglement in the system. Yet it is the preferred set from a physical point of view.

We emphasize that many measures discussed in Sec.~\ref{sec:preliminary-measures} implicitly or explicitly choose a set of free operations. Let us briefly comment on this choice for measures designed for spin systems. First, even a collective local rotation $U^{\otimes N}$ can formally affect the macroscopic quantumness of the state for a measure that is defined for a fixed operator (e.g., a collective operator like $A=S_z$). In practice\footnote{For example, \citet{Bjork_Size_2004,Park_Quantum_2016,Sekatski_General_2017,Kwon_Disturbance-Based_2016}}, however, one typically chooses an optimal direction of the collective observable, $A^{\prime} = \vec{n}\cdot \vec{S}$. This leads to the choice $A'=U^{\dag \otimes N} S_z U^{\otimes N}$ and hence collective rotations $U^{\otimes N}$ are implicitly considered to be free. % This can be made explicit by looking for the extensive quantity for which the variance of the state is maximal (for pure state), which is equivalent to optimizing over all global unitary rotations applied on the state while keeping a the collective operator fixed.
The next level is to consider all local unitary (LU) transformations (i.e., $U_1 \otimes \dots \otimes U_N$) as free. This is implicitly done in \citet{Frowis_Measures_2012,Shimizu_Detection_2005} as one chooses the optimal local operator that maximizes the variance of the state. Next, one could consider a single round of local operations (i.e., measurements) followed by local unitaries (LO+LU). In fact, the state $| \mathrm{GHZ}_{N/2} \rangle $, with variance $(N/2)^2$, can be prepared from a one-dimensional cluster state composed of $N$ qubits with variance $N + O(1)$, by measuring every second qubit in the $\sigma_x$ with subsequent local rotations on the remaining qubits depending on the measurement results \cite{Frowis_Measures_2012,Yadin_Quantum_2015}. Hence, under the assumption that LO+LU are free, cluster states become macroscopically quantum. Finally, \citet{Yadin_Quantum_2015} consider all LOCC as being free operations.

\subsection{How to determine the effective size in experiments}
\label{sec:meas-exper}

An important point is the applicability of the measures to real experiments. In addition of being convincing on a theoretical level, the effective size of a measure should ideally be extractable
from experimental data. In this section, we examine some proposed protocols where this is possible without a full state tomography.

\subsubsection{Macroscopic distinctness}\label{sec:macr-dist}
Macroscopic distinctness as defined by \citet{Korsbakken_Measurement-based_2007,Sekatski_Size_2014} can theoretically be measured in the lab. For the effective number of particles \cite{Korsbakken_Measurement-based_2007}, one needs to have access to single particles or at least small groups of them. The distinguishability is given by the optimal measurement. If this is not available in the lab, any measurement will give a lower bound on the effective size. In the case of classical detectors \cite{Sekatski_Size_2014}, it is necessary to have a detector that can be modeled in the way presented in Sec.~\ref{sec:Sekatski}, ideally with a tunable resolution.
For both approaches, one has to generate either \alive or \dead and measure the probability distribution as a function of number of measured particles or detector resolution. From this distribution, one can calculate the effective size. Similarly, the ``subjective'' part of \citet{Laghaout_Assessments_2015} can be measured by preparing the single components $| b_k \rangle $, but a special access to subsystems or tunable detectors is not necessary.
The quantumness, as discussed in Sec.~\ref{sec:macr-superp}, is measured independently.

\subsubsection{Macroscopic coherence}\label{sec:witn-macr-quant}
Already \citet{Leggett_Macroscopic_1980} noted that quantum states with large-scale quantum correlations are only distinguishable from mixtures if $O(N)$ particle correlations are measured.\footnote{An extreme example is the GHZ state, Eq.~(\ref{eq:39}), for which omitting a single particle is enough to hide all essential quantum features.} He argued that a way out is the unitary time evolution even for local or two-body Hamiltonians as the expansion of $\exp(-i H t)$ contains $O(N)$ correlation operators. Implicitly, this idea is present in the following findings.

The way \citet{Bjork_Size_2004} formalize the idea of interference utility is precisely in the spirit of \citet{Leggett_Macroscopic_1980}. A broad distribution in the spectrum of the Hamiltonian $H$ is directly connected to the maximal ``speed'' of evolution via the inequality \cite{Mandelstam_Uncertainty_1945,Fleming_unitarity_1973}
\begin{equation}
\label{eq:45}
|\left\langle \psi \right| e^{-i H t } \left| \psi \right\rangle |^{2} \geq \cos^2 \left( \Delta H t  \right)
\end{equation}
for all $\Delta H |t| \leq \pi/2$. This inequality tells us that witnessing a fast change of the state implies wide-spread coherence spread in the spectrum of $H$. Equation (\ref{eq:45}) can be directly generalized to mixed states \cite{Frowis_Fleming's_2008}, but the inequality generally becomes loose for low purity.

Luckily, the intimate connection between a sensitive notation of statistical distance measured by the Fubini-Study metric and the variance of the generator has a well-behaving extension to mixed states \cite{Wootters_Statistical_1981,Uhlmann_gauge_1991,Braunstein_Statistical_1994}. There, the equivalent metric is the Bures metric $ds_B$ and the variance is replaced by the quantum Fisher information. One then finds that
\begin{equation}
\label{eq:47}
ds_B = \frac{1}{2} \sqrt{\mathcal{F}(\rho,H)} dt,
\end{equation}
where $ds_B$ measures the infinitesimal distance between $\rho(t)$ and $\rho(t+dt)$ generated by $H$. This relation has far-reaching consequences for the foundations and applications of quantum mechanics. Here, we are interested in measuring lower bounds on the quantum Fisher information using Eq.~(\ref{eq:47}). This can be done in several ways. First, one can directly tighten Eq.~(\ref{eq:45}). One replaces the left-hand side by the fidelity, $F(\rho,\sigma) = [\mathrm{Tr} \sqrt{\sqrt{\rho}\sigma \sqrt{\rho}}]^2$, between the initial and final state and, in addition, the standard deviation on the right-hand side by $\sqrt{\mathcal{F}}/2$ \cite{Frowis_Kind_2012}. After choosing a fixed measurement with outcomes $\left\{ x \right\}$, the probability distribution before, $p(x)$, and after, $q(x)$, the evolution of duration $t$ can be used to bound the quantum Fisher information \cite{Frowis_Detecting_2016}
\begin{equation}
\label{eq:49}
\mathcal{F}(\rho,A) \geq \frac{4}{t^2} \arccos^2 \sum_x \sqrt{p(x) q(x)}.
\end{equation}
Second, one can derive a tighter version of the Heisenberg uncertainty relation\footnote{\citet{Frowis_Tighter_2015,Pezze_Entanglement_2009,Kholevo_Generalization_1974,Hotta_Quantum_2004}}
\begin{equation}
\label{eq:48}
\mathcal{F}(\rho,A)  \geq \frac{\langle i [A,B] \rangle^2}{(\Delta B)^2}.
\end{equation}
By measuring the variance of $B$ and the expectation value of $ i [A,B]$ one is able to find lower bounds on $\mathcal{F}(\rho,A)$.

Note the similarity between Eq.~(\ref{eq:48}) and Eq.~(\ref{eq:52}), whose conceptual closeness becomes even more evident when studying the proofs of the bounds. This highlights the connection between the ideas of \citet{Cavalcanti_Signatures_2006,Cavalcanti_Criteria_2008} and \citet{Frowis_Measures_2012}.

\subsubsection{Measuring loss of coherence}\label{sec:meas-effect-decoh}
The measure of \citet{Nimmrichter_Macroscopicity_2013} is specifically designed to be experimentally accessible. Every experiment that excludes a certain parameter regime of a collapse model has a value in the framework of Sec.~\ref{sec:NimmrichterHornberger}, which is directly measured by the amount of time over which quantum coherence is maintained.

\begin{figure}[htbp]
\centerline{\includegraphics[width=.7\columnwidth]{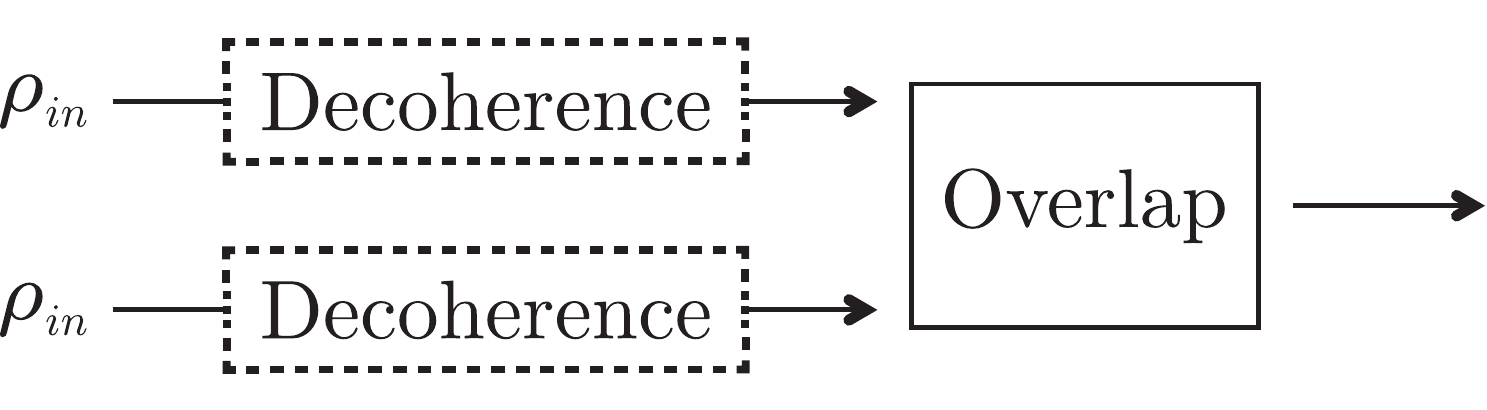}}
\caption[]{\label{fig:JeongOverlap} Basic scheme of \citet{Jeong_Detecting_2014} to experimentally access the measures of \citet{Lee_Quantification_2011,Park_Quantum_2016}. Two identical copies of a state $\rho$ are generated. Then, either both copies are subject to decoherence, Eq.~(\ref{eq:7}), or not. In both cases, the purity of $\rho$ is measured via an overlap measurement to determine $\mathrm{Tr}\rho^2$ (which is not the fidelity between the two states). Comparing the purity in the presence and absence of decoherence allows to estimate $\mathcal{I}(\rho)$. From \citet{Jeong_Detecting_2014}.}
\end{figure}

The contribution of \citet{Lee_Quantification_2011} is also strongly connected to the susceptibility under decoherence. The larger the loss of purity by applying a canonical noise channel, the more macroscopically quantum the state is according to the measure $\mathcal{I}$. \citet{Jeong_Detecting_2014} showed that $\mathcal{I}$ can be experimentally estimated without the need of a full state tomography.
The basic idea is to extract the purity from an overlap measurement of two identically prepared states (see Fig.~\ref{fig:JeongOverlap}). This is done by comparing the purity of the state before and after a short application by the decoherence channel described by Eq.~(\ref{eq:7}). An overlap measurement realizing $\mathrm{Tr}\rho^2$ can be implemented with a SWAP operation between the two systems, followed by a suitable measurement of both modes. For a single-mode photonic state, this can be realized using a beamsplitter and a photon-number resolving detector after one of the modes. Similarly, one can combine the two copies of the state with auxiliary systems and use controlled SWAP operations \cite{Jeong_Detecting_2014}. Furthermore, the scheme is adaptable to the measure of \citet{Park_Quantum_2016} for spin ensembles.

\subsubsection{Correlations for index $q$}\label{sec:correlations-index-q}
The index $q$ from \citet{Shimizu_Detection_2005}, Eq.~(\ref{eq:2}), is in principle measurable without full state tomography. We consider a scenario in which one has access to single particle measurements. Then, the goal is to measure two-body correlations in the spirit of Eq.~(\ref{eq:14}), but in several conjugate bases. This is a way to generalize the concept of Bell inequalities or entanglement witnesses to multipartite settings. By verifying $O(N^2)$ pairs with $O(1)$ correlations, one can conclude that the present state exhibits $q = 2$ \cite{Shimizu_Detection_2005}.

\subsection{Summary and conclusion}
\label{sec:summaryCht2}

Many different aspects of macroscopic quantumness are covered by a growing number of proposals. In view of our initial starting point --to formalize the idea of macroscopic distinctness in Schrödinger's cat example-- a possible conclusion is the following.

Measures based on an explicit \da structure clearly work well whenever \alive and \dead themselves are localized in a given spectrum in which the two states are maximally distinguishable (i.e., \alive and \dead behave ``classically'', see example \ref{ex:GHZ}). In this regime, a macroscopic superposition is a special case of a macroscopic quantum states as characterized by more general measures. However, it is open whether the original intention of an effective particle number is maintained for general states, in particular if \alive and \dead themselves are nonclassical (see example \ref{ex:Dicke}).

To measure macroscopic distinctness, it seems appropriate to find functions that evaluate the spread of probability distribution in the spectrum of a certain operator (or a set of possible operators). If the underlying quantum state is pure, the coherence between far distant parts is then a signature of macroscopic quantumness. The variance is proposed by many authors and is a good choice to classify the global structure of the probability distribution. More sophisticated measures (e.g., entropies and distant functions in combination with additional parameters) are able to resolve finer aspects.

It is important to have a proper scaling of the function that measures macroscopic distinctness for a given observable. For example, the variance of a product state in spin ensembles increases linearly for collective operators. However, this should not be seen as macroscopic quantum effect but as an accumulation of coherence on the single-particle level. In this example, a pure product state is then considered as a state carrying ``one unit of quantumness'' and hence a measure has to be appropriately rescaled.

The presented measures are applied to various physical systems. While the (rescaled) variance arguably makes sense for spins and photons, its applicability when dealing with spatial or superconducting degrees of freedom is open. On the other hand, for massive systems and superconducting devices, measuring the (potential) falsification of collapse models has a clear operational meaning.

The extension to mixed states should satisfy basic requirements from information theory. The convex-roof construction for measures defined for pure states generally fulfills these conditions, but it is not the only option. In the case of the variance, the convex-roof extension leads to the quantum Fisher information with the additional benefit of having tight and accessible lower bounds.

Currently, the quest of a well-motivated set of free operations for macroscopic quantumness will further help to classify and understand macroscopic quantumness. On the more practical side, there is a trend to make the presented measures applicable to experimental data.

%%% Local Variables:
%%% mode: latex
%%% TeX-master: "master"
%%% End:

\section{Limits for observing quantum properties in macroscopic states}
\label{sec:limits-observ-quant}

The measures that were presented in Sec.~\ref{sec:meas-macr-superp} provide various ways to characterize sets of macroscopic states and hence to study  the sensitivity of these states to different noise in a systematic way. This reveals certain limits and inherent difficulties to prepare, maintain and observe macroscopic states. In the following we will investigate these limitations, and summarize a number of results that were obtained in this context. In a certain sense, they suggest that the very same features that make a state macroscopic also make it susceptible to noise, difficult to prepare and maintain and almost impossible to measure.

We start by discussing how quantum states are affected by noise and decoherence in Sec.~\ref{sec:maint-macr-quant}. To this aim we first consider some general results on decoherence that are applicable to all quantum states, and then state more specifically how macroscopic quantum superposition states and general macroscopic quantum states are affected. In Sec.~\ref{sec:meas-detect-macr}, we then turn to the measurement of such states, and show how limited detector efficiency and resolution hinders the detection of macroscopic quantumness. We also discuss the closely related but slightly less demanding task of certifying quantum states (prior to the action of noise), which however requires (exponentially) growing resources for macroscopic superpositions. We shortly touch the issue of preparing states in Sec.~\ref{sec:prep-macr-quant}. Finally, in Sec.~\ref{sec:furth-limit-count}, we present a number of methods and approaches to circumvent some of the mentioned problems. Most of these methods are however only applicable in a limited sense, in particular for certain, specific kinds of noise. The only exception is quantum error correction, where we argue that encoded quantum superposition states can indeed be prepared, maintained and measured. Theses states are however not necessarily macroscopically quantum in a sense as discussed in Sec.~\ref{sec:meas-macr-superp}, but only on a coarse-grained level. We summarize in Sec.~\ref{sec:summary}.

\subsection{Maintaining macroscopic quantum states}
\label{sec:maint-macr-quant}

Let us start by discussing the limitations arising from the impossibility to maintain the state of the system by shielding it perfectly from the environment.

\subsubsection{Decoherence}
\label{sec:decoherence}

The decoherence program \cite{Zurek_Decoherence_1991, Zurek_Decoherence_2003} can be viewed as a general explanation why quantum features do not prevail at a macroscopic scale. Every quantum system interacts with its environment and becomes entangled. As the environment cannot be controlled, it has to be traced out and the system-environment entanglement manifests itself in decoherence of the system, i.e., a certain mixedness due to the lack of knowledge on the environment leading to a reduction of quantum coherences.

As we show below a very simple instance of the generic effective quantum-to-classical transition for states of bosonic fields follows from the seminal works of \citet{Husimi_Some_1940}, \citet{Glauber_Coherent_1963}, \citet{Sudarshan_Equivalence_1963} and others. Any state of a bosonic field $\rho$ can be faithfully represented by quasi-probability distributions P, W and Q \cite{Vogel_Quantum_2006}. In particular, the Glauber-Sudarshan P-representation is  an expansion of the state in the over-complete basis of coherent states
\be\label{eq:th noise}
\rho = \int  \text{P}(\alpha)\prjct{\alpha}d^2\alpha.
\ee
If P$(\alpha)\geq0$ is positive the state $\rho$ is a mere mixture of coherent states and is said to be classical. This terminology arises from the fact that the coherent states can be thought of as the most classical subset of the set of all possible state of a mode of a quantum field: they saturate the uncertainty relations for all pairs of quadratures, and also phases and number of photons. Furthermore, they are eigenstates of the forward part of the field operator, and they remain within the set of coherent states under passive operations. For instance coherence states are the only pure states of the field that do not generate entanglement when sent on a beam splitter. Negative values of the P-function indicate intrinsic quantum features. The P-function is related to the Husimi Q-function via a convolution with a Gaussian,
\be
\text{Q}(\alpha)= \int  \text{P}(\beta)e^{-|\beta-\alpha|^2}d^2\beta.
\ee

However, Q$(\alpha)=\frac{1}{\pi} \t{tr} \rho \prjct{\alpha}$ is a probability distribution and hence is always positive. A direct consequence of these two observations is that, regardless of the initial state $\rho$, a decoherence process $\mathcal{E}$ that acts on the P-representation of a state as a convolution renders the state classical -- the P-function of the state after the decoherence $\rho'= \mathcal{E}(\rho)$ is positive. It is easy to see that such a decoherence process is given by thermal noise, i.e., the diffusion of the state in phase-space generated by
\be\label{}
\mathcal{E}(\rho)= \int e^{-\ii (\lambda_1 X +\lambda_2 P)}\rho\, e^{\ii (\lambda_1 X +\lambda_2 P)}g_\sigma({\bm\lambda}) d^2{\bm \lambda},\ee
with a Gaussian $g_\sigma({\bm \lambda})=\frac{1}{2\pi\sigma^2}e^{-|{\bm \lambda}|^2/(2\sigma^2)}$, the  random variable ${\bm \lambda} = \binom{\lambda_1}{\lambda_2}$ with $\sigma=1$ and two conjugate quadratures $X$ and $P$.
A similar observation holds for spins. In this case quasi-probability distributions P$({\bf n})$, the expansion in spin-coherent states, and Q$({\bf n})$ are defined on the sphere \cite{Arecchi_Atomic_1972, Agarwal_State_1998}, and P is mapped to Q by spherical smoothing \cite{Agarwal_State_1998, Schmied_Habilitationsschrift_2017}. The noise that maps P to Q in this case corresponds to a random rotation of the state generated by $e^{\ii \,{\bm \lambda}\cdot {\bf S}}$ with the total spin operator ${\bf S}$ and a ``spherical Gaussian'' random variable ${\bm \lambda}$.

Decoherence is a generic mechanism that concerns quantum features in general, and i.e., it is not specific to  macroscopic quantum states or macroscopic quantum superpositions, discussed in this article. Nevertheless, it is commonly believed that macroscopic quantum superpositions are particularly affected by decoherence \cite{Zurek_Decoherence_2003}, and therefore are particularly fragile and hard to observe. Such beliefs are supported by a growing manifold of examples, but also generic statements, which we will discuss in this section.
Let us mention a few early results. \citet{Caldeira_Influence_1985} discussed the instability of superpositions of Gaussian wave packets in a harmonic potential weakly coupled to a bath. \citet{Yurke_Generating_1986} emphasized the same effect for the superposition of coherent states under loss. \citet{Milburn_Dissipative_1986} demonstrated that quantum signatures of an initial coherent state evolving in an anharmonic potential and coupled to a bath disappear faster when the energy of the initial state is larger. This list can be easily extended.
For instance, consider the effect of the thermal noise of Eq.~\eqref{eq:th noise} on a superposition of coherent states $\ket{\alpha}+ \ket{-\alpha}$. From the expansion of the coherent state state in the $X$-quadrature basis $\braket{\pm\alpha}{x}= \pi^{-1/4}e^{(x \mp \sqrt{2}\alpha)^2/2}$ (for real $\alpha$) it directly follows that for large $\alpha$ the coherence $\mathcal{E}(\coh{\alpha}{-\alpha})$ is damped by a factor $\propto e^{-\alpha^2 \sigma^2}$. Another prominent example is the GHZ state, for which
it easy to see that by interaction with independent local environments the coherences $(|0\rangle\langle 1|)^{\otimes N}$ also decay exponentially fast $\propto e^{-\gamma N t}$, as single-qubit coherences $|0\rangle\langle 1|$ decay as $\propto e^{-\gamma t}$. {See \cite{Simon_Robustness_2002, Cavalcanti_Open_2009, Aolita_Noisy_2010} for detailed studies of the effect of noise on the entanglement in GHZ and other graph states.}

Interestingly, one of the first attempts to provide an effective size of a macroscopic superposition \cite{Dur_Effective_2002} took the fragility of the coherences, i.e., the intrinsic quantum features of a superposition, as a starting point. Hence, any state that is macroscopic in this sense is by definition also fragile to noise.
Later attempts to provide measures for macroscopic quantumness do not contain a direct reference to the behavior of the state under noise and decoherence. Nevertheless, one can still relate the features responsible for macroscopic quantumness to the behavior of the states under certain noise processes as we now discuss.

\subsubsection{Fragility of macroscopic quantum superpositions}
\label{sec:fragility superpositions}

We first consider macroscopic quantum superpositions. In order to confirm that this is indeed a macroscopic quantum superposition, one needs to show that \alive and \dead are macroscopically distinct, and that the state is indeed quantum, i.e., a coherent superposition and not an incoherent classical mixture. Since these two aspects are independent we can formulate our question as follows:  how does the decay of the quantumness under noise is affected by the macroscopic distinctness of the state?  Let us consider the original situation imagined by Schr\"odinger where the macroscopic system $M$ is entangled with a microscopic atom $A$ as $\ket{\uparrow}_A\ket{\mathcal{A}}_M + \ket{\downarrow}_A\ket{\mathcal{D}}_M$, the quantumness of the state is then identified with the entanglement between the systems $M$ and the $A$.\footnote{The presence of the atom is very instructive but by no means necessary. In its absence the quantumness  can be identified as the distance of the state from the mixture, or as the amplitude of the coherence term $\coh{\mathcal{A}}{\mathcal{D}}$ in the density matrix.} We are interested in the decay of entanglement in the state after the action of some noise channel on the macroscopic system, which can always be represented by a unitary interaction of the system with the environment.

The example of the N-particle GHZ state, with \alive$=\ket{0}^{\otimes N}$ and \dead$=\ket{1}^{\otimes N}$, mentioned earlier is a good starting point. It is enough for the environment to measure a single particle in order  to extract the \emph{which-branch information} and collapse the superposition to a mixture losing all entanglement. Importantly, the probability that no particle is measured by the environment decreases exponentially with $N$. This makes the GHZ state such fragile. The crucial property is the ease to distinguish the two branches, which increases with $N$. As one could expect, such a link between fragility and macroscopic distinctness can be generally made for any measure based on the ease to distinguish between the two branches (cf.~Sec.~\ref{sec:cat-regard-purp} and \citet{Sekatski_Difficult_2014}).

The proposal of \citet{Korsbakken_Measurement-based_2007} (Sec.~\ref{sec:Korsbakken}) implies that a macroscopic superposition is fragile with respect to projectively measuring small parts of the system or simply losing these particles to the environment. This follows from the observation that if the two branches can be distinguished by measuring only a small number of subsystems, then even a tiny amount of loss leaks enough particle to the environment to allow it to fully collapse the state to one of the two branches.

Similarly, a superposition state that is macroscopic regarding classical measurements of an operator $A$ (see Sec. \ref{sec:Sekatski}), i.e., measurements that have in general some finite resolution and only disturb the system weakly, is fragile with respect to a dephasing noise generated by the same operator. This follows from the observation that such a noise channel can be represented as a weak measurement of the observable $A$ by the environment. Again for a superposition of macroscopically distinct states even a tiny amount of noise allow the environment to obtain full which-branch information and destroy the quantumness of the state.

This implies that a coherent superposition of two macroscopically distinct quantum states very quickly becomes a mixture under the effect of decoherence. Therefore the very same feature that defines macroscopic distinctness makes the maintenance of the quantumness of the superposition state extremely challenging.

We remark that a similar result was shown \cite{Sekatski_General_2017} for  states with a more general structure $\sum_n \sqrt{p_n}\ket{n}_A \ket{\mathcal{A}_n}_M$, for which the entanglement of formation is also increasingly fragile with respect to the measure of Sec.~\ref{sec:sekatski2017}.

\subsubsection{Sensitivity of macroscopic quantum states}
\label{sec:sens-regard-meas}

A separation between macroscopicity and quantumness for a given state is typically not made for general macroscopic quantum states. Therefore, the argument from the previous section cannot be easily extended to arbitrary states. Here, we present several ways to tackle the question about the fragility of macroscopic quantum states: (a) First, one can quantify the effect of noise on a state by measuring how fast the state becomes mixed under the effect of noise. (b) Second, the effect of noise on a state can be quantified by asking how much the state itself is susceptible or stable with respect to noise, i.e., how far does the state get from itself after the action of the noise. (c) Finally, one can directly analyze how the macroscopic quantumness of the state is affected by the noise. In other words one can bound the maximal size of the state that is the output of some noise process, or derive requirements on the noise that have to be satisfied in order to prepare states of a desired size.

\paragraph{Susceptibility regarding purity.}
\label{sec:decay-purity}

The purity of a state, $\mathrm{Tr}\rho^2$, measures how close a quantum state $\rho$ is to a pure state. Given a pure initial state  $\prjct{\Psi}$,    \citet{Shimizu_Stability_2002} considered the effect of weak local noise on the purity of the state by analyzing its decay rate
\begin{equation}
\label{eq:3}
\Gamma = -\frac{1}{2} \frac{d}{dt} \ln \mathrm{Tr}[\rho(t)^2]|_{\tau_c \ll t \ll 1/\Gamma},
\end{equation} where $\tau_c$ is a correlation time of the noise process. The local noise process is modeled as a local classical noise described by the Hamiltonian $H= \lambda \sum_x \, f(x) \hat a(x)$ with a random variable $f(x)$ and a local operator $\hat a(x)$ acting on the system at position $x$, or as local environmental induced decoherence described by $H=\lambda  \sum_x \hat f(x) \hat a(x)$ where $\hat f(x)$ is a local operator at position $x$ of the environment.
 \citeauthor{Shimizu_Stability_2002} found that, quite generically for both kinds of weak disturbances,
\begin{equation}
\label{eq:4}
\Gamma \approx \lambda^2 \sum_k g(k) \left\langle \Psi \right| \delta A_k^{\dagger} \delta A_k \left| \Psi \right\rangle ,
\end{equation}
where $\lambda$ is the small interaction strength, $g(k) = O(1)$ is the spectral intensity and $\delta A_k = \sum_x e^{-i k x}(\hat a(x) -\bra{\Psi}\hat a(x)\ket{\Psi})$ is a collective operator. Hence, the relative decoherence rate $\Gamma/N$ is always constant if the initial state has an $O(N)$ variance with respect to all local operators $A_k$. On the other hand, $\Gamma/N$ is system size dependent if there exists an $A_k \equiv A$ for which $(\Delta A)^2 = O(N^{1 + \epsilon})$ with $\epsilon > 0$ (cf.~the effective size $\max_{A: \text{local}}(\Delta A)^2/N$, Sec.~\ref{sec:QFI}).

Similar results were found for the phase space when the decoherence is generated by quadrature operators. \citet{Lee_Quantification_2011}  showed that their measure can be equivalently defined as the susceptibility of the state to the noise process given by loss of photons $\mathcal{I}(\Psi) \approx \Gamma$ (see Sec.~\ref{sec:LeeJeong}).

\paragraph{Susceptibility regarding change of state.}
\label{sec:susceptibility}

\citet{Shimizu_Stability_2002} define the stability of the state under local measurements through the correlations observed between two local observables $a(x)$ and $b(y)$ at two different locations. More precisely they say that a state $\rho$ is stable under local measurements if for any $\varepsilon>0$
the correlation between the observable is low enough $|P(b|a)-P(a)|\leq \varepsilon$ for sufficiently large distance $|x-y|$. \citet{Shimizu_Stability_2002} show that for this definition a quantum state is insensitive to local measurements if and only if it has the cluster property (i.e., states with $p = 1$ in Eq.~(\ref{eq:13})). Otherwise, the measurement of a single spin $l$ might significantly change the probability distribution for another spin $l^{\prime}$. This becomes evident when inspecting Eq.~(\ref{eq:14}): A large variance implies the presence some two-body correlations between different spins, hence the measurement results of $a^{(l)}$ and $a^{(l^{\prime})}$ are correlated.

The disturbance-based measure $M_{\sigma}(\rho)$ proposed by \citet{Kwon_Disturbance-Based_2016} (see Sec.~\ref{sec:kwon}) by definition quantifies how far the state gets from itself after the action of the map $\Phi_\sigma$, which corresponds to the coarse-grained measurement of an operator $A$. Since the map $\Phi_\sigma$ can be equivalently seen as a dephasing noise generated by the same operator $A$, it follows that macroscopic quantum states are particularly sensitive to such noise.

\paragraph{Fragility of macroscopic quantum states under noise.}
\label{Sec:limitQFI}

Let us start with the measure by \citet{Frowis_Measures_2012} based on the quantum Fisher information (see Sec.~\ref{sec:QFI}). In quantum metrology, the question about the maximally attainable quantum Fisher information in the presence of noise is central, as the quantum Fisher information is related to the precision of the estimation protocol. Hence, as summarized in Sec.~\ref{sec:quantum-metrology}, many tools were developed to upper-bound the quantum Fisher information of a protocol given a Hamiltonian and a noise process.\footnote{\citet{Fujiwara_Fibre_2008,Escher_General_2011,Demkowicz-Dobrzanski_elusive_2012,Sekatski_Quantum_2016}} Using these tools it can be shown\footnote{A. Lopez Incera \textit{et al.}, in preparation.} that the quantum Fisher information of any state after the action of generic local noise processes for any  operator $A$ can only scale as $O(N)$.\footnote{In the case of qubits the only exception is the Pauli noise given by $\mathcal{E}(\rho) = p \rho +(1-p) \sigma_{\bf n} \rho\, \sigma_{\bf n}$, under which the quantum Fisher information can still scale quadratically. As an example, recall that a Dicke state is an eigenstate of such noise process for $\sigma_{\bf n}=\sigma_Z$.} This limitation is rather severe. For instance, for the depolarizing noise\footnote{Local depolarizing noise is described by a completely positive trace preserving map acting on qubit $a$, ${\cal E}^{(a)}(p) \rho = p \rho + \tfrac{1-p}{4}\sum_{j=0}^3 \sigma_j^{(a)} \rho \sigma_j^{(a)}$, where $1-p$ is called the error probability. \label{fn:2}} with only 1\% error probability per particle, one is limited to $\mathcal{F}/(4N) \leq  10$. For the measure of \citet{Frowis_Measures_2012} this means that the effective size can not exceed ten. In other words the same quantum Fisher information can be attained by a state where spins are only entangled within groups of size ten.

Similarly, \citet{Park_Disappearance_2016} showed that thermalization suffices to destroy macroscopic quantum states. To this end, the authors argued that after thermalization the variance of all local operators behaves extensively, $(\Delta A)^2 \sim O(N)$ (as it is assumed in thermodynamics).

\citet{Carlisle_Limitations_2015} used the measure of \citet{Lee_Quantification_2011} (see Sec. \ref{sec:LeeJeong}) to assess the achievable size of macroscopic quantum superpositions in optomechanical set-ups. They showed that it is in general very hard to obtain macroscopic superpositions, and in fact require a large single-photon optomechanical coupling strength and postselection.

\subsection{Measuring and detecting macroscopic quantum states}
\label{sec:meas-detect-macr}

We will now discuss limitations to observe macroscopic quantum states with imperfect or size-limited measurement devices.

\subsubsection{Coarse-graining and control of measurements}
\label{sec:stab-dist-stat}

Another reason which is sometimes invoked to explain the apparent absence of quantum effects on macroscopic scale is the limited resolution of measurements apparatus.  The intuition here is that the detection of quantum effects such as quantum superpositions or entanglement on a macroscopic scale require high measurement precision. There is a long list of examples confirming this intuition.

Early works by \citet{Mermin_Quantum_1980} and \citet{Peres_Quantum_2002} analyzed the task of witnessing the entanglement of a singlet state (total spin zero) of two large spins by measuring the spin operators $S_{\bf n}$ on the two spins. Mermin showed that when the size of the spins increases one requires a better and better control on the angle of the measurements, the direction of the measured spin ${\bf n}$,  in order to demonstrate entanglement. Peres showed that the   resolution of the measurement $1/\sigma$ required to demonstrate entanglement also becomes more stringent with the size of a spin. A lack of resolution of the measurement of an operator $A=\sum_k a_k \prjct{k}$ refers to coarse-graining of the measurement outcomes. Formally, to account for a finite coarse-graining $\sigma$, the POVM elements of an ideal measurement $E_k=\prjct{k}$ are modified as
\be
E_k =\prjct{k} \to E_k^\sigma=\sum_{k'} n_\sigma(a_k-a_{k'})\prjct{k'},
\ee
with some distribution $n_\sigma(x)$ of width $\sigma$ (often taken to be a Gaussian distribution or a square function). Similar results for coarse-grained measurements of Stokes operators\footnote{The cases of photonic states discussed here is very similar to the case of large spins, as the Stokes operators for two bosonic modes $J_z= \frac{1}{2}(a^\dag a -b^\dag b)$, $J_x= \frac{1}{2}(a^\dag b +a \, b^\dag)$ and $J_y = \frac{-i}{2}(a^\dag - a \,b^\dag$) form the Schwinger representation of the spin operators.} have been obtained by \citet{Simon_Theory_2003} for multi-photon singlet states, and by \citet{Raeisi_Coarse_2011} for states where micro-macro entanglement is generated via parametric amplification of one photon from an entangled pair \cite{De_Martini_Multiparticle_2012,Sekatski_Towards_2009,Frowis_Cloned_2012}.

In the case of bosonic fields, coarse-grained measurements of quadratures $X_\theta$ are mathematically equivalent to decoherence as discussed in Sec.~\ref{sec:decoherence}. Gaussian coarse-graining of quadrature measurements
\be
\delta(X_\theta-x) \to \frac{1}{\sqrt{2\pi }\sigma}e^{-\frac{(X_\theta-x)^2}{2\sigma^2}}
\ee
can be modeled by inserting a noise channel, Eq.\eqref{eq:th noise}, that acts on the state just before the measurement. This is exactly the noise channel that diffuses the state in phase space. Hence coarse-graining of quadratures is sufficient to make all observable measurement statistics reproducible by ``classical'' states (those having a positive P-function).

In the case of spins it is a lack of control on the measurement angle ${\bf n}$ which is in direct correspondence to decoherence. Indeed the lack of control over the angle can be modeled by applying a random rotation on the state prior to the measurement. Again this is exactly the noise process that maps the P-function of the spin to its Q-function, demonstrating that a lack of control on the angle is sufficient to wash out non-classicality from the observed statistics.\footnote{Note this does not mean that one can not distinguish classical and non-classical states with such measurement. If the noise in the measurement is well characterized, it can be in principle deconvoluted from the observed statistics (cf.~Sec.~\ref{sec:cert-large-scale}).} In addition, the strength of the angular noise (i.e., the width of the ``spherical Gaussian'' that is convoluted with the  P-function) decreases with the size of the spin \cite{Schmied_Habilitationsschrift_2017}\footnote{This can be intuitively understood from the observation that the overlap between two spin coherent states at different angles decreases with the size of the spin $|\bra{\bf n}^{\otimes N} \, \ket{{\bf n}'}^{\otimes N}| = |{\bf n}\cdot {\bf n}'|^N$. Therefore non-classical features of the state (regions with a negative P-function) get more and more narrow with increasing $N$.}. \citet{Kofler_Conditions_2008} showed that a coarse-graining of the spin measurement approximately correspond to a lack of control on the angle, hence if the amount of coarse-graining $\sigma$ is larger then the square root of the total spin size $\sqrt{N}$ the observed measurement statistics can be explained by classical states. This, however, does not imply that no quantum features can ever be observed with measurement devices with limited stability or/and precision, as the system can be manipulated prior to the measurement or in-between subsequent measurements. We will come back to this in Sec.~\ref{SecCounterGrain}.

\subsubsection{Reference frames and the size of measurement apparatus}
\label{Sec:ReferenceFrame}

Precision and stability of measurement devices is a technical problem that can be in principle overcome by technological efforts. On the other hand there are  limitations that can not be overcome, for example the size of the measurement device, that are intrinsically limited (by the size of the Universe in the most optimistic case). It turns out that such limitation become relevant when one considers macroscopic quantum effects.

Using Heisenberg uncertainty relations and relativistic causality \citet{Kofler_Fundamental_2010} demonstrated that a measurement device of mass $M$ and size $R$ can only lead to a spin measurement that has an angular resolution of $\delta \theta \sim \sqrt{\frac{\hbar}{c R M}}$. Hence, any measurement suffers from some intrinsic amount of coarse graining due to its finite size.

Along the same lines it was recently shown in \citet{Skotiniotis_Macroscopic_2017} that the limited size of measurement devices forbids the observation of superposition states \da that are macroscopically  distinct in the center of mass position, total spin or energy. The authors adopt a formalism in which the total system consisting of the superposition state and the measurement device (reference frame) is closed and has to abide to the fundamental symmetries of nature given by the Galilean group. To explicitly account for the lack of any additional reference frame, a twirling map is applied on the total system. Under the assumption that the state of the reference frame is classical, it is then shown that in order to distinguish the superposition from the mixture the size of the measurement device has to be quadratically bigger than the size of the superposition.  On the other hand it is also shown that superpositions in relative degrees of freedom do not suffer from such limitations.

While these fundamental limitations are not a problem for microscopic or even mesoscopic experiments, it becomes highly relevant when considering true macroscopic superpositions. For example, \citet{Skotiniotis_Macroscopic_2017} show within a simple model that in order to observe a superposition state of the size of a cat one requires a reference frame of the size of the Earth.

\subsection{Preparation of macroscopic quantum states}
\label{sec:prep-macr-quant}
\label{sec:macr-quant-stat-1}

Let us now focus on the difficulties to prepare macroscopic quantum states. Several works have considered the question whether macroscopic superposition states or macroscopic quantum states can be ground states of ``physical'' Hamiltonians. If this would be the case, one could generate these states simply by means of cooling. However, this is not the case for all states. For example, it was shown that GHZ state, Eq.~\eqref{eq:39}, cannot be the unique ground states of a quasi-local Hamiltonian, but requires at least one global interaction term in the Hamiltonian that affects all particles \cite{Van_den_Nest_Graph_2008}. A similar result was found in \citet{Dakic_Macroscopic_2016} for general macroscopic superposition states. In particular, it was shown that local Hamiltonians that have macroscopic superposition states as unique ground states have a vanishing energy gap, which in turn requires the presence of a long-ranged interaction term in the Hamiltonian.

On the other hand, there exist unique ground states of two-body Hamiltonians that have a variance of $O(N^2)$, i.e., these states are macroscopically quantum according to several measures. One such example is an $N$-qubit Dicke state with $N/2$ excitations (see example \ref{ex:Dicke}). This state is the unique ground state of two-body Hamiltonians\footnote{As a simple exercise one can construct this Hamiltonian using the total spin operators ${\bf S}^2$, $S_z^2$ and $S_z$ \cite{Frowis_Certifiability_2013}.} and might be efficiently prepared by means of cooling, or even occur naturally. Notice, however, that for Dicke states the corresponding Hamiltonian is not local in the sense of arranging qubits on a lattice, but contains long-range two-body interaction terms.

\subsection{Further limitations and counter-strategies}
\label{sec:furth-limit-count}

Even though there are severe limitations and restrictions to prepare, maintain and measure macroscopic quantum states, there exist some counter-strategies and bypasses that allow one to circumvent these limitations to a certain extend.

\subsubsection{Certifiability of large-scale quantum states}
\label{sec:cert-large-scale}

One of the main goals of Sec.~\ref{sec:decoherence} was to understand how much of macroscopic quantumness can survive in a state after the action of some noise process. We saw that quite generically macroscopic quantum states are fragile, and thus very hard to observe in practice. Nevertheless, one can be less demanding and ask  whether the macroscopic character of the state \emph{prior} to the noise can be certified by measuring the final state and assuming that the noise process is perfectly known. If the noise process $\mathcal{E}$ is not too strong (more precisely, if its action on the vector space of operators $\mathcal{B}(\mathcal{H})$ has a trivial kernel), then the state of the system prior to the noise can always be reconstructed to any desired precision by collecting enough measurement statistics on the final state. But the crucial question then is how many times one has to repeat the measurements in order to collect enough statistics.

\citet{Frowis_Certifiability_2013} showed that any superposition state state \da that is macroscopic distinct by the criterion of \citet{Korsbakken_Size_2010} (see Sec. \ref{sec:Korsbakken}) is incertifiable whenever a (tiny) amount of noise local depolarization noise acts on the state. By incertifiability the authors mean that the number of repetitions that allows to distinguish the initial macroscopic superposition state \da from the orthogonal state $\ket{\mathcal{A}}-\ket{\mathcal{D}}$ (and hence also from the mixture $\prjct{\mathcal{A}}+ \prjct{\mathcal{D}}$) increases exponentially with its size. The reason for this is that the coherence between macroscopically distinct states $\mathcal{E}(\coh{\mathcal{A}}{{D}})$ is exponentially damped by local depolarizing noise, such that one needs exponentially many copies in order to distinguish between the macroscopic superposition and the corresponding mixture with a constant probability. This observation is directly related to the fragility of the macroscopic superposition states discussed in Sec.~\ref{sec:fragility superpositions}. In particular, the connection between macroscopic distinctness of \alive and \dead with respect to coarse-grained measurements of an operator $A$ in Sec.~\ref{sec:Sekatski} and the incertifiability of their superposition under dephasing noise generated by $A$ has been discussed by \citet{Sekatski_Difficult_2014}.

Conceptually, the incertifiability of macroscopic quantum superposition is quite a strong statement. It shows that in addition to be hard to maintain, even detecting traces of the superposition in the final state is extremely hard in practice, as it requires to increase the total duration of the experiment exponentially in the size of the superposition state.

On the other hand, \cite{Frowis_Certifiability_2013} also show that a quantum state is certifiable in presence of depolarizing noise if it is a unique ground state of a gapped quasi-local Hamiltonian. Certifiable here means that one can distinguish the initial state  from all orthogonal states with only polynomially many repetitions. While any state after the action of local depolarizing noise has linear quantum Fisher information (see Sec. \ref{Sec:limitQFI}), one can show that for the Dicke $N/2$ state, for instance, a quadratic Fisher information of the initial state can be certified with any desired statistical confidence (P-value) with only $O(N^4)$ repetitions of the measurement. This makes the Dicke state a promising candidate for the experimental detection of a macroscopic quantum state, however performing $O(N^4)$ repetitions can still be quite challenging for large $N$.

\subsubsection{Counter-strategies against noise}
\label{SecCounterNoise}

For particular noise channels, passive or active strategies might be available to maintain macroscopic quantum states or macroscopic quantum superpositions. Such strategies were particularly discussed in the context of quantum metrology, where states with a large quantum Fisher information are required to maintain a quantum scaling advantage. In \citet{Landini_Phase-noise_2014} the usage of particular states in a decoherence free subspace are discussed to protect the system against collective dephasing. Active quantum error correction is used in \citet{Dur_Improved_2014,Kessler_Quantum_2014,Arrad_Increasing_2014} to protect the system against a specific noise process, namely rank-one noise that is orthogonal to the sensing field. Fast quantum control is used in \citet{Sekatski_Quantum_2016} to maintain the usability in quantum metrology for any rank-one noise. The methods are similarly applicable to actively maintain a large quantum Fisher information under certain noise processes.

However, we emphasize that only some very specific noise processes can be dealt with in this way. The ultimate bounds for generic noise processes reported in Sec. \ref{Sec:limitQFI} still apply, making a quadratic quantum Fisher information generically inaccessible.

\subsubsection{Counter-strategies against coarse-graining}
\label{SecCounterGrain}

The results discussed in Sec.~\ref{sec:stab-dist-stat} always considered a restricted set of operators that can be measured with coarse-grained detectors: the spin in some direction $S_{\bf n}$ for spin systems and the quadratures $X_\theta$ for photons. Without any restriction, the impact of coarse-graining would be weak in general. To see this, imagine an observable with the same eigenbasis as before, but with a clever rearrangement of the eigenvalues. In this way, one can easily get rid of the difficulty to distinguish neighboring eigenstates of the original operator. While such a rearrangement of eigenvalues might seem rather abstract at first glance, something similar can be physically done by applying a unitary $U$ on the system just before it is measured  \cite{Kofler_Conditions_2008}. Later \citet{Jeong_Failure_2009} showed that a simple Kerr-nonlinear interaction Hamiltonian $H$ is sufficient to generate unitaries $U=e^{-i t H}$ that allow to observe quantum features (violation of macrorealism, see Sec. \ref{sec:legg-garg-ineq} for more details) with extremely coarse-grained detectors.

However, this strategy might add additional constraints. \citet{Wang_Precision_2013} analyzed the superposition of two coherent states $|\alpha\rangle +|-\alpha\rangle$ in the setting of coarse-grained quadrature measurements and the possibility to perform a Kerr-nonlinearity. The authors showed that while by using such a Kerr interaction it is possible to distinguish the superposition from a mixture even with a low resolution, the requirement on the control precision of the interaction (the exact value of the interaction time $t$) also increases with the size of the superposition $|\alpha|^2$. It is open whether these findings can be extended to general macroscopic quantum states.

A similar setting was analyzed for the task of observing macroscopic quantum states. In \citet{Frowis_Detecting_2016}, a method to detect large quantum Fisher information using detectors with limited resolution was presented. The main idea is to re-use the same operation as to prepare the state (e.g., a squeezing operation) in order to realize a modified measurement process. While the initial measurement device is coarse-grained, the squeezing operation allows one to increase the relevant resolution, and to detect a quantum Fisher information that, for existing set-ups, could be two orders of magnitude larger. A similar approach was used in \citet{Davis_Approaching_2016}. In both cases, the requirements on the stability of the squeezing operation was analyzed.

\subsubsection{Encoded macroscopic quantum states}
\label{sec:encod-macr-stat}

The results discussed in previous sections show that generic noise processes, in particular independent coupling of system particles to the environment, destroy macroscopic superpositions and macroscopic quantum states, even if each of the local noise processes is arbitrarily small. On the other hand, quantum error correction or fault-tolerant quantum computation \cite{Nielsen_Quantum_2000} can be used to actively protect quantum information. Hence, this gives us the tools to prepare, maintain and certify encoded macroscopic quantum states. For this, one has to accept a slight shift of the definition of macroscopic quantum states, which we discuss in this section.

In quantum error correction, quantum information is actively protected against the influence of noise and decoherence by encoding a logical two-level system into several physical two level systems, i.e., into a larger space. Each qubit of a quantum state $|\psi\rangle$ is thereby replaced by a logical qubit that consists of several physical qubits. For example, the state $|\phi\rangle=\alpha|0\rangle + \beta|1\rangle$ is replaced by $|\phi_L\rangle=\alpha|0_L\rangle + \beta|1_L\rangle$, where now the quantum information, i.e., the parameters $\alpha$ and $\beta$ are protected.
Codes can be designed to protect quantum information against different kinds of noise, most notable noise that is local in some sense, i.e., acting jointly in a correlated way only on a bounded, localized number of qubits.

Note that operations for encoding and error correction might themselves be noisy. The theory of fault-tolerant quantum computation \cite{Nielsen_Quantum_2000} tells us, however, that if noise in elementary single- and two-qubit operations is below a certain threshold value, then one can perform error correction in a fault-tolerant way. In fact, any quantum computation can be realized fault-tolerantly. In particular, an encoded macroscopic quantum superposition state $| \operatorname{L-GHZ}_N\rangle=(|0_L\rangle^{\otimes N}+|1_L\rangle^{\otimes N})/\sqrt{2}$ \cite{Frowis_Stable_2011} can be created, maintained and measured using noisy elementary operations, and under the influence of (quasi-)local decoherence. The same can be done for any other macroscopic quantum state via the mapping $| i \rangle \rightarrow \left| i_L \right\rangle , i = 0,1$.
Note that these encoded states are still susceptible against noise at the logical level, which is however exponentially suppressed by error correction if it results from noise on individual qubits.

We emphasis that mapping a macroscopic quantum state to its encoded version might drastically change its effective size according to some measures presented in Sec.~\ref{sec:meas-macr-superp}. For instance, measures that maximize over all local operators and do not consider an extension to quasi-local operators generally do not assign a large effective size to encoded states. The logical GHZ state $| \operatorname{L-GHZ}_N \rangle $, for example, does not have a large variance with respect to any local operator (which is necessary to ensure robustness), but it exhibits maximal variance for the sum of logical Pauli operators $\sigma_{z,L} = \left| 0_L \right\rangle\!\left\langle 0_L\right| - \left| 1_L \right\rangle\!\left\langle 1_L\right| $. Therefore, it depends on the precise definition of the measure (and hence on us) whether we call this state macroscopically quantum or not. Note that other measures (e.g., the measures of \citet{Korsbakken_Measurement-based_2007,Marquardt_Measuring_2008}) do not decrease under the encoding (cf.~example \ref{ex:cluster} in Sec.~\ref{sec:ex:spin-ensemble}).
Considering $|\operatorname{L-GHZ}_N\rangle$ as a toy example of a Schr\"odinger cat in a quantum superposition state between two macroscopically distinct states, one could say that the distinguishability between different constituents of the cat is here not at the level of the individual atoms, but at the level of molecules. In this sense, one may still call this a macroscopic quantum superposition.

\subsection{Summary}
\label{sec:summary}

In this section, we have reviewed many different findings that show that macroscopic quantum states are generally difficult to observe. First, such states are difficult to maintain because even weak perturbations leak enough information to the environment in order to lose the coherence between macroscopically distinct states. Second, a lack of resolution of the detectors or simply the finite size of the measurement devices also seems to make the observations of quantum features on the macroscopic scale difficult. Third, we have argued that the preparation of macroscopic quantum states might be difficult in the first place.

For some problems such as measurement precision or collective noise, counter-strategies have been conceived which promise significant improvement. However, they typically work only under certain constraints on the experimental control, and, at least for some examples, these constraints become hard to fulfill for macroscopic quantum states. Finally, we have seen that when considering encoded macroscopic quantum states, which are only macroscopically quantum on the logical level, it is possible to overcome many no-go results.

%%% Local Variables:
%%% mode: latex
%%% TeX-master: "master"
%%% End:

\section{Potentials of macroscopic quantum states}
\label{sec:potent-macr-quant}

In this section we review some selected applications and potentials of macroscopic quantum states. We will concentrate on applications in quantum information processing and quantum metrology, but also touch upon fundamental issues such as probing the limits of quantum theory. We will base our considerations on the characterization of macroscopic quantum states discussed in Sec. \ref{sec:meas-macr-superp}.

While the previous discussion was mainly concerned with the issue of how to classify macroscopic quantum states and how to determine their effective size, here we are concerned with more practical issues. With emergent quantum technologies the focus has shifted from fundamental considerations, e.g., on entanglement or non-locality, to practical applications of quantum theory in different tasks, ranging from quantum communication and quantum computation to high-precision measurements in quantum metrology. Entanglement is often said to play a key role in these applications. While this is certainly true, we do not aim for providing a review on quantum entanglement and its classification and quantification \cite{Horodecki_Quantum_2009}. We are rather concerned with large (macroscopic) quantum systems and their potential applications. Naturally, this is interlinked with certain issues of entanglement. We start however by considering the role of macroscopic quantum states in a more fundamental issue, namely for probing the limits of quantum theory.

\subsection{Probing the limits of quantum theory}
Quantum mechanics provides an accurate description of the microscopic world and is in fact the most accurate description of nature we have come up with so far. However at macroscopic scales, quantum effects typically can not be observed. So the question remains if quantum mechanics is indeed valid on all scales. Here we discuss two approaches that are related to this issue.

\subsubsection{Macrorealism and Leggett-Garg-like inequalities}
\label{sec:legg-garg-ineq}

In quantum mechanics, the superposition principle makes it impossible to assign definite properties to a system. As long as this principle holds only on microscopic scales (where it has been thoroughly confirmed), it does not contradict our classical perception of the macroscopic world. \citet{Leggett_Quantum_1985} formalize consequences from the basic assumption that a macroscopic object has well-defined properties independent of any observer. To do so, they introduce a second premise, namely, that there exist in principle measurements that do not disturb the measured system. With these two assumptions, the authors derive a bound --called Leggett-Garg inequality (LGI)-- on correlations between measurements performed at different times. Any experimental violation of an LGI implies the absence of one of the assumptions in nature. The so-called clumsiness loophole (i.e., the violation of the noninvasiveness of the measurement through lack of experimental control), can be in principle avoided with ideal negative measurements and/or carefully designed control measurements \cite{Wilde_Addressing_2012}. Lately, an LGI could be experimentally violated using Cesium atoms \cite{Robens_Ideal_2015} and superconducting devices \cite{Knee_Strict_2016}; see \citet{Emary_Leggett-Garg_2014} for a comprehensive review.

\begin{figure}[htbp]
\centerline{\includegraphics[width=\columnwidth]{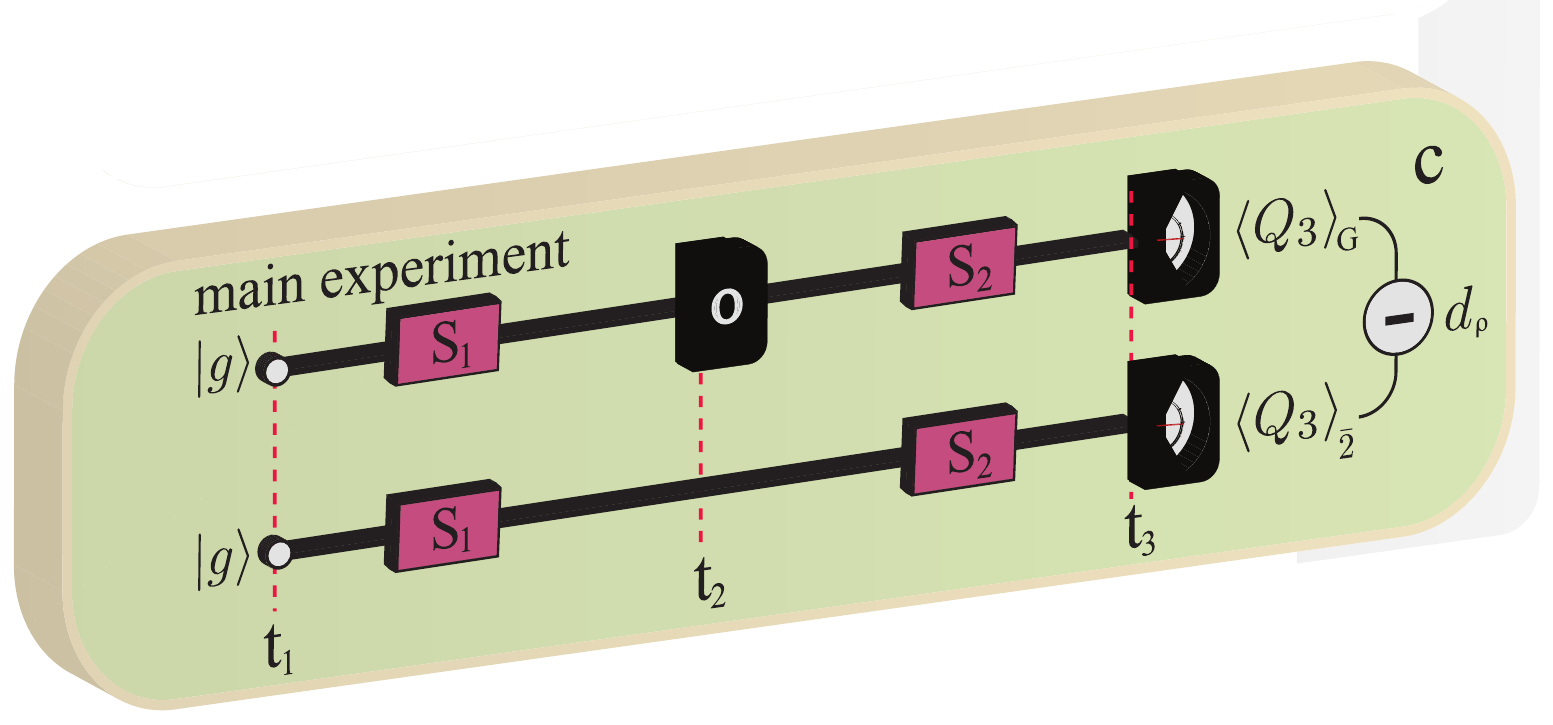}}
\caption[]{\label{fig:LeggettGarg} Basic scheme of violating an LGI or the NSIT condition. At time $t_1$, the system is initialized in state $| g \rangle $ and evolves under $S_1$ until time $t_2$. The free evolution is ideally designed to increase the spread of coherence in the measurement basis. Then, for half of the runs, a measurement is done ($O$, upper line); nothing is done for the other half (lower line). A subsequent free evolution $S_2$ is followed by a final measurement at time $t_3$. The protocol can be further relaxed by letting $O$ be any operation. Adapted from \citet{Knee_Strict_2016}.}
\end{figure}

Recently, alternative formulations of the same intuition were proposed \cite{Devi_Macrorealism_2013,Kofler_Condition_2013,Saha_Wigner's_2015}. In the following, we are particularly interested in the ``no-signaling in time'' (NSIT) condition by \citet{Kofler_Condition_2013}, which simply states that the probability distribution for a measurement performed at time $t = t_3$ should not depend on whether or not a measurement at an earlier time $t_2$ was done (see Fig.~\ref{fig:LeggettGarg})\footnote{The precise differences between the LGI and the NSIT are subject to ongoing investigations (see, e.g., \citet{Kumari_Probing_2017,Halliwell_Comparing_2017}). Here, we are more interested in the global concept rather than in differences between them.}. This condition can be even further relaxed. The operation at time $t_2$ does not have to be a measurement, but can be any sort of (small) disturbance \cite{Knee_Strict_2016,Frowis_Detecting_2016}.

\citet{Kofler_Classical_2007} show that \textit{any} pure quantum state potentially violates an LGI and the NSIT condition for the right choice of measurements. This is not surprising as projective measurements are highly invasive. On the other hand, limited measurement resolution prevents the violation of LGIs, given that the free time evolution between the measurements is linear \cite{Kofler_Classical_2007}. The reason of an LGI-violation is coherence of the quantum state in the measurement basis. The effect of the limited resolution is to disregard coherence between basis states with a spectral distance less than the scale of resolution. Therefore, violating an LGI with low measurement resolution implies coherence between far distant parts.

This directly links macroscopic quantumness to such tests of macrorealism, which becomes particularly evident by inspecting the measures of \citet{Kwon_Disturbance-Based_2016,Frowis_Measures_2012,Cavalcanti_Signatures_2006}. First, \citet{Kwon_Disturbance-Based_2016} directly define macroscopic quantumness as being highly susceptible to small influences. A low-resolution measurement $O$ in the basis of $A$ is precisely covered by the map $\Phi_{\sigma}$, Eq.~(\ref{eq:25}), when $\sigma$ is large (i.e, $\sigma \gtrsim O(N)$). Therefore, macroscopic quantum states in the sense of \citet{Kwon_Disturbance-Based_2016} are exactly those states that gives the largest LGI-violations. Second, measuring a large quantum Fisher information $\mathcal{F}(\rho,A)$ based on Eq.~(\ref{eq:49}) is directly connected to the NSIT protocol as in Fig.~\ref{fig:LeggettGarg}, where $O$ is an application of $\exp(-i \theta A)$ with $\theta$ small enough\footnote{A small disturbance is often necessary in case one wishes to exclude the clumsiness loophole with control measurements demonstrating (almost) no impact on ``semiclassical'' states.} to disturb only states with large $\mathcal{F}$ \cite{Frowis_Detecting_2016}. Last, macroscopic quantum coherence in the spirit of \citet{Cavalcanti_Signatures_2006} seems to be the kind of coherence necessary to violate an LGI with coarse-grained measurements as discussed by \citet{Lambert_Leggett-Garg_2016}.

Note that tests of macrorealism can also be done in bipartite scenarios in which measurements are space-like separated rather than time-like separated for a single system \cite{Reid_Signifying_2016}. As a result, a Bell-like inequality can be derived. The locality assumption replaces the second premise of noninvasiveness for the derivation of an LGI, which can be seen as a conceptual advantage. However, it is still necessary to require low-resolution measurements (``macroscopic degree of fuzziness'') in order to guarantee that a potential violation comes from entanglement (i.e., correlated coherence) between macroscopically distinct states \cite{Reid_Signifying_2016}.

\subsubsection{Validity of quantum mechanics at large scales - Collapse models}
\label{sec:valid-quant-mech}

One of the main motivations to study macroscopic quantum states, and in particular to try to generate them in the lab, is to show the validity of quantum mechanics at large scales - or the need for some alternative theory. While many, if not most, researchers believe that quantum mechanics is indeed valid on all scales, modifications of quantum theory have been suggested that lead to the absence of quantum effects such as superpositions of states on a macroscopic scale. Most prominently, collapse models \cite{Bassi_Models_2013} including gravitationally induced collapse \cite{Diosi_Models_1989, Penrose_Gravity_1996} or continuous spontaneous localization \cite{Ghirardi_Unified_1986, Ghirardi_Markov_1990, Gisin_Stochastic_1989} have been suggested. In these models, the time evolution is no longer given by the Schr\"odinger equation, but replaced by a master equation including a diffusion term that prevents massive systems to be in spatial superposition states. As discussed in Sec.~\ref{sec:NimmrichterHornberger}, these models are at the core of the measure proposed by \citet{Nimmrichter_Macroscopicity_2013}.

Collapse models have been extensively discussed in \citet{Bassi_Models_2013}, and a thorough review on the limits of quantum superpositions is provided in \citet{Arndt_Testing_2014}. We will hence only briefly comment on these aspects in the following. In \citet{Bose_Scheme_1999}, \citet{Romero-Isart_Quantum_2011}, \citet{Nimmrichter_Testing_2011} and \citet{Diosi_Testing_2015} for example,  the requirements to test different collapse models using superpositions of massive objects are investigated using quantum optomechanical systems, levitating nanospheres, matter-wave interferometry and classical mechanical oscillator respectively. 
The observation of quantum interference effects or spontaneous thermalization on a large scale allows one to put bounds on parameters in different collapse models (e.g., the strength/rate of the collapse), thereby allowing one to confirm or rule out these models in these parameter regimes. 
Note that in \citet{Sekatski_Macroscopic_2014}, a proposal is discussed to test collapse models by mapping the macro component of a photonic micro-macro state to an optomechanical system. This opens a way to connect measures for macroscopicity for photonic systems to measures for massive objects.

\subsection{Quantum metrology}
\label{sec:quantum-metrology}

We now turn to quantum metrology \cite{Giovannetti_Advances_2011,Toth_Quantum_2014,Pezze_Non-classical_2016} as a potential application of large-scale quantum systems. In quantum metrology, the goal is to determine an unknown parameter, e.g., the strength of a magnetic field, a frequency, a phase or a force, as accurately as possible with the given resources. To this aim, a quantum state of a certain number $N$ of systems is prepared. The state then undergoes an evolution that is governed by a Hamiltonian which depends on the unknown parameter $\vartheta$ (or possibly several parameters), and is then subsequently measured. The experiment is repeated $\nu$ times, and from the gathered measurement data an estimate for the unknown parameter $\vartheta$ is determined. One can distinguish between phase estimation and frequency estimation, where in the latter one has control over the evolution time where in the former this is not the case. There is a distinction between local metrology scenarios, where the value of the parameter is (almost) known and should be determined with increased accuracy, and Bayesian scenarios where the initial knowledge is expressed as a probability distribution which is updated. The local scenario deals with many repetitions, while the Bayesian scenario is a single-shot one. Depending on the concrete problem and scenario, the number of systems $N$, the evolution time $t$ and number of repetitions of the experiment $\nu$ are counted as resources. Bounds on the achievable accuracy can be found, and in many relevant cases optimal strategies (i.e., initial states, evolution time and measurements) can be determined \cite{Giovannetti_Advances_2011,Toth_Quantum_2014,Pezze_Non-classical_2016}.

A central quantity in in this context is the quantum Fisher information, which was already introduced in Sec. \ref{sec:QFI}. The quantum Fisher information bounds the achievable accuracy in the local metrology scenario via the Cramér-Rao bound \cite{Cramer_Mathematical_1945,RadhakrishnaRao_Information_1945}, where the precision scales inversely proportional to the quantum Fisher information. Certain quantum states have a quantum Fisher information that scales as $O(N^2)$, while classical states are limited to a quantum Fisher information of $O(N)$. This establishes a quadratic advantage of quantum strategies over classical ones. Sometimes, the ratio of the quantum Fisher information of a state $|\psi\rangle$ over the optimal classical state is called the metrological gain, which can be up to $N$ \cite{Pezze_Non-classical_2016}.

The quantum Fisher information is the basis of the measure by \citet{Frowis_Measures_2012} (see Sec. \ref{sec:QFI}), which is hence directly linked to the usefulness in metrological tasks. For pure state, the quantum Fisher information simplifies to four times the variance. Hence, much more measures are implicitly connected to pure-state quantum metrology (see Sec.~\ref{sec:conn-betw-meas-1}). The reason is the intimate connection between the coherent spread of state in the spectrum of an observable $A$ and the state's sensitivity to small changes induced by $\exp(-i \vartheta A)$ (cf.~Eq.~(\ref{eq:47})).

\subsection{Quantum computing}
\label{sec:quantum-computing}

Quantum computation is perhaps the holy grail of quantum information processing, and provides a long-term perspective with its applications in solving certain problems with an (possibly exponential) quantum speedup. Here we concentrate on one particular model for quantum computation, the so-called measurement-based quantum computation (MBQC) \cite{Briegel_Measurement-based_2009}, with the one-way model as the most prominent representative \cite{Raussendorf_One-Way_2001}.

\subsubsection{MBQC, entanglement and macroscopicity}
\label{sec:mbqc-entangl-macr}

In MBQC an entangled state serves as a resource, and is manipulated by single qubit measurements only. For a universal resource such as the 2D cluster state \cite{Briegel_Persistent_2001}, by definition any target state can be generated. An efficient generation (with polynomial overhead in auxiliary particles) is possible for all states that can be prepared using a quantum circuit with polynomially many single- and two-qubit gates. It follows that a universal resource for MBQC must contain all types of entanglement to an arbitrary amount \cite{Van_den_Nest_Fundamentals_2007}. In order to create a target state, local measurements transform and concentrate the entanglement of the (large) $N$-qubit cluster state into a (smaller) $M$-qubit target system.

Therefore, one might be tempted to argue that such universal resource states should also be macroscopically quantum. However, according to most of the definitions for macroscopic quantumness put forward in Sec.~\ref{sec:meas-macr-superp}, 2D cluster states (and other universal resources) are {\em not} macroscopically quantum (see example \ref{ex:cluster}). The reason is the lack of two-body correlations necessary for a large variance of some local operator. There are, however, higher-order correlations in the system (more precisely, five-body correlations) which can be converted into two-body correlations via LOCC.

As mentioned in example \ref{ex:cluster}, a 2D cluster state of size $N$ can be transformed into a GHZ state of $O(N)$ particles using only local measurements \cite{Briegel_Persistent_2001}.
Hence, a variance-based measure can increase under LOCC. Entanglement and macroscopicity, though seemingly related concepts, are therefore intrinsically different, as noted, e.g., in \citet{Frowis_Measures_2012,Yadin_Quantum_2015}. Hence a resource theory for macroscopic quantumness should take this into account, and cannot allow LOCC as free operations if the variance should be a proper measure (see Sec.~\ref{sec:conn-reso-theory} for further discussion). On the other side, quantum entanglement and quantum macroscopicity are not completely independent, see Sec.~\ref{sec:relat-mult-entangl}.

Cluster states are also highly robust against noise \cite{Hein_Entanglement_2006}, which is again in contrast to macroscopic quantum states. In particular, if one assumes that each of the qubits of the 2D cluster state interacts with an independent environment, one can show that the entanglement and other key features of the 2D cluster states are maintained up to a certain noise level, independent of the system size. Local depolarizing noise (see footnote \ref{fn:2}) with error probabilities of up to 10\% or more per particle can be tolerated such that the state remains distillable entangled, i.e., maximally entangled states between any pair of qubits can be generated from many copies. 
To generate entangled pairs between neighboring qubits, this can be achieved by measuring the surrounding qubits in the $Z$-basis, thereby decoupling them from the rest of the system and making it obvious that there is no dependence on the total size of the cluster.
We remark that 3D cluster states have been show to be universal for fault-tolerant (encoded) quantum computation using a 2D surface code, with an error threshold for single qubit depolarizing noise of about 0.75\% \cite{Raussendorf_Topological_2007,Briegel_Measurement-based_2009}.

This is in contrast to macroscopic superposition states such as the GHZ state, which are much more susceptible to noise (see Sec.~\ref{sec:decoherence}). It is interesting to note that a GHZ state obtained from a noisy 2D cluster state is as decohered as a GHZ state to which the same amount of noise has been directly applied. This can be easily seen as follows: A GHZ state of $N$ qubits can be generated by measurements on a subset of qubits of a noisy 2D cluster state of size $O(N)$. Even if we assume that all measured qubits have not been affected by noise, and measurements are perfect, single-qubit noise still acts on the remaining qubits that now form a GHZ state. Since measurements and noise operators act on different systems and hence commute, the effect of noise before the LOCC protocol is the same as after.

This discussion shows that usefulness for certain applications (such as quantum computation) or entanglement are different concepts than macroscopic quantumness. Even states that are rendered microscopic according to many measures can be valuable resources, and allow one to perform highly interesting task - such as fault-tolerant quantum computation. Note that the choice of free operations is crucial in this respect, as measurements play a central role in MBQC but are usually not considered to be free in the context of measures for quantum macroscopicity.

\subsubsection{States occurring in quantum computation and metrology} \label{sec:stat-occurr-quant}

In spin systems, large variance of local operators imply strong two-body correlations (i.e., entanglement for pure states). \citet{Shimizu_Necessity_2013,Ukena_Macroscopic_2005} conjectured that this kind of entanglement should be present in circuits for quantum algorithms that outperform classical computers. The first nontrivial step is to precisely formulate the statement. Every query of a quantum algorithm leads to different quantum states during the computation and not every instance is (exponentially) more difficult for a classical device. The authors work with a generalized version of the index $p$ (see Sec.~\ref{sec:index-p}) to deal with entire sets of different instances of an algorithm.

\citet{Shimizu_Necessity_2013,Ukena_Macroscopic_2005} found that $p = 2$ states generically appear in Shor's factorization algorithm \cite{Shor_Polynomial-Time_1999} and in Grover's search algorithm \cite{Grover_Quantum_1997}. Since $p = 2$ states are particularly sensitive to noise generated by local operators (see Sec.~\ref{sec:limits-observ-quant}), the findings emphasize the necessity for a well-designed error correction scheme. Furthermore, these results complement other findings about entanglement and nonclassicality in quantum enhanced algorithms. For example, \citet{Orus_Universality_2004, Kendon_Entanglement_2006} found a connection between computational speedup and large entropy of entanglement. Note that there exist states with $p = 2$ and small entropy of entanglement (e.g., the GHZ state) and states with $p = 1$ and large entropy of entanglement (e.g., eigenstates of chaotic systems \cite{Sugita_Correlations_2005}).

As discussed before, a large variance is an important property for pure states to be useful in parameter estimation. Hence, there is connection between quantum enhanced computation and sensing. However, this does not imply that particular instances of large-variance states are useful for computation and sensing at the same time. This connection was further investigated by \citet{Demkowicz-Dobrzanski_Quantum_2015}. The authors rephrased Grover's algorithm in a time-continuous fashion and expressed it as kind of estimation problem. Then, they used results from quantum metrology to bound the performance of the algorithm under some generic decoherence and loss channels. Like in quantum metrology, \citeauthor{Demkowicz-Dobrzanski_Quantum_2015} found a loss of the quadratic improvement in such situations. However, unlike in quantum metrology, where the Hamiltonian for parameter estimation is typically given by the problem and cannot be changed, a quantum algorithm is theoretically under full experimental control and, therefore, techniques like quantum error correction can be applied \cite{Dur_Improved_2014,Kessler_Quantum_2014,Arrad_Increasing_2014}.

\subsubsection{Quantum phase transitions}
\label{sec:quant-phase-trans}

Recent studies further highlight the importance of macroscopic quantum states in other quantum algorithms and paradigms \cite{Yuge_Superposition_2017}. But macroscopic quantum states are also expected to play a role in  ground states of strongly coupled systems, systems with topological order or topologically protected phases of matter.
There already exist some works that relate macroscopic quantum states and quantum phase transitions. One example is \citet{Hauke_Measuring_2016}, where it is shown that the quantum Fisher information can be obtained by means of the dynamic susceptibility, and can be used to detect entanglement during phase transitions. In \citet{Shitara_Determining_2016} it was shown how to obtain the quantum Fisher information from linear response functions.

\subsection{Summary}
\label{sec:summ-potent-appl}

Macroscopic quantum states are no longer only an interesting virtual possibility that illustrate puzzling features of quantum mechanics, as in the times of Schr\"odinger. Nowadays such states are thought of as valuable resources. The usefulness of states for certain tasks, their entanglement features and their classification of being macroscopically quantum with respect to certain measures are however different concepts. Though there are certain relations, the goal and merit of these concept vary.

We have highlighted in this section possible applications of macroscopic quantum states for testing the limits and validity of quantum mechanics at large scale - a question that is primary of fundamental interest. We have also discussed more practical applications, in particular in the context of quantum metrology, where the usefulness for metrology coincides with some concepts of quantum macroscopicity, in particular measures based on the variance or quantum Fisher information. The link is less obvious for applications for (measurement-based) quantum computation and to different aspects of entanglement. On the one side, the underlying free, allowed operations differ in these approaches. On the other side, macroscopic quantum states seem to appear in crucial steps of quantum algorithms.

%%% Local Variables:
%%% mode: latex
%%% TeX-master: "master"
%%% End:

\section{Implementations}
\label{sec:implementations}

We now review experiments reporting on the creation and detection of macroscopic quantum states. In particular, we focus in Sec.~\ref{subsection_photoexp} on photonic experiments, separating optical and microwave setups. We then review in Sec.~\ref{subsection_spinexp} experiments with spin systems, distinguishing setups where the spins are addressed individually and collectively. Sec.~\ref{subsection_massiveexp} is devoted to massive systems including atom interferometry and recent experiments with optomechanical systems. Superconducting experiments are mentioned in  Sec.~\ref{sec:superc-quant-interf}.
Finally, in Sec.~\ref{Comparison_photo_spin_mass_states}, we provide comparisons of the size of states based on experimental data. We summarize in Sec.~\ref{sec:summary-1}.

\subsection{Photonic experiments}
\label{subsection_photoexp}

Photonic setups can naturally be divided into two groups: The optical setups which mostly rely on sources based on parametric conversions and the microwave setups where strong light-matter interactions are used to shape SCS. We provide a quick presentation of experiments nicely illustrating the research activities of several groups and invite the reader to look at for more exhaustive review papers on optical\footnote{\textcite{De_Martini_Multiparticle_2012, Pan_Multiphoton_2012, Jeong_Characterizations_2015, Chekhova_Bright_2015}} and microwave\footnote{\textcite{Raimond_Manipulating_2001, Makhlin_Quantum_2001, Schoelkopf_Wiring_2008, Devoret_Superconducting_2013, Haroche_Nobel_2013}} setups, respectively.

\subsubsection{Optical photons}
\label{sec:optical-photons}

The central tool of optical experiments is spontaneous parametric down conversion, that is, a bulk crystal or a waveguide for example, with a second order non-linearity that is used to convert photons of a pump laser into photon pairs. These pairs have combined energies and momenta equal to the ones of the laser photons and are correlated in polarization. We can distinguish Type I and Type II down converters depending on the pair polarization. \\

\begin{figure}
\centerline{\includegraphics[width=6cm]{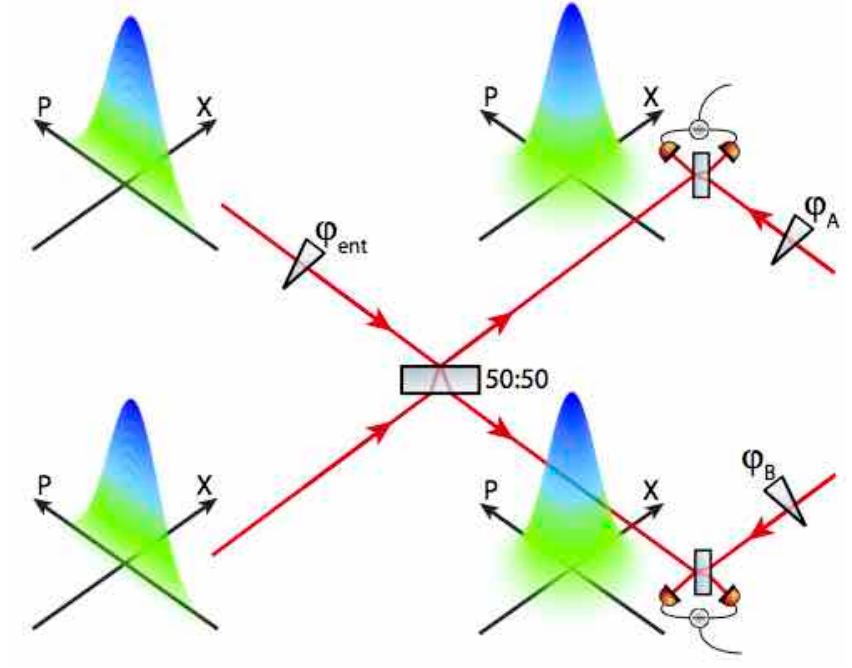}}
\caption[]{\label{fig:Eberle_Stable_2013} Schematic representation of the experiment reported in \citet{Eberle_Stable_2013} in which two single-mode squeezed vacuum states are combined on a beam-splitter before characterizing their correlations in phase space. From \citet{Eberle_Stable_2013}.}
\end{figure}

Type I down converters produce pairs with identical polarization and result in single-mode squeezed vacuum (see Eq.~\eqref{eq:1}) containing only even photon numbers. The experiment reported in \citet{Eberle_Stable_2013} has combined two of these squeezed vacuum modes on a beamsplitter before performing phase $ P$ and amplitude $ X$ quadrature measurements at each output ($A$ and $B$) of the beamsplitter, see Fig.~\ref{fig:Eberle_Stable_2013}. The results exhibit a $\approx 10$dB reduction of noise variances $[\Delta( X_A +  X_B)]^2$ and $[\Delta( P_A -  P_B)]^2$ with respect to the sum and difference of the corresponding quadratures for a vacuum state. These correlations in amplitude and anti-correlations in phase has been used to certify entanglement and can be used to quantify the size of the produced state, see Sec.~\ref{Comparison_photo_spin_mass_states}. Note that record squeezing, down to $\approx 15$dB, has recently been reported using type I down converters in \citet{Vahlbruch_Detection_2016}. \\ 

\citet{Iskhakov_Macroscopic_2011} used two co-linear type I down converters but in a Mach-Zehnder interferometer such that they can be excited coherently with orthogonally polarized pumps (see also \cite{Eisenberg_Quantum_2004} for a similar work). This led to polarization entanglement, that is, an Hamiltonian of the form $\sum_i (a_H^{(i)\dag} b_H^{(i)\dag} + a_V^{(i)\dag} b_V^{(i)\dag} +h.c.)$ where the bosonic operators $a_H^{(i)},$ $b_H^{(i)}$ correspond to two spatial modes with horizontal polarization and similarly for $a_V^{(i)}$ and $b_V^{(i)}.$ The sum means that the emission is multimode, that is, photons are created in different angular modes. By collecting tens of thousands of modes, hundreds of thousands of entangled photons have been successfully detected \cite{Iskhakov_Macroscopic_2011, Iskhakov_Polarization-Entangled_2012}. Note that, as a result of the multi-mode emission, the state of these photons corresponds essentially to independent two-qubit maximally entangled states $(a_H^{(i)\dag} b_H^{(i)\dag} + a_V^{(i)\dag} b_V^{(i)\dag})|0\rangle$ where $|0\rangle$ is the vacuum for all modes. For this reason, all measures discussed in Sec.~\ref{sec:preliminary-measures} consider this state not as macroscopically quantum even under otherwise ideal circumstances.

\begin{figure}
\centerline{\includegraphics[width=9cm]{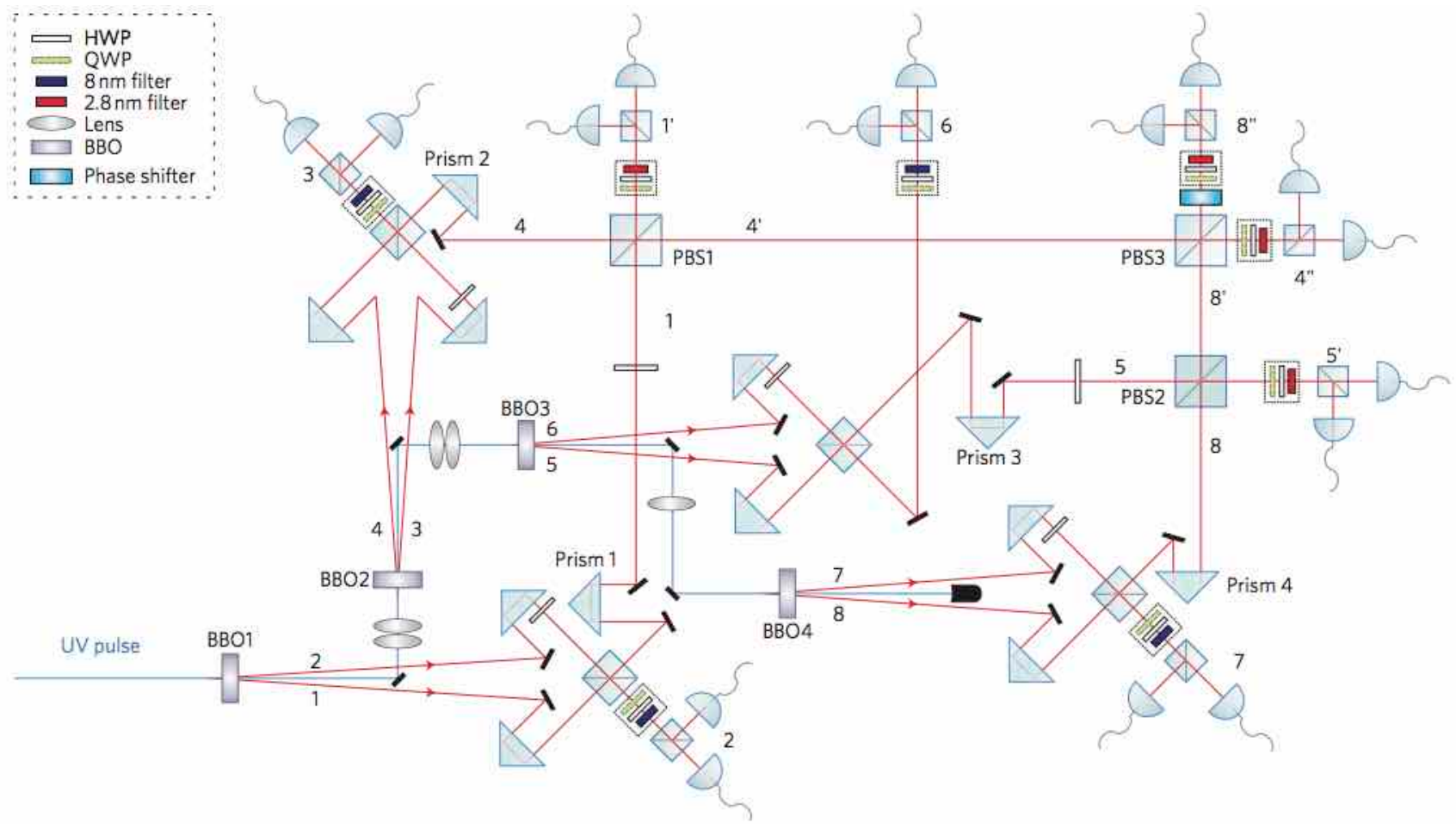}}
\caption[]{\label{fig:Yao_Observation_2012} Schematic representation of the experiment presented in \citet{Yao_Observation_2012} using four type II down conversion processes to create 8 photons GHZ states. Laser pulses with a short duration and a high repetition rate successively pass through four non-linear (BBO) crystals to produce four photon pairs, which are further combined using a half- and quarter-wave plate (HWP, QWP) and a polarization beamsplitter (PBS). The photons in mode 1 and 4 are then combined on PBS1, photons 5 and 8 on PBS2, and finally photons 4' and 8' on PBS3. The photons are detected by 16 single-photon detectors and a complete set of 256 eight-fold coincidence events are post-selected and registered to perform a tomography of the post-selected state. From \citet{Yao_Observation_2012}.}
\end{figure}

Intensive effort has been dedicated to the generation of photon pairs in single modes. \citet{Yao_Observation_2012} used type II converters pumped by short pulses together with narrow filters of the output photons to erase their frequency correlations. They combined the outputs of four of these down converters using linear optical elements and post-select events using photon counting devices, see Fig. \ref{fig:Yao_Observation_2012}. Using intense pumps, 8 photons GHZ states have been postselected with 70\% fidelity \cite{Yao_Observation_2012} and recently, similar techniques led to creation of up to 10 photons GHZ states with 57\% fidelity \cite{Wang_Experimental_2016}. 

\begin{figure}
\centerline{\includegraphics[width=6cm]{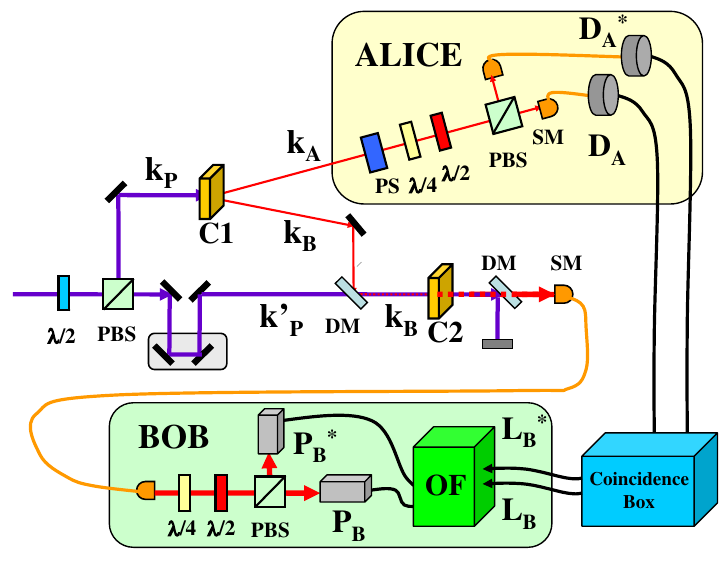}}
\caption[]{\label{fig:DeMartini_Entanglement_2008} Schematic representation of the setup used in \citet{DeMartini_Entanglement_2008} aiming to produce micro-macro entanglement via the amplification of micro-micro entanglement. A first type II process converts photons from pump laser into photon pairs with polarization entanglement. A photon from each pair is used to seed a second type II converter pumped by the same laser. The resulting entanglement is characterized by recording the coincidences between photon counting devices that are preceded by a set of waveplates ($\lambda/2,$ $\lambda/4$) and a polarization beamsplitter (PBS). DM: dichroic mirror, SM: single-mode fiber, PS: phase shifter. From \citet{DeMartini_Entanglement_2008}.}
\end{figure}

Aside from these experiments aiming to create macroscopic photonic states directly from spontaneous parametric down conversion, various techniques have been proposed to amplify the photon number in few-photon quantum states while preserving their quantum nature. The first experiment along this line has been reported in \citet{DeMartini_Entanglement_2008}. Polarization entangled photon pairs were first created from a type II converter and a photon from each pair was subsequently injected into a second type II converter, the latter playing the role of a phase covariant cloner \cite{Sekatski_Towards_2009}. This ideally creates a photonic state of the form
\begin{equation}
\label{eq_phasecovcloner}
\frac{1}{\sqrt{2}}\left(|1\phi^\bot\rangle_A |\Phi^\phi\rangle_B + |1\phi\rangle_A |\Phi^{\phi^\bot}\rangle_B\right)
\end{equation}
after amplification. $|\Phi^\phi\rangle_B$ and $|\Phi^{\phi^\bot}\rangle_B$ are two orthogonal states that are defined by
\begin{eqnarray}
\nonumber
&& |\Phi^\phi\rangle_B = \sum_{i,j=0}^\infty \gamma_{ij} \frac{\sqrt{(1+2i)!(2j!)}}{i!j!}|(2i+1)\phi; 2j \phi^\bot\rangle_B\\
\nonumber
&& |\Phi^{\phi^\bot}\rangle_B = \sum_{i,j=0}^\infty \gamma_{ij} \frac{\sqrt{(1+2i)!(2j!)}}{i!j!}|2j\phi; (2i+1) \phi^\bot\rangle_B
\end{eqnarray}
with $\gamma_{ij} =(-1)^i \cosh(g)^{-2} \left(\frac{\tanh(g)}{2}\right)^{i+j},$ g being the gain of the amplification. $|n\phi\rangle_A$ ($|n\phi^\bot\rangle_A$) corresponds to a $n$ photon Fock state with polarization $\frac{1}{\sqrt{2}}\left(H + e^{i\phi}V\right)$ $\left(\frac{1}{\sqrt{2}}\left(H - e^{i\phi}V\right)\right)$, H and V standing for horizontal and vertical polarizations respectively. Interestingly, $|\Phi^\phi\rangle_B$ contains $3\sinh^2 g+1$ photons on average with polarization $\frac{1}{\sqrt{2}}\left(H + e^{i\phi}V\right)$ and $\sinh^2 g$ photons with the orthogonal polarization whereas $ |\Phi^{\phi^\bot}\rangle_B$ contains $\sinh^2 g$ photons with polarization $\frac{1}{\sqrt{2}}\left(H + e^{i\phi}V\right)$ and $3\sinh^2 g+1$ photons with the orthogonal polarization. The state \eqref{eq_phasecovcloner} can thus been considered as a micro-macro entangled states in which a polarization mode contains about 3 times more photons than the orthogonal polarization mode \cite{DeMartini_Entanglement_2008}. In the experiment, $g = 4.4$ was used resulting in thousands of created photons.

The analysis of the ideal state regarding its macroscopic quantumness reveals differences between the proposed measures (cf.~example \ref{ex:Dicke} in Sec.~\ref{sec:examples}). For example, \citet{Sekatski_Size_2014} assigns only a large effective size when measuring the photon number in one polarization mode if a relatively low success probability $P_g \lesssim 0.74$ is accepted (higher $P_g$ could be accepted by taking the second mode into account and/or by changing the branching into $| \mathcal{A} \rangle $ and $| \mathcal{D} \rangle $). In this case, an effective size of $\approx 1000$ can be found for $P_g = 2/3$ and $g = 4.4$. All measures based on the variance of quadrature operators find a large effective size in the order of the photon number.
Note, however, that an experimental test showing this large macroscopic quantumness was not done so far. In particular, the coherence of the state (\ref{eq_phasecovcloner}) is difficult to prove. In this context, \citet{Raeisi_Coarse_2011} showed that coarse-grained measurements cannot reveal the quantum nature of these states, a property that is shared by many macroscopic quantum states, see discussion in Sec.~\ref{sec:stab-dist-stat}. \citet{DeMartini_Investigation_2015} argued that the quantumness of this state is experimentally shown in the low-$g$ regime with a mean photon number of up to twelve.\\

States with similar properties can be obtained with simpler amplification techniques. \citet{Sekatski_Proposal_2012} for example proposed to start with path-entangled state, that is, a micro-micro state of the form
$
\frac{1}{\sqrt{2}}(|0\rangle_A|1\rangle_B - |1\rangle_A|0\rangle_B)
$
where $|0\rangle_A$ and $|1\rangle_A$ are the vacuum and single photon Fock state for the spatial mode A and similarly for B. The idea then consists in amplifying the photon number in one mode through a displacement in phase space, $D(\alpha),$ where the amplitude $\alpha$ is considered to be real without loss of generality. This leads to
\begin{equation}
\label{eq:displaced_singlephoton}
\frac{1}{\sqrt{2}}(|0\rangle_A D(\alpha)|1\rangle_B - |1\rangle_A |\alpha\rangle_B)
\end{equation}
where $|\alpha\rangle_B = D(\alpha)|0\rangle_B$ is a coherent state for mode B. When the initial micro-micro state is seen in the rotated basis $\{|+\rangle = 2^{-1/2}(|0\rangle+|1\rangle),|-\rangle = 2^{-1/2}(|0\rangle-|1\rangle)\},$ the entangled state after the amplification involves two components $D(\alpha)|\pm\rangle_B$ whose mean photon numbers are separated by $2\alpha.$
This proposal has triggered two experiments \cite{Bruno_Displacement_2013, Lvovsky_Observation_2013}, both using spontaneous down conversion based sources to create photon pairs, the detection of a photon from each pair serving to herald the creation of its twin photon. The heralded photons were then sent into a balanced beamsplitter to create path-entanglement before undergoing a displacement operation. The latter was implemented with an unbalanced beamsplitter and coherent states. To facilitate the detection needed to reveal entanglement, the displacement operation was undone and the coherence and photon number probability distribution in each arm were obtained by photon counting counting techniques. In \citet{Bruno_Displacement_2013}, entanglement was recorded as the photon number is amplified and the authors succeeded to reveal entanglement for up to $\alpha^2=500$. The size of the target state \eqref{eq:displaced_singlephoton} has been discussed in \citet{Sekatski_Size_2014}, see Sec.~\ref{sec:Sekatski} for the corresponding measure. For a guessing probability $P_g=2/3$ for example, it has been shown that its effective size for $\alpha^2 = 500$ is the same than the one of $\frac{1}{\sqrt{2}}(|0\rangle_A |n\rangle_B - |1\rangle_A |0\rangle_B)$ with $n=38.$ \\

\begin{figure}
\centerline{\includegraphics[width=6cm]{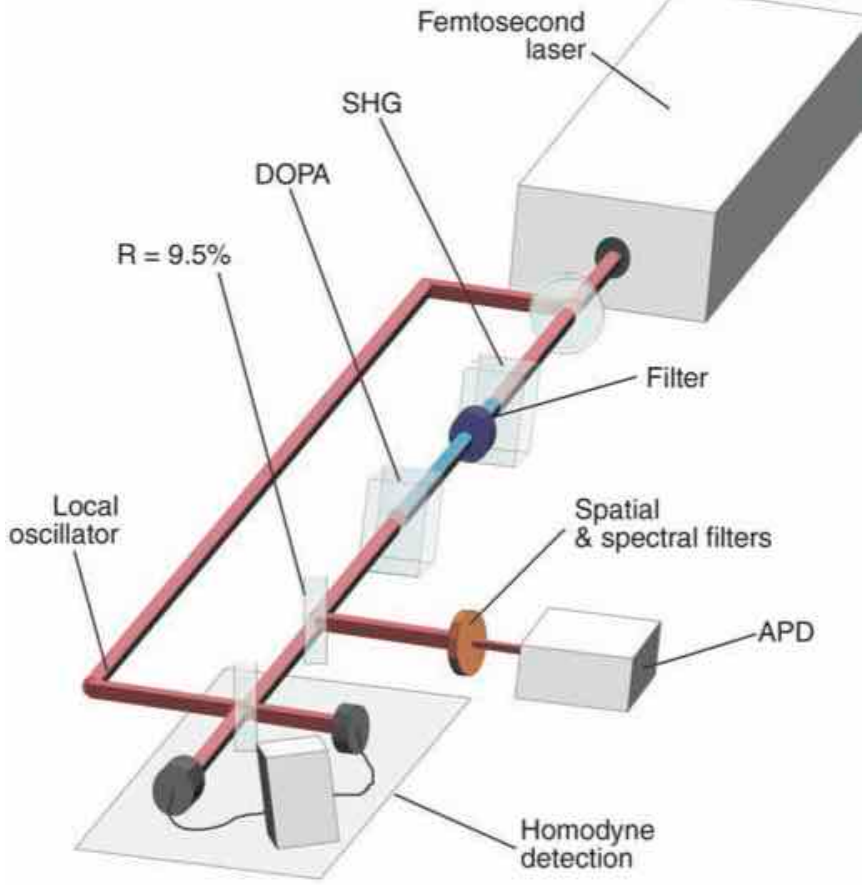}}
\caption[]{\label{fig:Ourjoumtsev_Generating_2006} Schematic representation of the setup used in \citet{Ourjoumtsev_Generating_2006} aiming to produce superposition of coherent states with opposite phases. A squeezed vacuum beam is first produced in a frequency-degenerate optical parametric amplifier (DOPA) by down-conversion of frequency-doubled (SHG)  laser pulses. A beamsplitter reflects less than 10\% of the squeezed beam toward a photon detector (APD) through a filtering system, whereas the transmitted beam is analyzed by a homodyne detection. A tomography of the photon subtracted squeezed vacuum state is performed and the obtained state is then be compared with a superposition of coherent states with opposite phases. APD : avalanche photo-diode. Adapted from \citet{Ourjoumtsev_Generating_2006}.}
\end{figure}

Let us finally mention conditional techniques that can also be used for example to create SCS (Eq. \eqref{eq:43}) as proposed in \citet{Danka_Generating_1997}. \citet{Ourjoumtsev_Generating_2006} for example reported on the creation of such a state by subtracting one photon from a squeezed vacuum state.
The latter was created by means of a frequency-degenerate optical parametric amplifier. Photons are then subtracted from the output in a probabilistic way, using a partially reflecting beamsplitter and a photon detector, see Fig.~\ref{fig:Ourjoumtsev_Generating_2006}. A successful photon subtraction projects the transmitted part into a state which is close to a SCS. A tomography revealed SCS-like states characterized by $|\alpha|^2=0.79$ and a fidelity of 70\% \cite{Ourjoumtsev_Generating_2006}. Larger sizes, up to $|\alpha|^2=3.2$ have been reported in \citet{Neergaard-Nielsen_Generation_2006, Gerrits_Generation_2010, Yukawa_Generating_2013, Sychev_Enlargement_2017} using similar protocols. Note that the usefulness of the photon subtraction has been discussed in \citet{Oudot_Two-mode_2015} where the size of squeezed vacuum states and SCS are compared with different measures. More recently, advanced conditional techniques have been used to create entanglement of the form $|+\rangle_A |\alpha\rangle_B + e^{i\varphi} |-\rangle_A |-\alpha\rangle_B$ \cite{Morin_Remote_2014, Jeong_Generation_2014}. Note also that iterative conditional techniques that could be used to obtained SCS with larger sizes have been implemented to create superpositions of squeezed coherent states \cite{Etesse_Experimental_2015}.

\subsubsection{Microwave photons}
\label{sec:microwave-photons}

\begin{figure}
\centerline{\includegraphics[width=6cm]{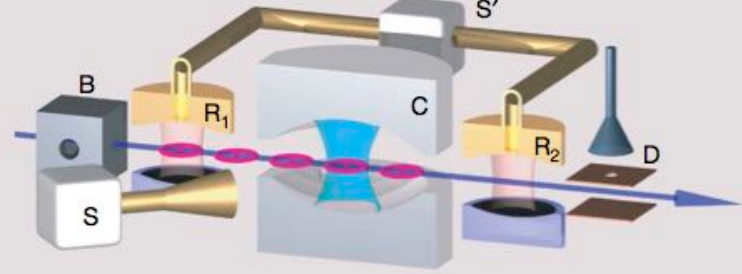}}
\caption[]{\label{fig:Deleglise_Reconstruction_2008} Schematic representation of the setup used in \citet{Deleglise_Reconstruction_2008} to produce various quantum states of light trapped in a high finesse cavity (C), including superposition of coherent states with opposite phases. A stream of atoms is prepared in box B and cross the R1-R2 cavities playing the role of a Ramsey interferometer in which the cavity C is inserted. The source S generates a coherent microwave pulse that can be used to inject into C coherent states with controlled amplitude and phase. Another pulsed source, S', feeds the interferometer cavities R1 and R2. Information is extracted from the field by state-selective atomic counting in D. Adapted from \citet{Deleglise_Reconstruction_2008}.}
\end{figure}

Fock states and superpositions of coherent states with opposite phases have been created by letting Rydberg atoms interact one by one with the electromagnetic field of a high finesse cavity \cite{Deleglise_Reconstruction_2008, Sayrin_Real-time_2011}. The principle relies on a dispersive light-atom interaction that is used to imprint the information about the photon number in the cavity into the phase of a superposition between two internal atomic states, the latter being measured through a Ramsey interferometer, see Fig.~\ref{fig:Deleglise_Reconstruction_2008}. Fock states have been prepared by first launching a coherent field in the cavity and by then letting it interact with atoms, achieving a quantum non-demolition measurement of the photon number that progressively projects the field onto a Fock state $|n\rangle$. Reconstruction of Fock state with up to $n=4$ photons have been reported in \citet{Deleglise_Reconstruction_2008} and up to $n=7$ photons in \citet{Zhou_Field_2012} using an additional quantum feedback procedure. \\

To generate SCS, a coherent field is first injected into the cavity before interacting with an atom prepared in a superposition of two internal states. This results in an atom-light entangled state in which the internal atomic states are correlated with a coherent state with different phases. The projection of the atom into the appropriate state ideally leaves the field in the desired superposition. \citet{Deleglise_Reconstruction_2008} reported on SCS states with $\alpha^2 = 3.5$ and a fidelity of 72\%. 

Similar techniques were used more recently with superconducting devices. \citet{Vlastakis_Deterministically_2013, Wang_Schrodinger_2016} reported on a set of multi-photon operations using a superconducting charge qubit called transmon, coupled to waveguide cavity resonators. The authors succeeded to obtain ideal strong-dispersive coupling, where the strengths of the off-resonant qubit-cavity interactions were several orders of magnitude larger than both the cavity and transmon decay rates. This allowed them to create and detect SCS with unprecedented sizes, that is, $\alpha^2 \approx 7.8$ and a visibility of $\approx 0.57$ \cite{Vlastakis_Deterministically_2013}. A similar setup has also been used to implement two-mode SCS of the form $\ket{\alpha, \beta} + \ket{-\alpha, -\beta}$ with $|\alpha |^2 = 9.0,$ $|\beta|^2 = 7.0$ with a visibility of $0.58$ \cite{Wang_Schrodinger_2016}.\\

\subsection{Spin experiments}
\label{subsection_spinexp}

Regarding experiments with spin systems, we can distinguish between setups in which the spins can be addressed individually or collectively. As before, we quickly present techniques nicely illustrating experiments along these two lines. \citet{Leibfried_Quantum_2003, Blatt_Entangled_2008,Ritsch_Cold_2013} are more exhaustive reviews related to trapped ions and cold atoms respectively.

\subsubsection{Spins with individual addressing}
\label{sec:spins-with-indiv}

One of the most advanced systems where (pseudo-)spins can be addressed individually are trapped-ion systems. The spin states are often encoded in a ground state and a metastable electronic state of each ion. Laser coupling between these internal states and the vibrational mode produces a collective spin flip that can be used as an entangling gate \cite{Sorensen_Quantum_1999}. From high fidelity quantum gates \cite{Benhelm_Towards_2008}, GHZ states (see Eq.~\eqref{eq:39}) with up to 14 ions \cite{Monz_14-Qubit_2013} have been created. The diagonal elements of the corresponding density matrice have been measured directly by fluorescence measurements while the off-diagonal elements have been accessed via the amplitude of parity oscillations. The measurements allowed to infer the fidelity of GHZ states for different ion numbers. Fidelities larger than 95\% and 80\% have been observed for up to 4 and 8 ions GHZ states respectively while 14-ion GHZ states have been measured with a fidelity above 50\%, which is in principle sufficient to violate a Bell inequality \cite{Lanyon_Experimental_2014}. 

\subsubsection{Spins with collective addressing}
\label{sec:spins-with-coll}

\begin{figure}
\centerline{\includegraphics[width=7cm]{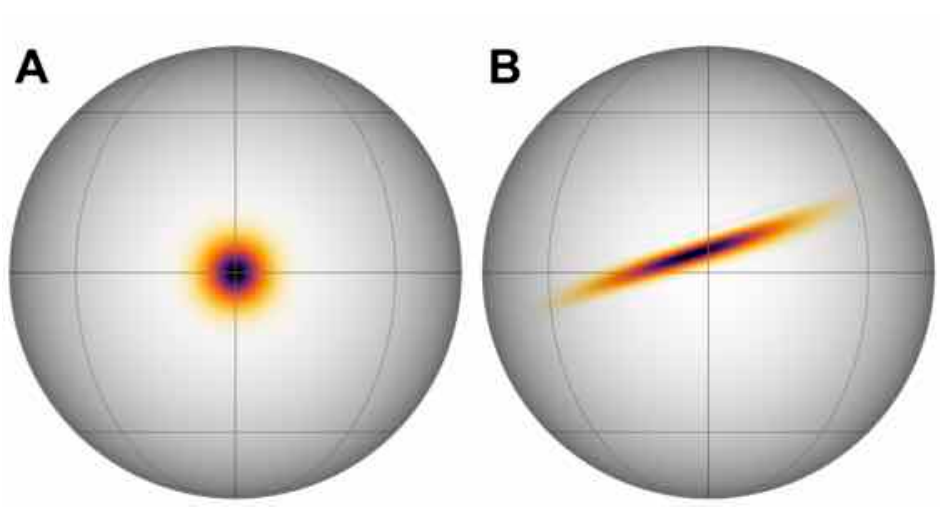}}
\caption[]{\label{fig:Schmied_Bell_2016} Spherical projections of the Wigner function on the Bloch sphere for 100 spins which help understanding how the techniques used in \citet{Riedel_Atom-chip-based_2010, Schmied_Bell_2016} lead to spin squeezing. (A) an initial coherent spin state is prepared along the x direction. (B) Controlled elastic collisions in state dependent potentials lead to an effective interaction of the from $S_z^2.$ The corresponding unitary is nothing else than a rotation around z whose angle depends on the projection on the z axis. This results in a spin squeezed state whose squeezing and anti-squeezing directions can be measured by projective measurements are along the vertical (+z) spin axis preceded by the appropriate rotation. Adapted from \citet{ Schmied_Bell_2016}.}
\end{figure}

An example of spin systems where the spins cannot be addressed individually is given by Bose-Einstein condensates where the internal states of the atoms constituting these condensates can be initialized with optical pumping techniques, coherently coupled through controlled elastic collisions and readout using absorption imaging, see Fig.~\ref{fig:Schmied_Bell_2016}. This allowed one to create spin squeezed states with 1250 atoms \cite{Riedel_Atom-chip-based_2010} with a squeezing parameter $\xi^2 \approx -2.5$dB, see  Eq. \eqref{eq:42} for the definition. Note that the detection of spin squeezing is  connected to quantum correlations between the spins \cite{Kitagawa_Squeezed_1993}. Recently, the techniques reported in \citet{Riedel_Atom-chip-based_2010} have been used to prove that the internal correlations between 480 atoms in a spin-squeezed states are strong enough to violate a Bell inequality \cite{Schmied_Bell_2016}.
Note also that spin-squeezed states have been created with larger squeezing parameters. \citet{Gross_Nonlinear_2010} reported on spin squeezing with 2300 atoms and $\xi^2 \approx -8.2$dB in a Bose-Einstein condensate, through Feshbach control of interactions in an optical trap. Note finally that similar states have also been obtained with trapped ions \cite{Bohnet_Quantum_2016}, with room-temperature \cite{Vasilakis_Generation_2015} and cold \cite{Hosten_Measurement_2016} atoms trapped in cavities. The latter succeeded to report spin-squeezing with $5\times 10^5$ atoms with squeezing parameter $\xi^2 \approx -20.1$dB. \\

\subsection{Experiments with massive systems} \label{subsection_massiveexp}

Experiments aiming to bring a massive system in a quantum superposition of well distinct positions include matter interferometry and quantum optomechanics. While intensive efforts are devoted to the former for more than 30 years, the latter is nowadays attracting a lot of attention and impressive results have been obtained in the last decade. Matter interferometry has been reviewed in \citet{Cronin_Optics_2009, Arndt_Testing_2014} and the most recent progress in optomechanics can be found in \citet{Yin_Optomechanics_2013, Meystre_Short_2013, Aspelmeyer_Cavity_2014}.\\

\begin{figure}
\centerline{\includegraphics[width=3cm]{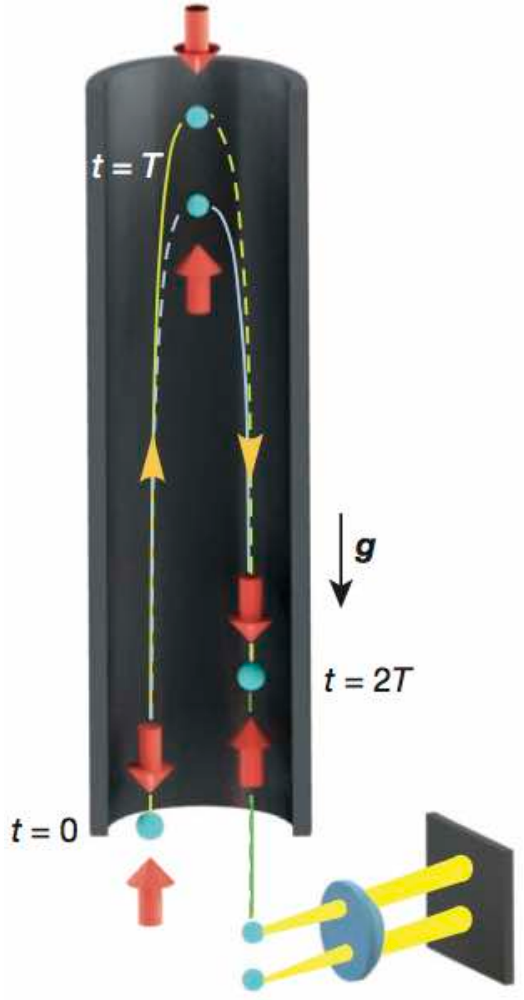}}
\caption[]{\label{fig:Kovachy_Quantum_2015} Schematic representation of the setup used in \citet{Kovachy_Quantum_2015} to coherently split the wave packet of single atoms up to $\approx 54$cm. An ultra-cold atom cloud is launched vertically in magnetic shield using an optical lattice. At time $t=0,$ a first sequence of laser pulses split the cloud into a superposition of states with different momenta. A time $T$ later, the wave packet is spatially separated, and a laser sequence reverses the momenta of each superposed component. At time $2T,$ the clouds spatially overlap, and laser pulses make them interfere. Adapted from \citet{Kovachy_Quantum_2015}.}
\end{figure}

\subsubsection{Matter interferometry}
\label{sec:matt-interf}

Interferometry with widely delocalized and more and more massive objects is an active research domain since mid 80's \cite{Gould_Diffraction_1986, Keith_Diffraction_1988}. Impressive results have been obtained along this line by the trapped ion community \cite{Wineland_Nobel_2013}. The starting point of these experiments is the use of laser cooling techniques to bring an ion down to its motional ground state $\ket{0}.$ Using a displacement whose direction depends on the spin state of the ion's outermost electron, an ion prepared in a superposition of spin states $(\ket{\uparrow}+\ket{\downarrow})$ ends up in a superposition of two positions \cite{Monroe_Schroedinger_1996}. More precisely, the spin-depend force promotes an initial state $(\ket{\uparrow}+\ket{\downarrow})\ket{0}$ into $\ket{\uparrow}\ket{\alpha}+\ket{\downarrow}\ket{-\alpha}$ where the coherent states $\ket{\pm \alpha}$ represent the amplitude and phase of the ion motion in its local harmonic trapping potential. The two positions can then be recombined to form the analog of an interferometer which allows one to access the coherence of the entangled state $\ket{\uparrow}\ket{\alpha}+\ket{\downarrow}\ket{-\alpha}$ through the coherence of the spin state superposition.
In \citet{Kienzler_Observation_2016}, such a superposition with $\alpha \approx 5.9$  has been reported, which effectively corresponds to a spatial separation between the superposed locations dozen of times larger than the extent of local position fluctuations.\\

To date, the widest delocalization has been obtained by launching a Bose-Einstein condensate made with about $10^5$ Rubidium atoms in a $10$m high atomic fountain, see Fig.~\ref{fig:Kovachy_Quantum_2015}. Once launched, a sequence of pulses is applied to control the atom momenta so that the wave packet of each atom is split and recombined coherently to form the analogue of a Mach-Zehnder interferometer. In the experiment presented in \citet{Kovachy_Quantum_2015} the wave packets get separated during a drift time $\approx 1$s after which they reach their maximum separation of up to $\approx 54$cm. The wave packet of each atom is then recombined to spatially overlap after another drift interval $\approx 1$s. The contrast of the interference is determined by measuring the variation of the normalized number of atoms in one of the two outputs of the interferometer. Interestingly, the contrast of $\approx 28\%$ reported in \citet{Kovachy_Quantum_2015} is incompatible with an explicit collapse model based on quantum gravity \cite{Minar_Bounding_2016} in the parameter regime that was initially proposed by the fathers of this collapse model \cite{Ellis_Quantum_1989}. Note however, that Refs. \cite{Stamper-Kurn_Verifying_2016,Kovachy_Kovachy_2016} clarified that the experiment reported in \citet{Kovachy_Quantum_2015} did not have a stable phase reference for the interferometer, which is required to constrain models that would introduce overall phase noise.  In subsequent work \cite{Asenbaum_Phase_2017}, the same group introduced a second, spatially displaced interferometer as a stable phase reference. This experiment demonstrated phase stability of interferometers with $16$cm arm separation and it remains to be clarified if the observed interference is compatible with the collapse model presented in Ref. \cite{Ellis_Quantum_1989} .\\

%doubts have been raised about the origin of the interference reported in \citet{Kovachy_Quantum_2015} as independent Bose-Einstein condensate can interfere through a second order coherence process \cite{Stamper-Kurn_Verifying_2016,Kovachy_Kovachy_2016}.\\

\begin{figure}
\centerline{\includegraphics[width=9cm]{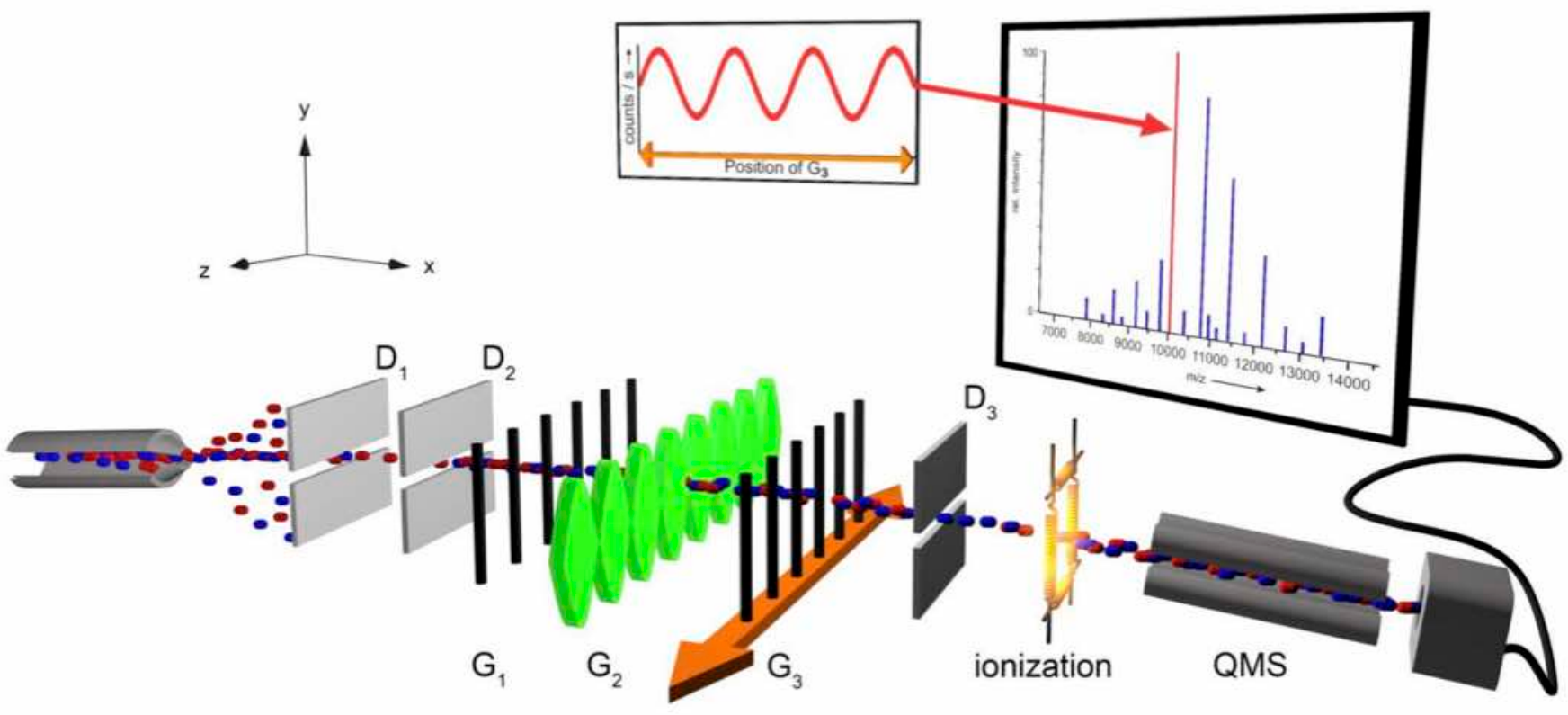}}
\caption[]{\label{fig:Eibenberger_Matter-wave_2013} Schematic representation of the setup used in \citet{Eibenberger_Matter-wave_2013} to make interferometry with massive molecules combining 810 atoms with a molecular weight exceeding $10^4$ amu. Three narrow slits $D_1$-$D_3$ select a particular particle velocity following a specific parabola in the gravitational field. $G_1$-$G_2$ are gratings. $G_2$ is a standing wave used to change the phase of the interferometer through an optical dipole force. The transmitted molecules are detected via a ionization technique (QMS : quadrupole mass spectrometer) after $G_3$ which can be shifted to sample the interference fringes. 
From \citet{Eibenberger_Matter-wave_2013}.}
\end{figure}

Despite using a Bose-Einstein condensate, the interference reported in  \citet{Kovachy_Quantum_2015} depends only on the wavelength of a single atom. The experiment is essentially single atom interferometry with $\approx 10^5$ interferences performed at each experimental run. The relevant mass is thus limited to that of a single atom, that is 87 amu for $^{87}$Rb. However, larger masses are required to test a broad class of collapse models. This provides motivation to perform interferences with macromolecules and clusters. One of the first results along this line was obtained with a Talbot-Lau type interferometry using fullerenes \cite{Arndt_Wave-particle_1999}, a carbon molecule with 720 amu. The basic principle of such an experiment is shown in Fig.~\ref{fig:Eibenberger_Matter-wave_2013}. First molecules from a thermal source are evaporated into a vacuum chamber and a velocity selection is done using narrow slits. The selected molecules are then sent into an Talbot-Lau interferometer which is made with three gratings. The first grating preselects a molecular transverse coherence. Diffraction at the second grating then produces a molecular density pattern at the location of the third grating through the Talbot effect. If the molecular pattern and the third grating mask are aligned, the transmission is high. When the third grating is shifted by half a grating period, the total transmitted signal is minimal. The signature of molecular interference is thus obtained by counting the molecule number as a function of the position of the third grating. An advanced version of the Talbot-Lau interferometer %\cite{Haslinger_Universal_2013} 
currently holds the mass record in matter-wave interference, with molecules combining several hundreds of atoms with a molecular weight of thousands \cite{Gerlich_Quantum_2011} and even tens of thousands amu \cite{Eibenberger_Matter-wave_2013}.\\

\begin{figure}
\centerline{\includegraphics[width=6cm]{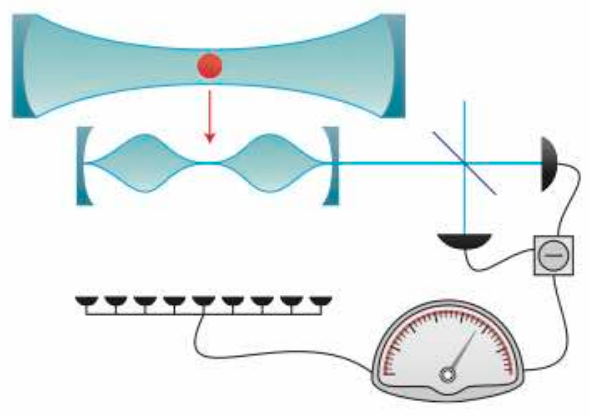}}
\caption[]{\label{fig:Arndt_Testing_2014} Schematic representation of the setup envisioned for matter interferometry with a nanosphere. A cooled sphere is first optically trapped before being released to let its wave-packet expand. The sphere then enters a second cavity where a pulsed interaction is performed using a quadratic optomechanical coupling. A homodyne measurement of the output field phase performs a quadratic measurement of the sphere position and prepares it in a quantum superposition of two positions whose spread depends on the measurement outcome. Then, the sphere is again released before its center-of-mass position is measured and interference fringes are observed. From \citet{Arndt_Testing_2014}.}
\end{figure}

To conclude this section, let us mention that matter-interferometry experiments are nowadays envisioned with nanosphere exceeding $10^7$amu. \citet{Romero-Isart_Large_2011} for example propose to trap a dielectric sphere in the standing wave of an optical cavity, see also \citet{Chang_Cavity_2010, Barker_Cavity_2010} for related proposals. The mechanical motion of the sphere's center of mass is predicted to be a high-quality mechanical oscillator due to the absence of thermal contact and dissipation arising from clamping. This is expected to facilitate laser cooling. The cooled levitating object can then be released by switching off the trap to let the wave-function expand. A quadratic measurement of the mechanical position finally creates a scenario similar to matter wave interferometry experiments, see Fig.~\ref{fig:Arndt_Testing_2014}. While significant experimental progress has been realized to trap and cool such a nanosphere\footnote{\citet{Gieseler_Subkelvin_2012, Kiesel_Cavity_2013, Millen_Cavity_2015, Ranjit_Attonewton_2015, Arita_Rotation_2015}}, we are not aware of experiments reporting on quantum interference with such a system.\\

\subsubsection{Quantum optomechanics}
\label{sec:quant-optom1}

\begin{figure}
\centerline{\includegraphics[width=6cm]{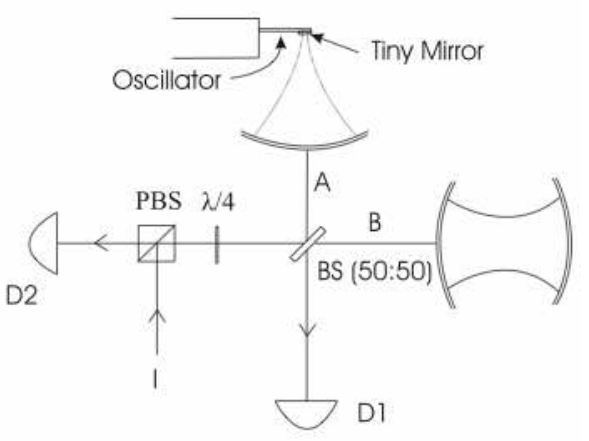}}
\caption[]{\label{fig:Marshall_Towards_2003} Schematic representation of the interferometer envisioned for studying the creation and decoherence of a mirror in a spatial superposition. A cavity is placed on each arm of a Michelson interferometer. One of the two cavities is made with a high-quality mechanical resonator whose motion is affected by a single photon through its radiation pressure. A photon entering in the interferometer leads, after half a mechanical period, to an entangled state in which the photon is stored in cavity B and the mechanical oscillator is in its motional ground state and the photon is stored in A and the mechanical motion is excited. After a full mechanical period, the photon mode and the mirror position disentangled, resulting in a maximum photon interference in the absence of oscillator decoherence. Recording the photon interference for multiple mechanical periods allows one to infer the mechanical decoherence. From \citet{Marshall_Towards_2003}.}
\end{figure}

While many aspects of quantum cavity optomechanics started to be explored theoretically in the early 90's \cite{Fabre_Quantum-noise_1994, Mancini_Quantum_1994}, proposals \cite{Bose_Preparation_1997, Bose_Scheme_1999} have been done in the late 90's to create a superposition of mechanical states with a distance of the order of the mechanical zero-point fluctuation where the effects of unconventional decoherence might be observed \cite{Marshall_Towards_2003, Kleckner_Creating_2008}. The basic idea is to use a Michelson interferometer with a high-finesse cavity in each arm, see Fig.~\ref{fig:Marshall_Towards_2003}. One of the two cavities is made with a movable mirror, that is, a high-quality oscillator with a mechanical frequency larger than the cavity decay rate such that it can initially be prepared in its motional ground state by sideband cooling. A single photon is then launched in the Michelson interferometer and its energy is stored coherently in both cavities. The radiation pressure force shifts the mirror position and a maximal displacement is achieved after half a mechanical period. After a full mechanical period, the mirror comes back at its original position and in the absence of decoherence, full interference is expected. The mirror decoherence, however, alters the interference of the photon. In other words, by observing the photon interference, we can infer the mirror decoherence rate. For an optomechanical coupling rate larger than the mechanical frequency, the maximum displacement is expected to be larger than its zero-point motion at half the mechanical period. If in addition, the device operates in the strong coupling regime so that the photon can be stored long enough, this device could be used as a test bench for unconventional decoherence models. While massive oscillators with eigenfrequencies in the kilohertz regime were initially envisioned, first experiments now succeeded in entering the field of optomechanics in the quantum regime using lighter and more rigid mega or gigahertz oscillators. This includes ground state cooling of the mechanical motion\footnote{\citet{OConnell_Quantum_2010, Meenehan_Pulsed_2015, Teufel_Sideband_2011, Chan_Laser_2011}}, electromechanical entanglement \cite{Palomaki_Entangling_2013} or  squeezing of a micromechanical state \cite{Wollman_Quantum_2015}. As far as we can tell however, no experimental realization so far entered a regime relevant for ruling out unconventional decoherence models. \\

\subsection{Superconducting quantum interference devices}
\label{sec:superc-quant-interf}

\begin{figure}
\centerline{\includegraphics[width=.9\columnwidth]{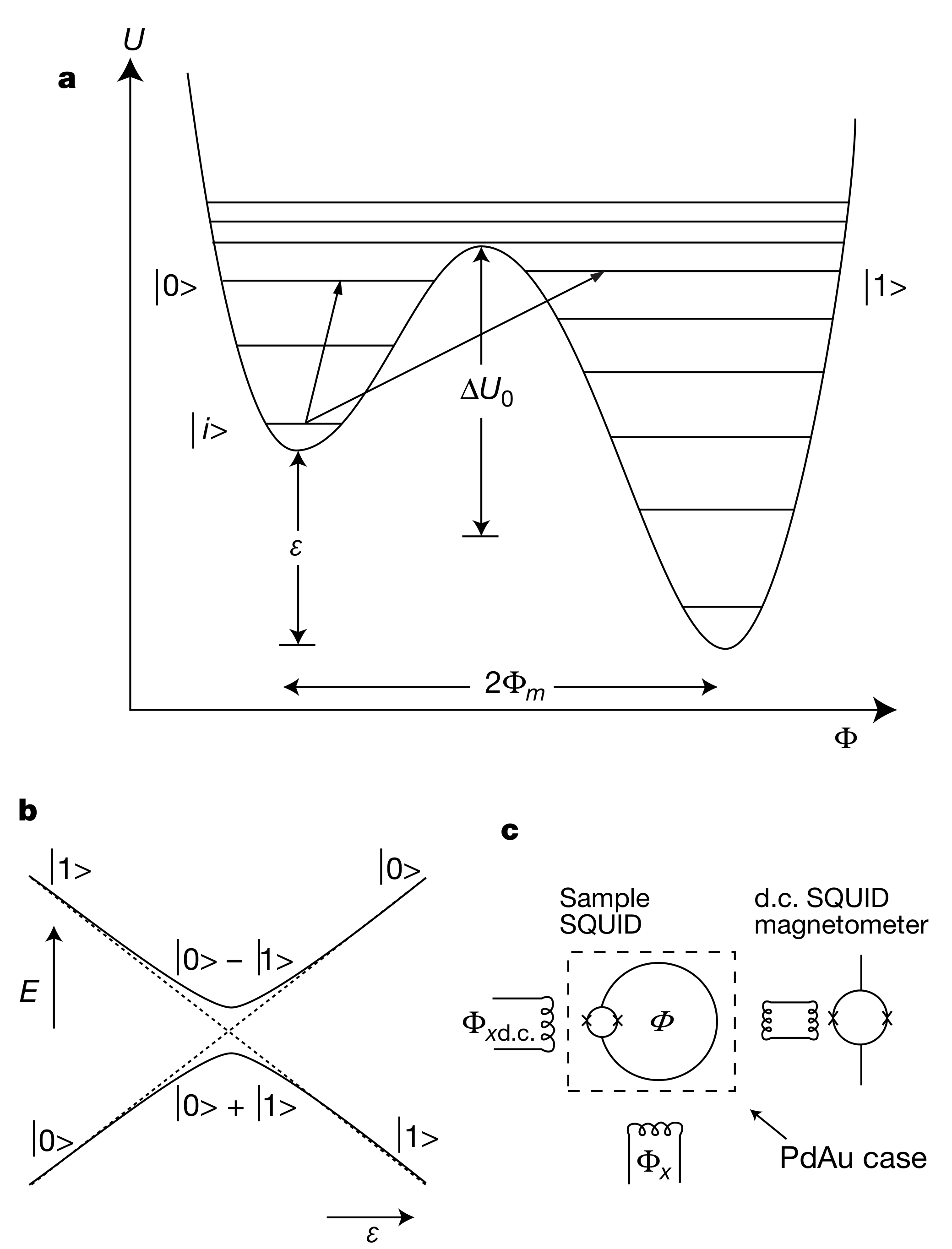}}
\caption[]{\label{fig:Friedman} Illustration of the physical principle behind the experiment reported in \citet{Friedman_Quantum_2000} aiming to create coherent superpositions between clockwise and anti-clockwise circulating currents using a SQUID. (a) Illustration of eigenenergies of the device as a function of the current flux. By controlling the parameters of the potential, the SQUID can be prepared in the eigenstate  $| i \rangle$ corresponding to a current flux with a well defined circulation. Driving the transitions from $| i \rangle $ to excited states $| 0 \rangle $ and $| 1 \rangle$ and controlling the parameters $\epsilon$, $\Delta U_0$ of the double-well, the state ideally ends up in a superposition of clockwise and anti-clockwise circulating current. (b) The coherence of this superposition state is revealed thought an avoided crossing between the systems eigenenergies when $\epsilon$ is tuned. (c) Schematics of the experiment with the SQUID inside the dashed box, the external control and an additional SQUID that operates as a magnetometer to probe the energy of the neighboring SQUID. From \citet{Friedman_Quantum_2000}.}
\end{figure}

The question of whether or not macroscopic quantumness can be realized in superconducting devices \cite{Friedman_Quantum_2000,vanderWal_Quantum_2000,Hime_Solid-State_2006} was intensively discussed.\footnote{In particular, in \citet{Leggett_Macroscopic_1980,Leggett_Testing_2002,Dur_Effective_2002,Bjork_Size_2004,Korsbakken_Measurement-based_2007,Marquardt_Measuring_2008,Korsbakken_Electronic_2009,Korsbakken_Size_2010,Nimmrichter_Testing_2011}}  While superconductivity itself was argued to be a classic example of a microscopic quantum effect \cite{Leggett_Macroscopic_1980}, there have been attempts to create macroscopic superpositions of clockwise and anti-clockwise circulating currents in superconducting quantum interference devices (SQUIDs). In its simplest version, a SQUID is a superconducting ring with conductance $L$ interrupted by a single Josephson junction with capacitance $C$ and critical current $I_c$, which allows electrons to pass through by tunneling (see \citet{Souletie_hasard_1987} and references therein for a detailed description of different variant of SQUIDs). In \citet{Friedman_Quantum_2000}, the single junction is replaced by two parallel junctions (see Fig.~\ref{fig:Friedman} (c)). Even though many electrons (up to 10$^{10}$) are involved, the only relevant degree of freedom is the total flux in the ring. This gives rise to a simple phenomenological model that is mathematically equivalent to a particle with effective mass $C$ in a one-dimensional double-well structure potential. The energy eigenfunctions that are localized in one well corresponds to currents circulating either clockwise or anti-clockwise. By controlling the height of the potential barrier as well as the difference between the left and right local minima (see Fig.~\ref{fig:Friedman} (a)), experimenters prepared the device in a localized ground state (state $| i \rangle$ in Fig.~\ref{fig:Friedman} (a)) and drove it to localized excited states (either $| 0 \rangle $ or $| 1 \rangle$). By tuning the local minimum ($\epsilon$), the authors observed an avoided crossing in the energy spectrum which is an indication of the coherence between clockwise and anti-clockwise circulating currents (see Fig.~\ref{fig:Friedman} (b) for the theoretical prediction). Note that other experiments, for example, \citet{vanderWal_Quantum_2000,Hime_Solid-State_2006} differ in the details, but also work with a double-well potential and prove the coherence via an avoided crossing.

\subsection{Comparing the size of states describing photonic, spins and massive systems}
\label{Comparison_photo_spin_mass_states}

In this section, we summarize estimates of the effective size of states detected experimentally. Ideally, this is done as much as possible directly from the experimental data with a minimum of additional modeling. In this sense, the first section \ref{sec:disc-size-flux} misses this ideal as all discussions so far mainly focus on the theoretical model and less on the experimental results. Even more importantly, it shows the difficulties to find an agreement when evaluating the size of a given experiment or state  with different measures. In Sec.~\ref{sec:comp-size-observ}, we present results for several experiments by applying the frameworks of \citet{Frowis_Measures_2012,Nimmrichter_Testing_2011}.

\subsubsection{Discussions on the size of flux states in SQUID systems}
\label{sec:disc-size-flux}

Here, we summarize different contributions that aim for assigning an effective size to various SQUID experiments. While there are clearly differences that arise from conceptual disagreement\footnote{For example, the disagreement between \citet{Leggett_Testing_2002,Korsbakken_Electronic_2009,Marquardt_Measuring_2008}.}, we note that measures are partially defined for different scales\footnote{In particular, \citet{Bjork_Size_2004} (which was the square root of, e.g., \citet{Korsbakken_Measurement-based_2007} in spin examples, Sec.~\ref{sec:ex:spin-ensemble}); or \citet{Nimmrichter_Testing_2011}.} This further complicates a comparison between the measures.

\textcite{Leggett_Testing_2002} claims about the SQUID experiments that ``any reasonable'' measure for macroscopic quantum states would assign an effective size that is in the order of the number of involved electrons. For the experiments \citet{Friedman_Quantum_2000,vanderWal_Quantum_2000}, this is $10^{10}$ (in units of Bohr magneton in the case of Leggett's extensive difference). \citet{Knee_Strict_2016} found an extensive difference of roughly $1.3\times 10^5$ in units of Bohr magneton for their experiment. 

This number is strongly contrasted by \textcite{Marquardt_Measuring_2008} using their framework. Analyzing only \textcite{vanderWal_Quantum_2000}, the authors find an effective size between one and two, that is, it suffices to apply at most two basic steps to map, say, the clockwise circulating state to the anticlockwise one. Similarly low is the number found by \citet{Nimmrichter_Testing_2011}, who assign an effective size of $\mu \approx 5$ to the experiment of \citet{Friedman_Quantum_2000}. Again different results are obtained by \textcite{Bjork_Size_2004}, who apply their proposal to \citet{Friedman_Quantum_2000}. They take the width of the coherent superposition of the wavefunction living in both wells and divide this by the width of the ground state localized in only one well. They find that the spread of the apparent macroscopic quantum state is only about 33 times larger than that of the ``classical'' ground state.

To apply the approach of \textcite{Korsbakken_Measurement-based_2007}, a microscopic model of the experiment is necessary to answer the question of how many electrons are effectively different in the two branches of the superposition. A detailed analysis is provided in \citet{Korsbakken_Electronic_2009,Korsbakken_Size_2010}, which leads to effective sizes of roughly 3800 - 5750 for \citet{Friedman_Quantum_2000}, 42 for \citet{vanderWal_Quantum_2000} and 124 for \citet{Hime_Solid-State_2006}. The authors explain the difference to Leggett's result by taking into account the fermionic nature of the electrons, which reduces the number of effectively different electrons.

It is not astonishing that these different results provoked many discussions. The quantitative results of \textcite{Leggett_Testing_2002} are critically seen by \textcite{Bjork_Size_2004,Korsbakken_Measurement-based_2007,Korsbakken_Electronic_2009,Marquardt_Measuring_2008}. \textcite{Leggett_Note_2016} remarks that the results of \textcite{Korsbakken_Electronic_2009} strongly depend on the choice some characteristic quantities such as the Fermi velocity. To illustrate his claims, he discusses the hypothetical example of a dust particle in the superposition of two macroscopically different momenta. By introducing ``reasonable'' characteristic scales, he shows that the approach of \textcite{Korsbakken_Measurement-based_2007} lead to trivially low effective size.\footnote{Similar arguments hold for the analysis of \citet{Nimmrichter_Testing_2011}.}

To continue a critical dialogue, we would like to add that the proposals of \textcite{Korsbakken_Measurement-based_2007,Marquardt_Measuring_2008}  depend on the splitting of the total wave function into ``dead'' and ``alive'', which can lead to ambiguous situations (see example \ref{ex:Dicke} in Sec.~\ref{sec:examples}). Furthermore, the choice of the ``basic step'' in the framework of \textcite{Marquardt_Measuring_2008} seems to be intuitive, but needs further justification. The approach of \textcite{Bjork_Size_2004} has a clear operational meaning, but the connection to the idea of an effective size as followed by the other works is unclear. A similar argument holds for the collapse model used by \citet{Nimmrichter_Testing_2011}.
The unresolved answers regarding the interpretation of the experimental evidence should be seen as a motivation to further improve the theory of macroscopic quantumness.

\subsubsection{Comparing the size of observed states}
\label{sec:comp-size-observ}

Some measures presented in Sec.~\ref{sec:preliminary-measures} are applicable to real experimental situations. In particular the measure of \citet{Frowis_Measures_2012} has been used in \citet{Frowis_Lower_2017} to compare the effective size $N_{\text{eff}}$ of various experimental photonic and spin states. In \citet{Nimmrichter_Macroscopicity_2013}, the effective size $\mu$ of states obtained in various experiments with massive and SQUID systems has been evaluated using the measure presented in the same manuscript. We quickly summarize the main results of these two studies separately. Note that the two measures are defined for different scales and hence a direct comparison between $N_{\text{eff}}$ and $\mu$ is meaningless.

The measure of \citet{Frowis_Measures_2012} is based on the quantum Fisher information. This is a promising mixed-state extension of the variance because of tight and accessible lower bounds (see Sec.~\ref{sec:witn-macr-quant}). For (spin)squeezed states, the left-hand side of the tighter Heisenberg uncertainty relation, Eq.~(\ref{eq:48}), becomes identical to $\xi^{-2}$, Eq.~(\ref{eq:42}). For experiments targeting superpositions of two ``classical'' states (e.g., GHZ state or SCS), Eq.~(\ref{eq:49}) can be used to bound the quantum Fisher information by witnessing large susceptibility to small external influences. Measuring the coherence terms $C = 2 |\left\langle 0 \right| ^{\otimes N} \rho \left| 1 \right\rangle ^{\otimes N}|$ for the GHZ and $C = 2| \left\langle \alpha \right| \rho \left| -\alpha \right\rangle| $ for the mono-mode SCS, allows to derive (approximate) lower bounds $N_{\mathrm{eff}}(\mathrm{GHZ}) \geq C^2 N$ and $N_{\mathrm{eff}}(\mathrm{SCS}) \gtrsim 4 C^2 |\alpha|^2 + 1$, respectively. Since these quantities are frequently measured, a first estimate can often be directly done with the data given in the publication. For example, the published data for the 8 photon GHZ state \cite{Wang_Experimental_2016} allows the rough estimate $N_{\text{eff}} \approx 2.3$.

\begin{figure}
\centerline{\includegraphics[width=8.5cm]{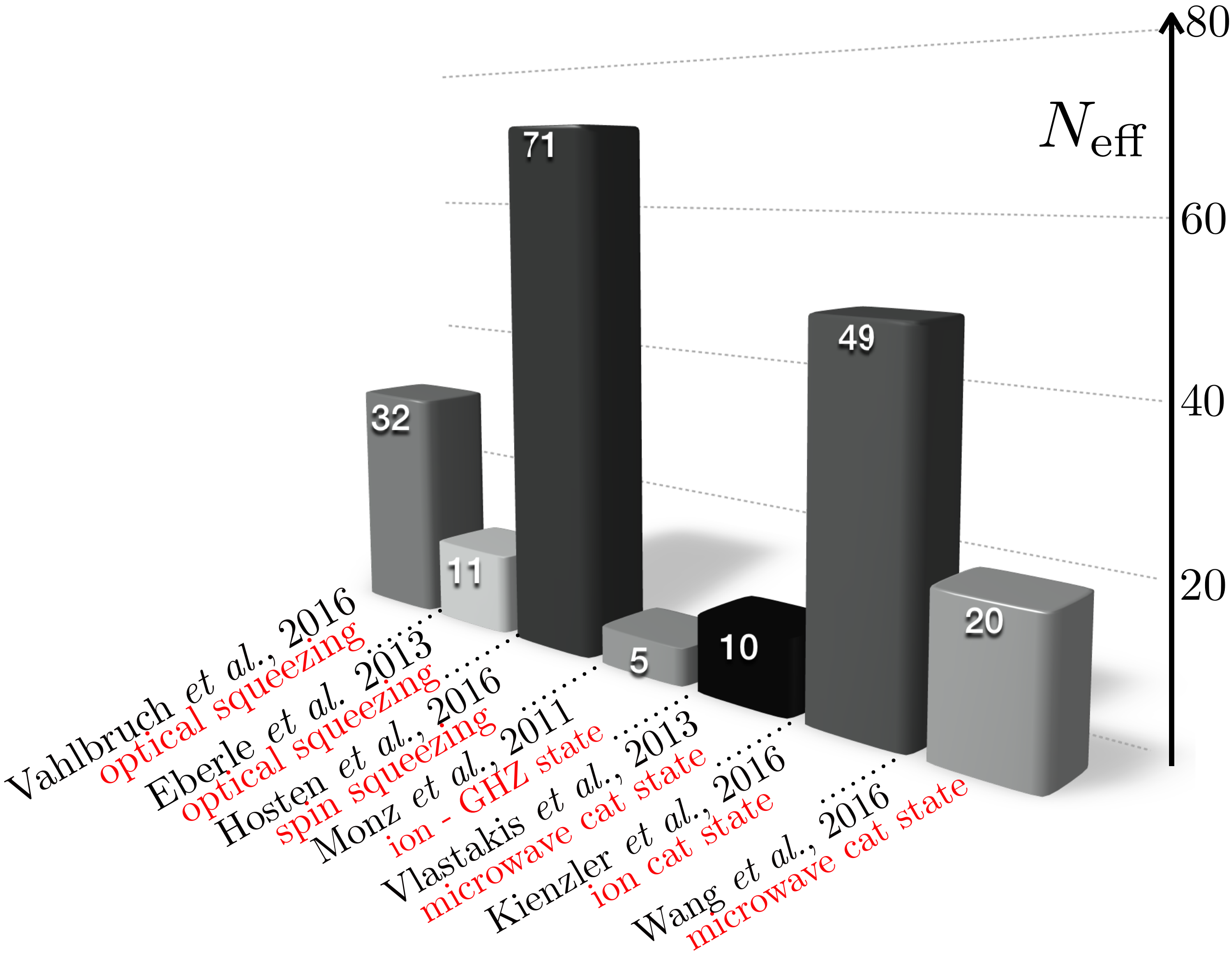}}
\caption[]{\label{fig:EffSizeSpinsPhotons} Summary of effective sizes for several experiments with (pseudo-)spins and photons. In various experimental setups, the macroscopic quantumness measured with the quantum Fisher information \cite{Frowis_Measures_2012} could be significantly increased over recent years. The numbers are taken from \citet{Frowis_Lower_2017}.}
\end{figure}

A complete analysis of several experiments was done in \citet{Frowis_Lower_2017}, see Fig.~\ref{fig:EffSizeSpinsPhotons}.
For example, the spin-squeezing experiment of \citet{Hosten_Measurement_2016} results in an effective size of $\approx 71$, which is comparable with an ideal GHZ state made out of 71 particles. Note the work of \citet{Kienzler_Observation_2016} realizes a spatial superposition of a \textit{single} atom, which is in apparent contradiction to our initial premise that only large systems can have a large effective size. However, in this case, the system size is taken as the mean number of phononic excitations in a harmonic trap, which is similar to the treatment of single-mode photonic systems.

\begin{figure}[h!]
\centerline{\includegraphics[width=8.5cm]{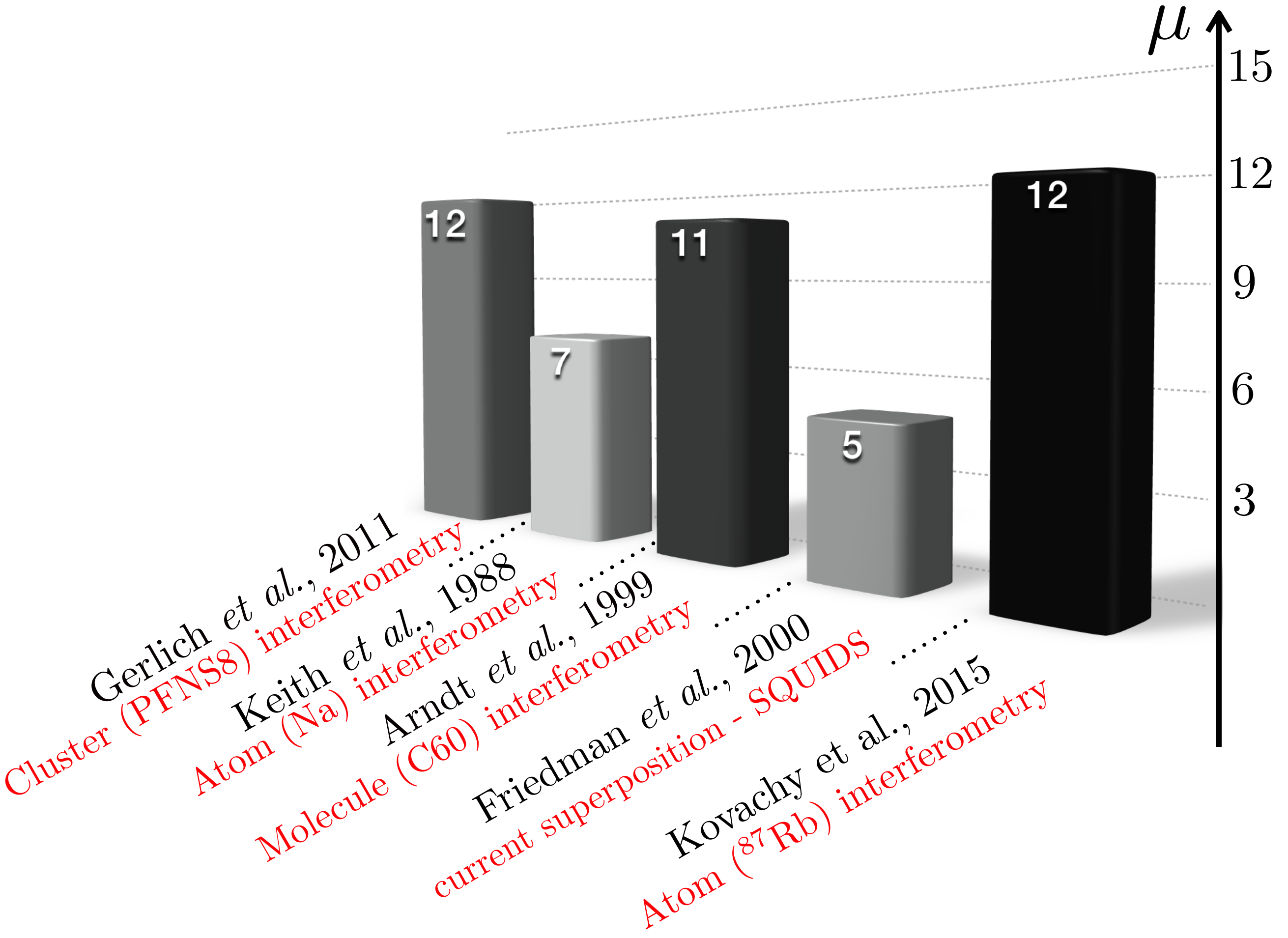}}
\caption[]{\label{fig:EffSizeMass}. Effective size of various states associated to massive systems evaluated with the measure in \citet{Nimmrichter_Macroscopicity_2013}. The numbers are taken from \citet{Nimmrichter_Macroscopicity_2013} except for the experiment reported in \citet{Kovachy_Quantum_2015} where the calculation is shown in the main text. }
\end{figure}

The measure of \citet{Nimmrichter_Macroscopicity_2013} is based on the capability of a state to test modifications of quantum theory. In case of interference experiments with objects of total mass $M$ whose expansion is much smaller than the path separation of the interferometer, the modification of the master equation shown in Eq. \eqref{eq:ME} leads to a decay of the coherence which scales as $(m_e/M)^2$ where $m_e$ is a reference mass taken as the mass of an electron. By comparing this decay with the actual period $t$ (in second) during which the coherence is maintained and taking the contrast $f$ of the observed interference pattern into account, a simple approximate expression is obtained for evaluating the effective size
\begin{equation}
\mu= \log_{10} \Big[\frac{1}{|\ln f|} \left(\frac{M}{m_e}\right)^2 t\Big].
\end{equation}
For example, for the experiment reported in \citet{Kovachy_Quantum_2015} where the wave packet of $^{87}$Rb atoms (86.91 a.m.u.) gets separated over 54cm during a drift time of $t=2.08$s and led to an interference pattern with a contrast of 28\%, the above expression gives to $\mu=10.6.$ Note that higher values have been obtained in the same experiment with smaller spatial separations. In particular, $\mu=12.3$ has been obtained for a separation of $\sim1$cm where the observed contrast is of 97.5\%. This value is comparable to interferometry experiments with cluster and molecules \cite{Arndt_Wave-particle_1999, Gerlich_Quantum_2011}. By describing the superposition of currents with displaced Fermi spheres of Cooper pairs, Nimmrichter and Hornberger found $\mu=5.2$ for the SQUID experiment reported in \citet{Friedman_Quantum_2000}, see \citet{Nimmrichter_Macroscopicity_2013} for details. These results are shown graphically in Fig. \ref{fig:EffSizeMass}.

\subsection{Summary}
\label{sec:summary-1}

Many attempts have been realized to create and detect macroscopic quantum states. Photonic experiments have been implemented both in the optical domain using parametric processes and in the microwave domain in the framework of cavity quantum electrodynamics. While the size of optical SCS has been improved by a factor $\approx$4 in the last decade now achieving $\alpha^2 \approx 3.2$, unprecedented sizes $\alpha^2 \approx 7.8$ has been achieved in the microwave domain. However, optical squeezed state holds the largest effective size for photonic states, according to a measure based on the quantum Fisher information \cite{Frowis_Measures_2012}. The same measure witnesses unprecedented sizes in spin systems, in particular in Bose-Einstein condensates where spin squeezed states have been obtained with very high squeezing parameters. Matter-wave interferometry and quantum optomechanics with more and more massive systems are also at the core of intensive experimental efforts.

According to a measure based on the capability of a state to test modifications of quantum theory \cite{Nimmrichter_Macroscopicity_2013}, very large effective size have been obtained almost 20 years ago, with molecule interferometry. Interestingly, such a measure also witness large effective size for single-atom interferometry where atomic wave packets gets coherently separated for seconds. As noted in \citet{Arndt_Testing_2014}, there is plenty of room for improvement and unprecedented sizes could be obtained in a near future using e.g. levitating nano-spheres. 

%%% Local Variables:
%%% mode: latex
%%% TeX-master: "master"
%%% End:

\section{Discussion and outlook}
\label{sec:discussions-outlooks}

In this review we have summarized different approaches to qualify and quantify the notion of macroscopic quantum superpositions and macroscopic quantum states. While we have seen that there are a multitude of proposals that differ strongly in their intuition and its formalization, many of the measures agree on some core features. In particular, it seem that the variance with respect to linear observables plays a key role in this respect. It is probably to early to say that we have obtained an agreement on a specific measure, or even all relevant or desirable features of such measures, but there has been significant progress in the last couple of years. As we have discussed in detail, there are many facets of the problem, and perhaps there is no single measure that takes all these aspects into account. Establishing a resource theory for quantum macroscopicity might seems to be an attractive avenue, but a good choice of free operations remains a challenge. Nevertheless, we are now much closer to be capable of judging experiments, or providing guidelines in which direction to go.

The latter is of particular importance given the multitude of fundamental limitations to prepare, maintain and measure macroscopic quantum states that have been identified (see Sec. \ref{sec:limits-observ-quant}). While certain macroscopic quantum states, in particular certain kinds of macroscopic superposition states, seem to be notoriously difficult to maintain and certify even within the framework of standard quantum mechanics, other states were identified where such limitations do not apply. It is still very challenging to perform experiments with such macroscopic quantum states, however there seem to be no principle obstacles to prevent us from observing quantum effects on much larger scales than today. In some cases only a relatively small change is required: working with relative degrees of freedom rather than absolute degrees of freedom suffices, or preparing two copies of a state enables one of them to act as a kind of self-reference.
Also encoded macroscopic states pose an interesting perspective. While the notion of macroscopic quantumness might slightly change, they possess the desired features on a coarse-grained level. If these insights can be harnessed for applications such a quantum metrology or quantum computation remains to be seen.

At the level of experiments, recent years have seen tremendous progress with different physical setups, bringing us ever closer to a true macroscopic regime. Which system is most suited to demonstrate truly macroscopic quantum effects depends on the goal one has in mind. However, we are now not only able to test the fundamental principles of quantum mechanics or its validity at larger and larger scales, but also harness some of its features for practical applications. Its perhaps the mixture of fundamental interest and the possibility of practical applications that makes the study of macroscopic quantum states so appealing. The road ahead still promises many challenges, but also new insights and surprises. While it seems that it is impossible to ever realize the thought experiment of Schr\"odinger -- performing experiments where the spirit of his proposal is maintained might at least plausible.

\section*{Acknowledgments.} We thank all colleagues with whom we have discussed about this topic over the last years. This work was supported by the European Research Council (ERC-AG MEC), the Swiss National Science Foundation through grant No. P300P2\_167749,~200021\_149109 and PP00P2\_150579 and through Quantum Science and Technology (QSIT) and the Austrian Science Fund (FWF) through project P28000-N27.

%%% Local Variables:
%%% mode: latex
%%% TeX-master: "master"
%%% End:

\bibliographystyle{apsrmp4-1}
\bibliography{References}

\end{document}